\documentclass[reprint,aps,prd,nofootinbib,superscriptaddress]{revtex4-2}
\usepackage[utf8]{inputenc}
\usepackage{amsmath}
\usepackage{amsfonts}
\usepackage{amsthm}
\usepackage{amssymb}
\usepackage{graphicx}
\usepackage[usenames,dvipsnames]{color}
\usepackage{hyperref}
\usepackage{mathrsfs}
\usepackage{orcidlink}

\theoremstyle{definition}

\theoremstyle{remark}

\newcommand{\norm}[1]{\left\Vert#1\right\Vert}
\newcommand{\abs}[1]{\left\vert#1\right\vert}
\newcommand{\set}[1]{\left\{#1\right\}}

\newcommand{\eps}{\varepsilon}

\newcommand{\tr}{\textrm{tr}}
\newcommand{\expec}[1]{\left\langle#1\right\rangle}
\newcommand{\del}{\boldsymbol{\nabla}}

\def\mathbi#1{\textbf{\em #1}}
\def\vk{\mathbi{k}}

\def\actaa{\ref@jnl{Acta Astron.}}      
\def\apj{\textrm{ApJ}}                 
\def\apjl{\textrm{ApJ}}                
\def\apjs{\textrm{ApJS}}               
\def\aap{\textrm{A\&A}}                

\def\jcap{\textrm{J. Cosmology Astropart. Phys.}}
\def\mnras{\textrm{MNRAS}}             
\def\prd{\textrm{Phys.~Rev.~D}}        
\def\pre{\textrm{Phys.~Rev.~E}}        
\def\prl{\textrm{Phys.~Rev.~Lett.}}    
\def\nat{\textrm{Nature}}              

\def\physrep{\textrm{Phys.~Rep.}}   

\usepackage[normalem]{ulem}

\begin{document}

\title{Gravitational Turbulence: the Small-Scale Limit of the Cold-Dark-Matter Power Spectrum}%
\author{Yonadav Barry Ginat\,\orcidlink{0000-0003-1992-1910}}%
\email{yb.ginat@physics.ox.ac.uk}%
\affiliation{Rudolf Peierls Centre for Theoretical Physics, University of Oxford, Parks Road, Oxford, OX1 3PU, United Kingdom}%
\affiliation{New College, Holywell Street, Oxford, OX1 3BN, United Kingdom}%
\author{Michael L. Nastac}
\affiliation{Rudolf Peierls Centre for Theoretical Physics, University of Oxford, Parks Road, Oxford, OX1 3PU, United Kingdom}%
\affiliation{St.~John's College, St.~Giles, Oxford, OX1 3JP, United Kingdom}%
\author{Robert J. Ewart}
\affiliation{Rudolf Peierls Centre for Theoretical Physics, University of Oxford, Parks Road, Oxford, OX1 3PU, United Kingdom}%
\affiliation{Balliol College, Broad Street, Oxford, OX1 3BJ, United Kingdom}%
\affiliation{Department of Astrophysical Sciences, Princeton University, Peyton Hall, Princeton, NJ 08544, United States}%
\author{Sara Konrad}
\affiliation{Institute for Theoretical Physics, Heidelberg University, Germany}
\author{Matthias Bartelmann}
\affiliation{Institute for Theoretical Physics, Heidelberg University, Germany}
\author{Alexander A. Schekochihin}
\affiliation{Rudolf Peierls Centre for Theoretical Physics, University of Oxford, Parks Road, Oxford, OX1 3PU, United Kingdom}%
\affiliation{Merton College, Merton Street, Oxford, OX1 4JD, United Kingdom}%

\begin{abstract}
The matter power spectrum, $P(k)$, is one of the fundamental quantities in the study of large-scale structure in cosmology. Here, we study its small-scale asymptotic limit, and show that for cold dark matter in $d$ spatial dimensions, $P(k)$ has a universal $k^{-d}$ asymptotic scaling with the wave-number $k$, for $k \gg k_{\rm nl}$, where $k_{\rm nl}^{-1}$ denotes the length scale at which non-linearities in gravitational interactions become important. We propose a theoretical explanation for this scaling, based on a non-perturbative analysis of the system's phase-space structure. Gravitational collapse is shown to drive a turbulent phase-space flow of the quadratic Casimir invariant, where the linear and non-linear time scales are balanced, and this balance dictates the $k$ dependence of the power spectrum. A parallel is drawn to Batchelor turbulence in hydrodynamics, where large scales mix smaller ones via tidal interactions. The~$k^{-d}$ scaling is also derived by expressing~$P(k)$ as a phase-space integral in the framework of kinetic field theory, which is analysed by the saddle-point method; the dominant critical points of this integral are precisely those where the time scales are balanced. The coldness of the dark-matter distribution function---its non-vanishing only on a~$d$-dimensional sub-manifold of phase-space---underpins both approaches. The theory is accompanied by $1\mathrm{D}$ Vlasov--Poisson simulations, which confirm it.
\end{abstract}

\maketitle


\section{Introduction}
\label{sec:introduction}
One of the basic observable quantities in the study of the large-scale structure of the Universe is the two-point correlation function of the density field, whose Fourier transform is the power spectrum, $P(k,t)$ \cite{Peebles1980,Dodelson:2003ft}. The two-point correlation function is of fundamental importance, for it allows us to probe theories of the early Universe, dark matter, inflation, and to study gravity \cite{Peebles1980,Desjacquesetal2018,Planck2020,TegmarkZaldarriaga2002}.
In this paper, we will explore the small-scale asymptotic behaviour of $P(k)$ (we omit the explicit time dependence when it is not confusing to do so), in the limit~$k\gg k_{\rm nl}$, where the (inverse) non-linear scale~$k_{\rm nl}$ is defined by~$k_{\rm nl}^3 P_{\rm lin}\left(k_{\rm nl}\right) = 2\pi^2$ \cite{Cabassetal2023}, with $P_{\rm lin}$ denoting the linearly evolved over-density power spectrum. The small-scale limit of $P(k)$ is theoretically important for the understanding of the gravitational $N$-body problem in the large-$N$ limit~\cite{HeggieHut2003}, and the formation of large-scale structure, non-linear clustering and self-similarity~\cite{Peebles1980,Hamiltonetal1991}, but also for the understanding of the nature of dark matter and gravitational back-reaction of small scales on large ones---both relativistic~\cite{Ginat2021} and in the context of the effective field theory of large-scale structure~\cite{Baumannetal2012} (for reviews and references see, e.g.,~\cite{Baldauf2020,Cabassetal2023,Ivanov2023}) and the general bias expansion~\cite{Desjacquesetal2018}. Even before modifying gravity, it is important to know what non-linear phenomena occur in the standard theory. 

Data from cosmological dark-matter-only simulations are consistent with $P(k,t)$ developing a $k^{-3}$ tail on small scales \cite[e.g.,][fig. 6]{Springeletal2018} in $3\mathrm{D}$, and a $k^{-1}$ tail in $1\mathrm{D}$, as shown in Ref.~\cite[][figure 6]{TaruyaColombi2017} (cf.~\cite{Jenkinsetal1998,Springeletal2005,Moczetal2020,ChenPietroni2020}). The emergence of a power-law tail---and the simplicity of its exponent---hint that a fundamental physical reason for it must exist, ultimately stemming from the nature of the gravitational interaction of cold dark matter. Here, we will describe the mechanism that produces this asymptotic scaling, by studying the mass distribution in the velocity space as well as in position space.

We restrict ourselves to the strictly collisionless case where the particle mass $m\to 0$, while the total particle number $N \to \infty$, so that $M \equiv Nm$ remains fixed (and so does the volume). In this limit, the phase-space distribution of particles is well described by the Vlasov equation~\cite{Gilbert1968,Peebles1980,Chavanis2012,LazaroviciPickl2017,TaruyaColombi2017,Halleetal2019,Rampf2021}:\footnote{The collisionless Vlasov equation, of course, ignores dissipation via collisions (equivalently, finite-$N$ effects). However, we will find that due to turbulence, such dissipation will inevitably be accessed; this point will be discussed further in appendix \ref{coll_numerics}.}
\begin{equation}\label{eqn:Vlasov}
\frac{\partial f}{\partial \eta} + \mathbf{v}\cdot\frac{\partial f}{\partial \mathbf{x}} + \mathbf{g} \cdot\frac{\partial f}{\partial \mathbf{v}} = 0,
\end{equation}
where $f$ is the distribution function (the one-point probability density in phase-space), $\eta$ is the conformal time, defined by $\mathrm{d}t = a\mathrm{d}\eta$, where $a$ is the scale factor of the background (which is taken to be a Friedmann--Lema\^{i}tre--Robertson--Walker, FLRW, space-time), $\mathbf{x}$ and $\mathbf{v}$ are the co-moving position and peculiar velocity, and the gravitational field $\mathbf{g}$ is self-consistently determined by Poisson's equation
\begin{equation}\label{eqn:Poisson}
\del \cdot \mathbf{g} \equiv -\nabla^2 \Phi = -4\pi Ga^2\left[\int f(\mathbf{x},\mathbf{v},\eta) \mathrm{d}^dv - \overline{\rho}_m\right],
\end{equation}
where $\overline{\rho}_m$ is the mean background matter density and~$\Phi$ is the gravitational potential. The gravitational field $\mathbf{g}$ thus defined gives $\mathrm{d}(a\mathbf{v})/\mathrm{d}t$, when evaluated along a particle's trajectory, and $\mathbf{v}$ is defined so that the $4$-velocity in space-time co-ordinates $x^\mu = (\eta,\mathbf{x})$ is $u^\mu = (1-\Phi/c^2,\mathbf{v})/a$.
This system of equations applies on scales much smaller than the horizon, for particles much slower than the speed of light,~$c$. Henceforth, we will work in general spatial dimensions~$d \in \set{1,2,3}$, to enable comparison between our theory and our~$d=1$ simulations (or future $d=2$ ones).

Dark matter is assumed to be cold initially, with a thermal velocity $v_{\rm th} \to 0$, so that the initial gravitational potential energy is much larger than the initial thermal energy (this assumption is excellent for our Universe \cite{Greenetal2004,LoebZaldarriage2005,Kunzetal2016,Thomasetal2016,Gilmanetal2020,Planck2020,Ilicetal2021}).
Despite being cold, dark matter is inherently kinetic, and cannot be described by fluid equations adequately on non-linear scales, because these equations cease to be valid after streams cross~\cite{Menci2002,Kozlikinetal2021,Rampf2021}. 

Below we will derive the $k \gg k_{\rm nl}$ limit of $P(k)$ in two complementary ways. First, we will study the problem in phase-space, and show that the small-scale asymptotics arise when the time scales involved in the Vlasov equation balance with each other in a particular way, to be explained below. Two ingredients will comprise this argument: the balance of time scales in the Vlasov equation during gravitational collapse, and the conservation of the second Casimir invariant \cite{Knorr1977,Diamond_Itoh_Itoh_2010,LesurDiamond2013,Nastacetal2023,Nastacetal2024}
\begin{equation}\label{eqn:C2 definition}
C_2 \equiv \iint f^2 \mathrm{d}^dx\mathrm{d}^dv.
\end{equation}
In plasma physics, this invariant is sometimes referred to as `enstrophy' \cite{Knorr1977}, `phasestrophy' \cite{Diamond_Itoh_Itoh_2010}, or `$f$-strophy' \cite{AlexKTnotes}.
The Vlasov equation conserves an infinite number of phase-space invariants---not only $C_2$---but we will use $C_2$ here because it is directly related to the power spectrum (\emph{vide infra}). We will show that the spectrum characterises a turbulent cascade (i.e., flow) of $C_2$ from large scales to small ones. The system's evolution is non-linear, and involves a mixing in phase-space with gravitational collapses on many scales. Gravitational turbulence is different from other types of turbulence in that the flux of $C_2$ is not constant over the range of scales of interest, but rather will turn out to be scale-dependent, and yet there is a universal scaling r\'{e}gime of the phase-space power spectrum; this is in contrast to the usual situation in turbulent systems \cite{Kolmogorov1941,Batchelor1959,Nastacetal2023,Nastacetal2024}, where there exists an inertial range of scales with a constant-flux cascade of an energy-like quadratic invariant.


For the second, alternative approach, we use non-perturbative kinetic field theory (KFT) for cosmic structure formation (for a review, see \cite{KonradBartelmann2022}). Here,
$P(k)$ is expressed as an integral over the initial particle positions and velocities (weighted by the initial-condition probability distribution) of the characteristic function of the displacement field; this integral will turn out to have an explicit $k$ dependence, and we will utilise this to perform an asymptotic saddle-point analysis.

The two approaches complement each other, both relying on the same assumptions, but highlighting their r\^{o}les in different ways. We remark that using phase-space expressions ensures that the validity of our results extends beyond the bounds of configuration-space-based approaches, such as Lagrangian \cite{Zeldovich1970,ShandarinZeldovich1989,ChenPietroni2020} or Eulerian \cite{Menci2002} techniques; in particular, it is regular at stream-crossing, and accounts for free streaming automatically. Indeed, as we already mentioned above, the inherently kinetic phenomenon of multi-scale structure of the distribution of dark matter, developing by virtue of strongly non-linear interactions, suggests that a type of turbulence in phase-space is involved. 
As early as \cite{ShandarinZeldovich1989,Gaite_2012}, it was realised that the phenomena of gravitation and turbulence might be linked---here we make the analogy precise and characterise this gravitational turbulence.

The rest of this paper is organised as follows: in \S \ref{sec:Phase-Space Cascade}, we formulate the first approach, based on the Vlasov--Poisson system, and use a phase-space analysis of gravitational collapse to derive the small-scale asymptotics of the phase-space power spectrum, from which $P(k)$ may be computed, and then describe the phase-space cascade that it supports. In \S \ref{sec:saddle-point approach}, as promised above, we derive the same asymptotic scaling of $P(k)$ again, but via a saddle-point analysis of an integral expression for $P(k)$. We test our theory by comparing it with numerical Vlasov--Poisson simulations throughout the paper. Our conclusions are discussed in \S \ref{sec:discussion} and summarised in \S \ref{sec:summary}.


\section{Phase-space turbulence}
\label{sec:Phase-Space Cascade}
We start by describing the initial condition in \S \ref{subsec:cold}---a cold stream or a collection of streams---and show what its evolution looks like, and then derive the equation that governs the phase-space Fourier transform of~$f$ in~\S \ref{subsec:Batchelor Fourier}. We then analyse the time scales involved in the Vlasov--Poisson system in~\S\ref{subsec:critical balance}, and show how gravitational collapse via the Jeans instability~\cite{Jeans1902} imposes stringent conditions on these time scales, and on the form of the power spectrum, in \S\ref{subsubsec: Jeans stream}. We describe in~\S\ref{subsec:phase space cascade} the phase-space turbulence that characterises the dynamics. To do that, we derive a transport equation for the integrand of $C_2$ in the Fourier dual of the phase-space. This has a (non-linear) source term, which is found to receive contributions from all larger scales in~\S\ref{subsec: source CDM}. We present simulation results throughout this section, to test the theory.

\subsection{Cold streams}
\label{subsec:cold}
We assume that the system has an initial condition consisting of a superposition of streams, each one of the form
\begin{equation}\label{eqn:initial condition f}
f(\eta=0,\mathbf{x},\mathbf{v}) =  f_{\rm in} (\mathbf{x},\mathbf{v})\equiv \frac{\rho_{\rm in}(\mathbf{x})}{(2\pi v_{\rm th}^2)^{d/2}} \mathrm{e}^{-\frac{\left[\mathbf{v}-\mathbf{u}_{\rm in}(\mathbf{x})\right]^2}{2v_{\rm th}^2}},
\end{equation}
where~$v_{\rm th}^2 \ll \min\set{u_{\rm in}^2,\int \abs{\Phi_{\rm in}}f_{\rm in} \mathrm{d}^dx \mathrm{d}^dv}$, where $\Phi_{\rm in}$ is derived from $\rho_{\rm in}$ via equation \eqref{eqn:Poisson}; that is, the initial condition is very cold---the thermal energy is negligible in comparison with the gravitational potential energy of the system or the kinetic energy of mean flows. In equation \eqref{eqn:initial condition f}, $\rho_{\rm in}$ and $\mathbf{u}_{\rm in}$ are typically Gaussian random fields \cite{Peebles1980,Desjacquesetal2018}, and $\mathbf{u}_{\rm in}$ is a gradient flow \cite{Dodelson:2003ft}, i.e.,~$\del \times \mathbf{u}_{\rm in}=0$. We also choose the co-ordinates so that $\int \rho(\mathbf{x}) \mathbf{u}(\mathbf{x}) \mathrm{d}^dx = 0$, where $\rho(\mathbf{x})$ is the density and $\rho \mathbf{u} \equiv \int \mathbf{v}f\mathrm{d}^dv$. The distribution \eqref{eqn:initial condition f} is essentially a single stream, and, in fact, its Maxwellian shape may be approximated by a Dirac delta-function. The thermal speed $v_{\rm th}$ is chosen to be the smallest velocity scale in the problem, so the entire analysis of this paper focuses on the limit where~$ sv_{\rm th} \ll 1$, where $s$ is the Fourier conjugate to velocity (see \S \ref{subsec:Batchelor Fourier}). 
The distribution function remains a collection of streams as it evolves in time, by Liouville's theorem~\cite[e.g.][]{Arnoldetal2006}: locally almost everywhere in phase-space, i.e., in a sufficiently small neighbourhood of almost any point~$(\mathbf{x}_0,\mathbf{v}_0)$ where~$f\neq 0$, it can be written as~$\rho_{\mathbf{x}_0,\mathbf{v}_0}(\mathbf{x},\eta) \delta^{\rm D}\left(\mathbf{v}-\mathbf{u}_{\mathbf{x}_0,\mathbf{v}_0}(\mathbf{x},\eta)\right) + O(v_{\rm th})$, where $\delta^{\rm D}$ is the Dirac delta-function; it remains so as long as collisions or finite-$N$ effects may be ignored. The local functions~$\rho_{\mathbf{x}_0,\mathbf{v}_0}(\mathbf{x},\eta)$ and~$\mathbf{u}_{\mathbf{x}_0,\mathbf{v}_0}(\mathbf{x},\eta)$ might in general differ from the mean density~$\rho(\mathbf{x},\eta)$ and the mean velocity~$\mathbf{u}(\mathbf{x},\eta)$.

To study the evolution of the initial condition \eqref{eqn:initial condition f}, we have conducted a suite of Vlasov--Poisson simulations in~$1\mathrm{D}$ on a static background~($a(\eta)=1$), using the Gkeyll code \cite{JUNO2018110}, which is originally a Vlasov solver for kinetic plasmas; by setting the vacuum permittivity to~$\eps_0 < 0$, we can use the code to study gravity (as opposed to electrostatics). Details on the numerical method and simulation set-up may be found in appendix~\ref{appendix: simulation set up}. The simulation's units are chosen so that~$\tau_0^{-2} \equiv 4\pi GM/L = 1$ (unit of time),\footnote{Note that Poisson's equation \eqref{eqn:Poisson} implies that $G$ has different dimensions in~$1\mathrm{D}$ from $3\mathrm{D}$.}~$k_0 \equiv 2\pi/L = 1$ (unit of length) and~$v_0 \equiv (k_0\tau_0)^{-1} = 1$ (unit of speed), where~$L$ is the length of the simulation `box' (with periodic boundary conditions) and $M$ is the total mass in the box.

\begin{figure*}
\centering
\includegraphics[width=\textwidth]{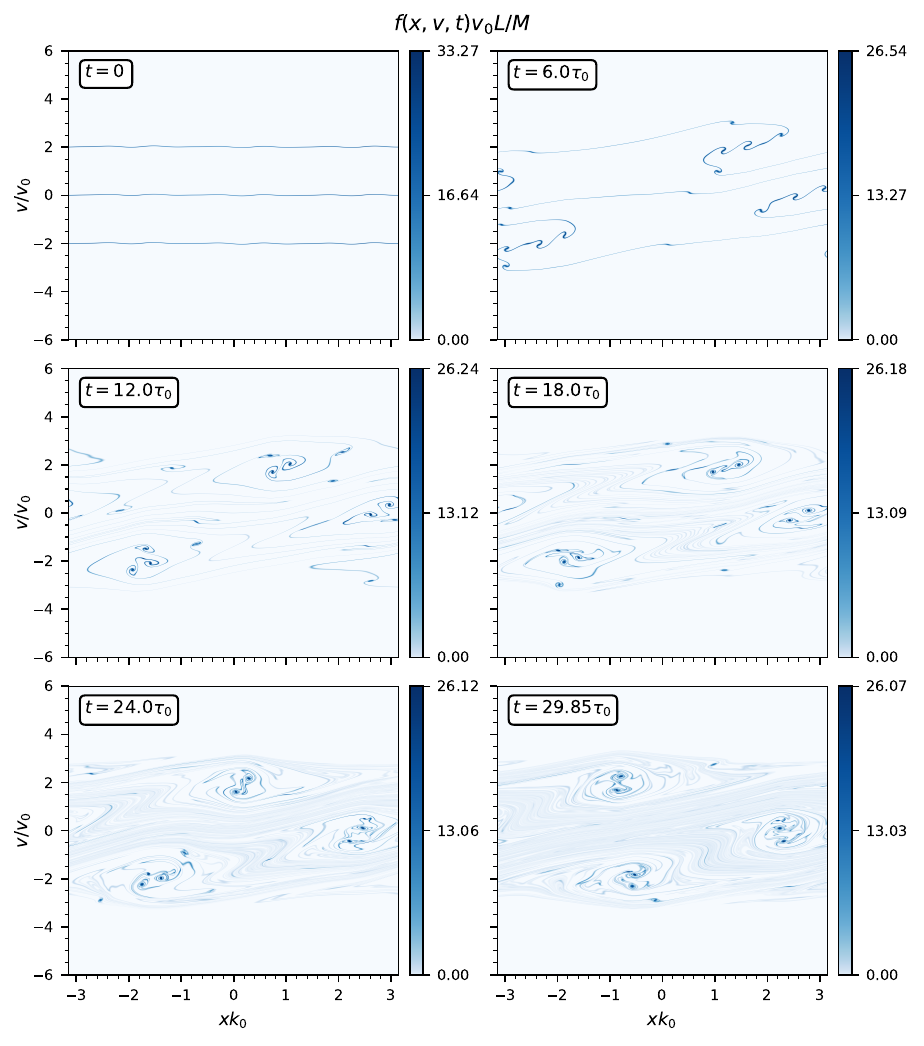}
\caption{Colour plots of the distribution function showing the time evolution of three cold streams, with $v_{\rm th}/v_0 = 5\times 10^{-3}$. The colour-bar in each panel extends from $0$ to $\max f$ at the corresponding time in the simulation. See text and appendix \ref{appendix: simulation set up} for details. A video is available in this \href{https://ybginat.com/index.php/gravitational-turbulence/}{link}.}
\label{fig:results time evolution}
\end{figure*}

The time evolution of a cold system of three streams---three copies of equation \eqref{eqn:initial condition f}---is displayed in figure \ref{fig:results time evolution}.
This figure shows that each stream is distorted quickly, by rotating and twisting in phase-space.
Evidently, this motion generates small-scale structure, which we consider to be a type of turbulence in phase-space. As usual, in order to characterise the turbulence and its spectrum of fluctuations, one must identify an invariant quantity that cascades to small scales, what the cascade's time scale is, and also the flux of that invariant quantity \cite{Frisch_1995} (which will not be scale-independent in our case). We will do so in \S\S \ref{subsec:phase space cascade}, \ref{subsec:critical balance}, and \ref{subsec: source CDM}, respectively.

That the turbulence is in phase-space and not merely in position space is manifest in figure \ref{fig:results time evolution}: already at~{${t=6\tau_0}$}, the system can no longer be described as three streams almost anywhere, so standard cosmological perturbation theory would already be inadequate, and the root-mean-square (rms) peculiar velocity $v_{\rm rms} \equiv \left[\int \mathrm{d}v\left(v-u(x)\right)^2f(x,v)/\rho(x)\right]^{1/2}$ is much larger than~$v_{\rm th}$, and is of order $v_0$. However, in phase-space, we see that the topology of three single lines is preserved. 

As explained above, the fact that the system is a collection of streams implies that \emph{locally in phase-space}, one can describe each stream with fluid equations: inserting~$\rho_{\mathbf{x}_0,\mathbf{v}_0}(\mathbf{x},\eta) \delta^{\rm D}\left(\mathbf{v}-\mathbf{u}_{\mathbf{x}_0,\mathbf{v}_0}(\mathbf{x},\eta)\right)$ into the Vlasov equation gives the continuity and Euler equations for~$\rho_{\mathbf{x}_0,\mathbf{v}_0}$ and $\mathbf{u}_{\mathbf{x}_0,\mathbf{v}_0}$, respectively, and the divergence of the latter yields the Raychaudhuri equation \cite{BertschingerJain1994},
which describes gravitational collapse under the Jeans instability \cite{Jeans1902,Henon1973,Zeldovich1970,FillmoreGoldrecih1984,Bertschinger1985,Barnesetal1986,Binney2004,ColombiTouma2014,White2014,Rampf2021}. Written in the frame of reference that moves with a given stream, it reads
\begin{equation}\label{eqn:Raychaudhuri}
\frac{\mathrm{d}\theta}{\mathrm{d}\eta} + \mathcal{H}\theta + \frac{\theta^2}{3}+ \sigma^{ij}\sigma_{ij} = \del \cdot \mathbf{g} + \frac{\Sigma}{\rho_{\mathbf{x}_0,\mathbf{v}_0}} + 2\omega_i\omega^i,
\end{equation}
where $\mathcal{H}$ is the conformal Hubble constant,~$\theta \equiv \del \cdot \mathbf{u}_{\mathbf{x}_0,\mathbf{v}_0}$ is the stream's divergence,~$\boldsymbol{\omega} \equiv \del \times \mathbf{u}_{\mathbf{x}_0,\mathbf{v}_0}/2$ is its vorticity, $\sigma_{ij} \equiv \left[\partial_i(u_{\mathbf{x}_0,\mathbf{v}_0})_j + \partial_j (u_{\mathbf{x}_0,\mathbf{v}_0})_i - 2\theta \delta_{ij}/3\right]/2$ is its shear (rate of strain), $\mathrm{d}/\mathrm{d}\eta$ is the Lagrangian derivative along the stream,~$\Sigma/\rho_{\mathbf{x}_0,\mathbf{v}_0}$ is a pressure-related term that is~$O(v_{\rm th}^2)$, and $\mathbf{g}$ is the gravitational field felt by the stream. We will use equation \eqref{eqn:Raychaudhuri} in \S\ref{subsubsec: Jeans stream} to estimate the time scale $\tau_{\rm J}$ of gravitational collapse, due to the Jeans instability.

\subsection{Spectral transport equations}
\label{subsec:Batchelor Fourier}
The evolution of the system is inherently a multi-scale process, so it is more convenient to study equations \eqref{eqn:Vlasov}-\eqref{eqn:Poisson} in Fourier space.
\subsubsection{Fourier transform}
\label{subsubsec: Fourier definitions}
Let us define the Fourier transform (marked by a circumflex) as follows:
\begin{equation}\label{eqn: Fourier transform definition}
\hat{f}(\mathbf{k},\mathbf{s}) \equiv \iint f(\mathbf{x},\mathbf{v}) \mathrm{e}^{\mathrm{i}\mathbf{k} \cdot \mathbf{x} - \mathrm{i}\mathbf{s} \cdot \mathbf{v}}\mathrm{d}^dx \mathrm{d}^dv,
\end{equation}
and similarly for all other functions of $(\mathbf{x},\mathbf{v})$. For~${d>1}$, we denote $k \equiv \abs{\mathbf{k}}$, $s \equiv \abs{\mathbf{s}}$, \emph{etc}.
Under this Fourier transform, the Vlasov--Poisson system (\ref{eqn:Vlasov}--\ref{eqn:Poisson}) becomes
\begin{align}
& \frac{\partial \hat{f}}{\partial \eta} + \mathbf{k} \cdot \frac{\partial \hat{f}}{\partial \mathbf{s}} + \mathrm{i} \mathbf{s} \cdot \int \frac{\mathrm{d}^d k'}{(2\pi)^d} \hat{\mathbf{g}}(\mathbf{k}')\hat{f}(\mathbf{k}-\mathbf{k}',\mathbf{s}) = 0 \label{eqn: Vlasov Fourier space},\\ &
k^2 \hat{\Phi} = -4\pi G a^2\hat{\rho},
\end{align}
where $\hat{\rho}$ is the Fourier-transformed density.
The second Casimir invariant, defined by equation \eqref{eqn:C2 definition}, is given in Fourier space by Parseval's theorem:
\begin{equation}\label{eqn: C2 Fourier}
C_2 = \frac{1}{(2\pi)^{2d}}\iint |\hat{f}|^2 \mathrm{d}^dk\mathrm{d}^ds.
\end{equation}
The integrand $|\hat{f}^2|$ is directly related to the density power spectrum. Indeed, let the \emph{phase-space power spectrum} be
\begin{equation}\label{eqn: F definition}
\hat{F}(\mathbf{k},\mathbf{s}) \equiv \langle |\hat{f}|^2 \rangle,
\end{equation}
where $\langle \cdot \rangle$ is an ensemble average over many random realisations of the initial conditions; then the density power spectrum is
\begin{equation}
P(k) \equiv \frac{1}{V}\expec{\hat{\rho}(\mathbf{k})\hat{\rho}^*(\mathbf{k})} = \frac{\hat{F}(\mathbf{k},0)}{V},
\end{equation}
where $V$ is the spatial volume.
\begin{widetext}
\subsubsection{Transport of $C_2$}
Let us now see how the integrand of the second Casimir invariant \eqref{eqn: C2 Fourier} evolves:
multiplying equation~\eqref{eqn: Vlasov Fourier space} by~$\hat{f}^*$ and taking the real part, we find

\begin{equation}\label{eqn: f^2 transport equation}
	\frac{\partial  |\hat{f}|^2}{\partial \eta} + \mathbf{k} \cdot \frac{\partial |\hat{f}|^2}{\partial \mathbf{s}} + \mathrm{i}\mathbf{s}\cdot \int \frac{\mathrm{d}^dk'}{(2\pi)^d} \left[\hat{\mathbf{g}}(\mathbf{k}')\hat{f}^*(\mathbf{k},\mathbf{s})\hat{f}(\mathbf{k}-\mathbf{k}',\mathbf{s}) - \hat{\mathbf{g}}^*(\mathbf{k}') \hat{f}(\mathbf{k},\mathbf{s})\hat{f}^*(\mathbf{k}-\mathbf{k}',\mathbf{s}) \right] = 0.
\end{equation}
Integrating equation \eqref{eqn: f^2 transport equation} over all $\mathbf{k}$ and $\mathbf{s}$ yields the conservation of $C_2$, as it should.

Let $\delta g_r$ be the amplitude of a typical fluctuation in $\mathbf{g}$ on scales $r = 1/k$ and smaller, \emph{viz.},
\begin{equation}\label{eqn:delta g r definition}
	\delta g_{r}^2 \equiv  \frac{2}{V}\int \frac{\mathrm{d}^dk'}{(2\pi)^d} ~\expec{|\hat{\mathbf{g}}(\mathbf{k'})|^2}\left(1-\mathrm{e}^{\mathrm{i}\mathbf{k}'\cdot \mathbf{r}}\right) \sim \frac{1}{V}\int_{k'>k} \!\!\!\!\!\!\mathrm{d}^dk' ~\expec{|\hat{\mathbf{g}}(\mathbf{k'})|^2}.
\end{equation}
Intuitively, $\delta g_r$ describes the amplitude of a 
typical tidal field on a scale $r$.
We take~$\mathbf{g}$ to be a sufficiently continuous field, so that for sufficiently small values of~$r$,
\begin{equation}\label{eqn:delta g r with lambda}
	\delta g_r \sim \kappa r^\lambda + \textrm{h.o.t.}
\end{equation}
with H\"{o}lder exponent~$0< \lambda \leq 1$, and~$\kappa$ a constant coefficient. A smooth gravitational field has~$\lambda = 1$, because one may Taylor-expand~$\mathbf{g}$, meaning that the fluctuations are dominated by tidal forces (and $\kappa$ has dimensions of [time]$^{-2}$).\footnote{The case~$\lambda > 1$ is also smooth, but highly atypical, where the tidal forces vanish; while this could happen in isolated points with exactly zero over-density, we ignore it here.} 
We prove in \S\ref{subsec:critical balance} that $\lambda = 1$ follows from Poisson's equation \eqref{eqn:Poisson}, in conjunction a balance of time scales, in $d\leq 3$ spatial dimensions, but, for the purposes of the qualitative discussion that follows, we will take $\lambda = 1$ now, \emph{a priori}.

Let us now take the large-$k$ limit of equation \eqref{eqn: f^2 transport equation}. In this limit, for a smooth gravitational field ($\lambda = 1$), the last term on the left-hand side is dominated by two contributions: (i) $k' \ll k$ and (ii) $\abs{\mathbf{k}-\mathbf{k}'} \ll k$. These are sometimes referred to as `squeezed' triangles; case (i) is also known as the Batchelor limit \cite{Batchelor1959,AdkinsSchekochihin2018,Sreenivasan2019,Nastacetal2023}. This approximation is justified by its subsequent success in producing a self-consistent solution (a spectrum that is steep enough for $\lambda$ to be $1$). In figure~\ref{fig:three point function}, it is validated for the simulation described in \S\ref{subsec:cold}.

\begin{figure*}
	\centering
	\includegraphics[width=0.49\textwidth]{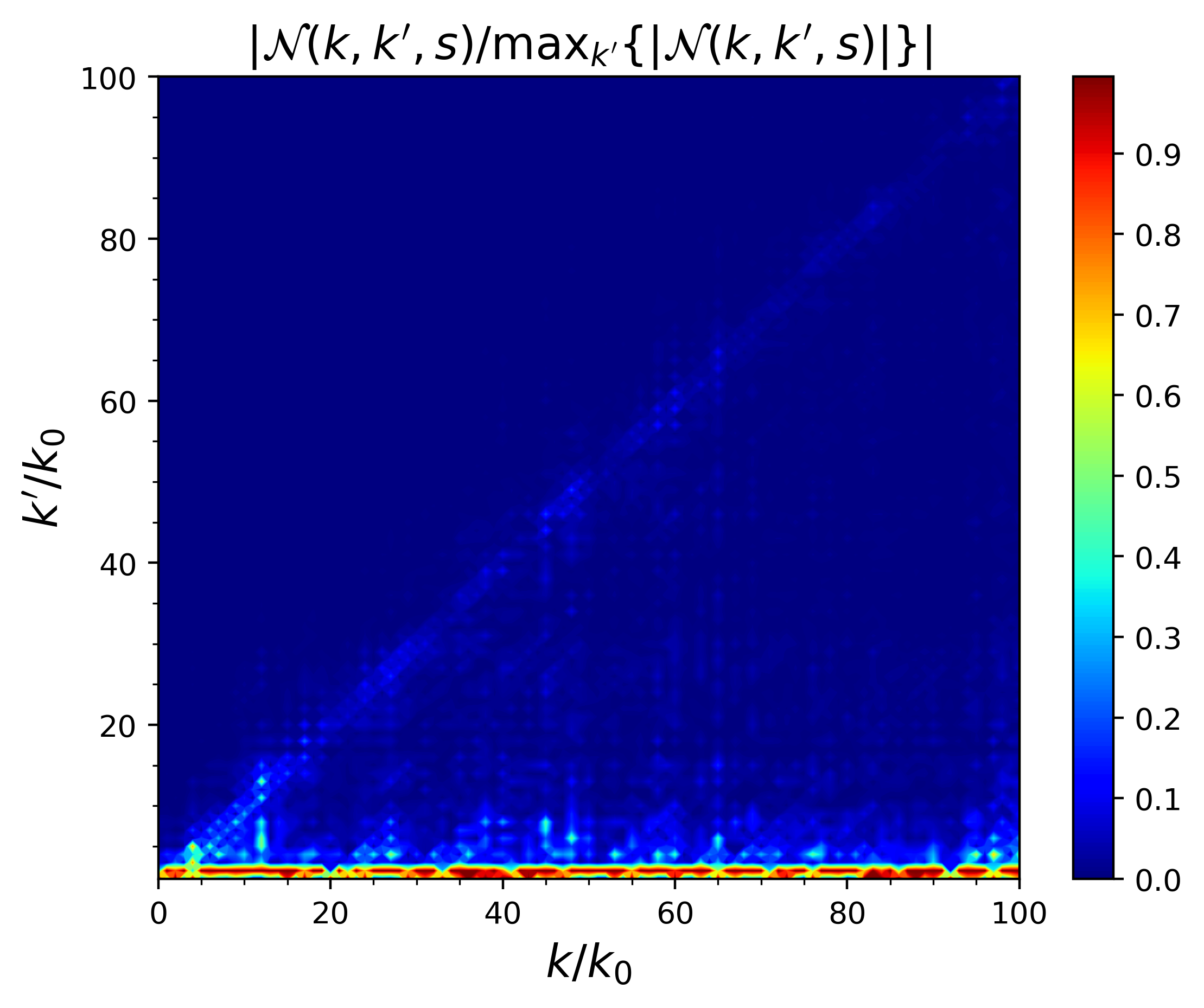}
	\includegraphics[width=0.49\textwidth]{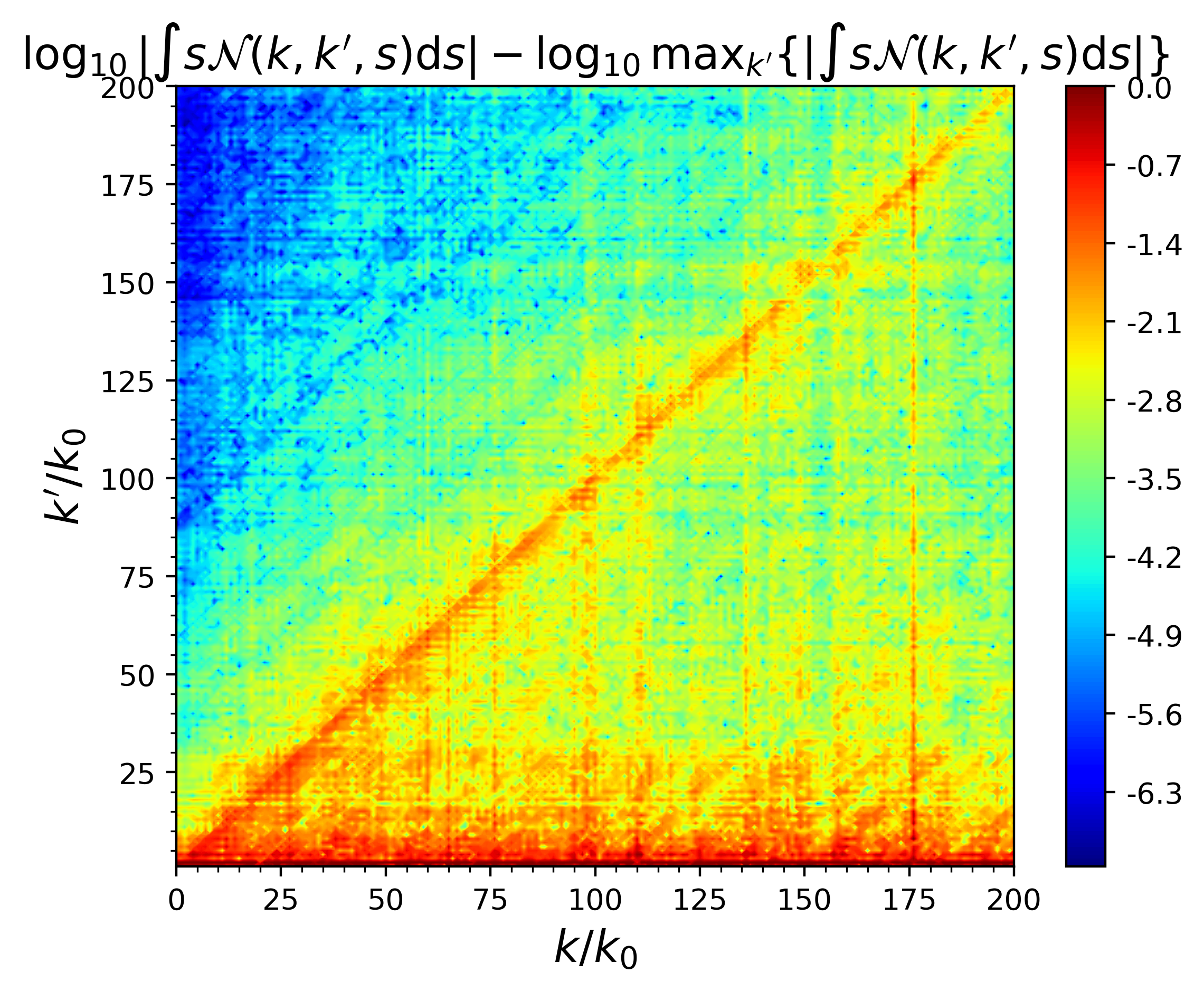}
	\caption{Plots of the contributions to the non-linear term in equation~\eqref{eqn: f^2 transport equation}, for the simulation shown in figure \ref{fig:results time evolution}. Defining $\mathcal{N}(\mathbf{k},\mathbf{k}',\mathbf{s}) \equiv \expec{\Im \left[ \hat{\mathbf{g}}(\mathbf{k}')\hat{f}(\mathbf{k}-\mathbf{k}',\mathbf{s})\hat{f}^*(\mathbf{k},\mathbf{s})\right]}_{\rm T}$, the \emph{left} panel displays $\abs{\mathcal{N}}/\max_{k'}\set{|\mathcal{N}|}$, at $sv_0 = 0.52$, as a function of both $k$ and $k'$ (the average is calculated as in \S \ref{subsec:phase space cascade}). The Batchelor peak is very prominent at low $k'$, and another peak is visible on the diagonal $k' = k$, corresponding to the limit (ii) of $\mathcal{N}$. In this paper we only require the limit (ii) when integrated over $s$ (in \S \ref{subsec: source CDM} below); to gauge the validity of the approximation in that case, we show in the \emph{right} panel shows the integral of $s\mathcal{N}$ over $0\leq s \leq k\tau_0/2$ (in logarithmic scale), normalised in the same way, exhibiting the same two prominent peaks. }
	\label{fig:three point function}
\end{figure*}

Taylor-expanding $\hat{f}(\mathbf{k}-\mathbf{k}',\mathbf{s})$ in the Batchelor limit turns the non-linear term in \eqref{eqn: f^2 transport equation} into
\begin{align}
	& \nonumber \mathrm{i}\mathbf{s}\cdot \int \frac{\mathrm{d}^dk'}{(2\pi)^d} \left[\hat{\mathbf{g}}(\mathbf{k}')\hat{f}^*(\mathbf{k},\mathbf{s})\left(\hat{f}(\mathbf{k},\mathbf{s}) - \frac{\partial \hat{f}}{\partial k^i}k'^i \right) - \hat{\mathbf{g}}^*(\mathbf{k}') \hat{f}(\mathbf{k},\mathbf{s})\left(\hat{f}^*(\mathbf{k},\mathbf{s}) - \frac{\partial \hat{f}^*}{\partial k^i}k'^i \right) \right] + \textrm{h.o.t.} \\ &
	\nonumber \simeq -\mathrm{i}\int \frac{\mathrm{d}^dk'}{(2\pi)^d} s^j\left[\hat{g}_j(\mathbf{k}')k'^i\hat{f}^*(\mathbf{k},\mathbf{s})\frac{\partial \hat{f}}{\partial k^i} - \hat{g}_j^*(\mathbf{k}') k'^i\hat{f}(\mathbf{k},\mathbf{s})\frac{\partial \hat{f}^*}{\partial k^i}\right] \\ &
	= -\mathrm{i}\int \frac{\mathrm{d}^dk'}{(2\pi)^d} s^j\hat{g}_j(\mathbf{k}')k'^i\left[\hat{f}^*(\mathbf{k},\mathbf{s})\frac{\partial \hat{f}}{\partial k^i} + \hat{f}(\mathbf{k},\mathbf{s})\frac{\partial \hat{f}^*}{\partial k^i}\right]  \equiv -s^j \Phi^i_j\frac{\partial |\hat{f}|^2}{\partial k^i},\label{eqn:derivation of Batchelor term}
\end{align}
where we have defined $\Phi^i_j \equiv \mathrm{i}\int k'^i\hat{g}_j \mathrm{d}^dk'/(2\pi)^d$---the matrix {${\Phi^i_j = \delta^{in}\partial_n\partial_j\Phi}$} is the Hessian matrix of the gravitational potential, i.e., the tidal matrix. The transition from the first line to the second in \eqref{eqn:derivation of Batchelor term} is valid because, in the centre-of-mass frame, the leading-order terms are proportional to $\int \mathrm{d}^dk' \hat{\mathbf{g}}(\mathbf{k}') = 0$.
Thus, the part of the non-linear term in the limit (i) reduces to a tidal interaction: on a given (small) scale corresponding to a (large) wave-number $k$, the distribution function $f$ is distorted by the gravitational field at the same (large) energy-containing scale.

We now have, from equations \eqref{eqn: f^2 transport equation} and \eqref{eqn:derivation of Batchelor term},
\begin{equation}\label{eqn:Batchelor f^2 transport equation}
\frac{\partial |\hat{f}|^2}{\partial \eta} + \mathbf{k}\cdot\frac{\partial |\hat{f}|^2}{\partial \mathbf{s}} - s^j\Phi^i_j\frac{\partial |\hat{f}|^2}{\partial k^i} = -\textrm{(ii)},
\end{equation}
where (ii) represents the part of the non-linear term in equation \eqref{eqn: f^2 transport equation} in the limit (ii), $|\mathbf{k}-\mathbf{k}'| \ll k$; upon substituting $\mathbf{k}'' = \mathbf{k}-\mathbf{k}'$ in equation \eqref{eqn: f^2 transport equation}, one has
\begin{align}
\nonumber \textrm{(ii)} & \simeq \mathrm{i}\mathbf{s}\cdot \int_{k'' \ll k} \frac{\mathrm{d}^dk''}{(2\pi)^d}\left[\hat{\mathbf{g}}(\mathbf{k}-\mathbf{k}'')\hat{f}^*(\mathbf{k},\mathbf{s})\hat{f}(\mathbf{k}'',\mathbf{s}) 
-\hat{\mathbf{g}}^*(\mathbf{k}-\mathbf{k}'')\hat{f}(\mathbf{k},\mathbf{s})\hat{f}^*(\mathbf{k}'',\mathbf{s})\right] + \mbox{h.o.t.}\nonumber \\ &
\simeq \frac{\mathrm{i}\mathbf{s}}{V}\cdot\left[\hat{\mathbf{g}}(\mathbf{k})\hat{\overline{f}}(k,\mathbf{s})\hat{f}^*(\mathbf{k},\mathbf{s}) - \hat{\mathbf{g}}^*(\mathbf{k})\hat{\overline{f}}^*(k,\mathbf{s})\hat{f}(\mathbf{k},\mathbf{s}) \right]\label{eqn:limit ii},
\end{align}
where $\hat{\overline{f}} \equiv (2\pi)^{-d}V\int_{k''< k}\mathrm{d}^dk''~\hat{f}(\mathbf{k}'',\mathbf{s})$ contains the low-wave-number contributions to $f$.
\end{widetext}

Equation \eqref{eqn:Batchelor f^2 transport equation} is a transport equation in Fourier space, with a source $-\textrm{(ii)}$.
Taking its average yields an evolution equation for the power spectrum \eqref{eqn: F definition}:
\begin{equation}\label{eqn: F transport equation Batchelor}
\frac{\partial \hat{F}}{\partial \eta} + \mathbf{k}\cdot\frac{\partial \hat{F}}{\partial \mathbf{s}} - s^j\frac{\partial}{\partial k^i}\left\langle\Phi^i_j|\hat{f}|^2\right\rangle = \hat{S},
\end{equation}
with the source
\begin{equation}\label{eqn:source definition}
\hat{S} = -\frac{\mathrm{i}\mathbf{s}}{V}\cdot\left\langle\hat{\mathbf{g}}(\mathbf{k})\hat{\overline{f}}(k,\mathbf{s})\hat{f}^*(\mathbf{k},\mathbf{s}) - \hat{\mathbf{g}}^*(\mathbf{k})\hat{\overline{f}}^*(k,\mathbf{s})\hat{f}(\mathbf{k},\mathbf{s}) \right\rangle.
\end{equation}

We will estimate~$\hat{S}$ in~\S\ref{subsec: source CDM} below. Note, that $\hat{S}$, as given by equation \eqref{eqn:source definition}, need not vanish when integrated over all $\mathbf{k}$ and $\mathbf{s}$, because equation \eqref{eqn: F transport equation Batchelor} is only valid in the large-$k$ limit: non-squeezed contributions would matter at small $k$. Contrast this to equation~\eqref{eqn: f^2 transport equation}, where there is no `source' of $C_2$; effectively, equation \eqref{eqn: F transport equation Batchelor} represents not an injection of total $C_2$ into the system, but rather a transfer of $C_2$ from large scales to small ones.


The time scales appearing on the left-hand side of equation \eqref{eqn: F transport equation Batchelor} are the advection (linear) time scale associated with $\mathbf{k}\cdot\left(\partial/\partial \mathbf{s}\right)$, and the non-linear (gravitational) time scale associated with $s^j\Phi^i_j \partial/\partial k^i \sim s\,\delta g_r$.
We will discuss these time scales in~\S\ref{subsec:critical balance}. Then, we will describe the left-hand side of~\eqref{eqn: F transport equation Batchelor} in~\S\ref{subsec:phase space cascade}, and the right-hand side in~\S\ref{subsec: source CDM}.

\subsection{Critical balance}
\label{subsec:critical balance}
There are two (conformal) time scales in the Vlasov equation \eqref{eqn:Vlasov}: linear,~$\tau_{\rm l} \sim r/\delta v$, and non-linear,~$\tau_{\rm nl} \sim \delta v/\delta g_r$, where $\delta g_r$ is defined in equation \eqref{eqn:delta g r definition} and $\delta v$ is a velocity-difference scale.
In the Fourier-transformed Vlasov equation \eqref{eqn: Vlasov Fourier space} the linear (or phase-mixing) time scale is
\begin{equation}\label{eqn: linear time-scale definition}
\tau_{\rm l} \equiv \frac{s}{k},
\end{equation}
and the non-linear time scale is
\begin{equation}\label{eqn:non-linear time-scale definition}
\tau_{\rm nl} \equiv \frac{1}{s\,\delta g_r},
\end{equation}
where $k= 1/r$ and $s = 1/\delta v$. In the Batchelor approximation \eqref{eqn:Batchelor f^2 transport equation}, the non-linear time scale is $\tau_{\rm nl}^{-1} \sim s \norm{\Phi^i_j}/k$, which is the same as \eqref{eqn:non-linear time-scale definition} (up to an order-unity constant), in the case of a smooth gravitational field.

In the highly non-linear r\'{e}gime, the two time scales must balance each other: if $\tau_{\rm l}$ were much shorter than~$\tau_{\rm nl}$, so that only the first two terms of equation \eqref{eqn: Vlasov Fourier space} dominated, linear phase mixing would drive velocity gradients up until $\tau_{\rm nl}$ shrank to the same order of magnitude as $\tau_{\rm l}$. Conversely, if $\tau_{\rm nl}$ were much shorter than $\tau_{\rm l}$, so the third term in equation \eqref{eqn: Vlasov Fourier space} dominated, then non-linear mixing would drive spatial gradients of $f$, and, therefore, of~$\mathbf{g}$, up until $\tau_{\rm l}$ and $\tau_{\rm nl}$ matched. This so-called \emph{critical balance} may be thought of as a type of dominant-balance asymptotic argument for the Vlasov equation, where one 
allows the system enough time to establish this balance (see \cite{GoldreichSridhar1995,NazarenkoSchekochihin2011,Schekochihin2022,Nastacetal2023,Nastacetal2024} for some examples of critical balance in various areas of physics).


Critical balance amounts to setting $\tau_{\rm l} \sim \tau_{\rm nl}$. These are the only time scales in equation \eqref{eqn: F transport equation Batchelor}, so the \emph{cascade time} $\tau_{\rm c}$, which is the time scale for the transfer of $C_2$ from scale to scale, must also be
\begin{equation}\label{eqn:critical balance time}
\tau_{\rm c} \sim \tau_{\rm l} \sim \tau_{\rm nl}.
\end{equation}
The shared value $\tau_{\rm c}$ of the two time scales $\tau_{\rm l}$ and $\tau_{\rm nl}$ is also known as \emph{critical-balance time}. 
We will use this equality of time scales in \S\ref{subsubsec: Jeans stream} to fix the scalings of the phase-space power spectrum.
Given equation \eqref{eqn:critical balance time}, for each length scale $r=1/k$, there exists a corresponding (inverse) velocity scale
\begin{equation}\label{eqn:critical balance}
s_{\rm c}(k) \equiv \sqrt{\frac{k}{\delta g_r}}.
\end{equation}
It is also convenient to define $k_{\rm c}(s)$ as the inverse function of~$s_{\rm c}$.

The argument that the two time scales must balance applies only to modes well inside the horizon, such that~$k \gg \mathcal{H}/c$ (where~$\mathcal{H}$ is the conformal Hubble constant). Indeed, the conformal time in any asymptotically de-Sitter cosmology (i.e., one with a positive cosmological constant) is bounded from above by some value~$\eta_{\max}$~\cite{Hawking_Ellis_1973}; so, for modes with $k$ too small, linear phase mixing can only generate velocity gradients up to~$s_{\max} \simeq k \eta_{\max}$. If the amplitude of~$\hat{f}(k,s_{\max})$ is not large enough for the non-linear term to become important, then the evolution of such a mode will always be primarily linear. Alternatively, this can be understood as a failure of the requirement that $k \gg \mathcal{H}/c$: at late times $\mathcal{H} \sim 1/(\eta_{\max} - \eta)$, so it will eventually exceed $kc$; if $s$ does not grow sufficiently by then, that mode can never become non-linear. 
Here we are interested in the~$k \to \infty$ limit, so it is safe to ignore this nuance. Additionally, due to the finite age of the Universe and hierarchical structure formation, the decrease of~$\tau_{\rm nl}$ until it matches~$\tau_{\rm l}$ might not have happened yet for all values of~$k$, as structures on the largest scales have yet to collapse. Again, this does not affect the $k \to \infty$ limit, and we may simply take~$k \gg \max\set{k_{\rm nl},\mathcal{H}/c}$, where $k_{\rm nl}$ is defined in the Introduction. 

\subsubsection{Non-smooth gravitational field is impossible}
Let us now show that the gravitational field must be smooth in a critically balanced Vlasov--Poisson system. To do this, let us parameterise $\hat{F}$: to leading order in the large-$(k,s)$ limit, let
\begin{equation}\label{eqn: power spectrum parameterisation}
\hat{F}(\mathbf{k},\mathbf{s}) \sim \begin{cases}
F_1 k^{\gamma}s^{\xi}, & \mbox{if } s \lesssim s_{\rm c}(k) \ll v_{\rm th}^{-1}, ~k \gg k_{\rm nl}, \\
F_2 s^{\delta}k^{\sigma}, & \mbox{if } k \lesssim k_{\rm c}(s),~s \gg v_{\rm rms}^{-1},
\end{cases}
\end{equation}
for some $\gamma, \delta < 0$, $\xi, \sigma \in \mathbb{R}$. For $P(k)\propto\hat{F}(k,0)$ to be finite, $\hat{F}$ must be independent of $s$ in the limit $s \to 0$, $k_{\rm c}(s) \ll k \ll k_{\rm c}(v_{\rm th}^{-1})$, whence $\xi = 0$. Since $\overline{f}$ must have a finite velocity variance, $\sigma =0$. 
Furthermore, matching along the critical-balance line \eqref{eqn:critical balance} gives us
\begin{equation}\label{eqn:matching critical balance line}
F_1 k^\gamma \sim F_2 \left[s_{\rm c}(k)\right]^\delta,
\end{equation}
which will shortly lead to a relationship between~$\gamma$ and~$\delta$.

The exponent $\gamma$ may be related directly to the H\"{o}lder exponent $\lambda$ in equation \eqref{eqn:delta g r with lambda} by Poisson's equation \eqref{eqn:Poisson}. If the gravitational field were not smooth, i.e., $\lambda < 1$, then by the Paley--Wiener theorem and equations~\eqref{eqn:Poisson} and~\eqref{eqn:delta g r definition},~$F(k,0) \propto k^{2-d}\delta g_r^2$, whence, from equation \eqref{eqn: power spectrum parameterisation},
\begin{equation}\label{eqn:gamma and lambda for non-smooth field}
\gamma = 2-d-2\lambda.
\end{equation}
Furthermore, by equation \eqref{eqn:critical balance}, $s_{\rm c}(k) \propto k^{(1+\lambda)/2}$, whence equation \eqref{eqn:matching critical balance line} yields
\begin{equation}\label{eqn:gamma delta and lambda for non-smooth field}
2\gamma = (1+\lambda)\delta
\end{equation}
(this equation also applies when $\lambda = 1$).
Eliminating $\lambda$ from equations \eqref{eqn:gamma and lambda for non-smooth field} and \eqref{eqn:gamma delta and lambda for non-smooth field} gives
\begin{equation}\label{eqn:gamma and delta for non-smooth field}
\delta = \frac{4\gamma}{4-d-\gamma}.
\end{equation}
For $\lambda < 1$, equation \eqref{eqn:gamma and lambda for non-smooth field} implies that $\gamma > -d$. Then by equation \eqref{eqn:gamma and delta for non-smooth field} we also have, $\delta > -d$, for $d \in \set{1,2,3}$.

However, if $\gamma$ and $\delta$ are both greater than $-d$, the power spectrum $\hat{F}$ does not decay sufficiently fast to be critically balanced: consider the amount of $C_2$ at a scale $k^{-1}$ (or smaller), which is the variance of $f$ over all scales up to $k=1/r$, \emph{viz.}, analogously to \eqref{eqn:delta g r definition},
\begin{equation}\label{eqn:detla f r}
\delta f_r^2 = \frac{1}{(2\pi)^{2d}Vv_{\rm rms}^d}\int_{k'>k}\!\!\!\!\!\!\mathrm{d}^dk' \int \mathrm{d}^d s~ \hat{F}(\mathbf{k}',\mathbf{s}).
\end{equation}
If $\delta > -d$, then the $s$ integral is dominated by the ultra-violet cut-off, $v_{\rm th}^{-1}$, so
\begin{equation}
\delta f_r^2 \sim \frac{F_2}{Vv_{\rm rms}^dv_{\rm th}^{d+\delta}} \int_{k'>k}\!\!\!\!\!\!\mathrm{d}^dk',
\end{equation}
and the $k'$ integral will also be dominated by the ultra-violet cut-off. If this were true, the dynamics of $f$ at the length scale $k^{-1}$ could not be critically balanced as most of $C_2$ at that scale would be concentrated away from the critical-balance line $(k,s_{\rm c}(k))$. Even if one introduced such a configuration initially, it would, by the argument of \S\ref{subsec:critical balance}, dynamically evolve into a critically balanced state.
Likewise, considering the amount of $C_2$ stored at velocity scale $s^{-1}$ (or smaller), would imply that the system is not in critical balance at any $s \ll v_{\rm th}^{-1}$.

Thus, critical balance, in conjunction with Poisson's equation, precludes the possibility of a non-smooth gravitational field, whence it must be the case that $\lambda=1$ in equation~\eqref{eqn:delta g r with lambda}, and, therefore, $\gamma$, $\delta \leq -d$.

\subsubsection{Smooth gravitational field}
If $\delta g_r \sim \kappa r$, as it must be for a smooth field, then
\begin{align}
\label{eqn:critical balance time smooth} \tau_{\rm c} & \sim \frac{1}{\sqrt{\kappa}}, \\
s_{\rm c}(k) & = \frac{k}{\sqrt{\kappa}}\label{eqn:critical balance smooth s_c(k)}, \\
k_{\rm c}(s) & = \sqrt{\kappa}s,\label{eqn:critical balance k_c(s)}
\end{align}
and the critical-balance time is a scale-independent constant.

\subsection{Jeans collapse}
\label{subsubsec: Jeans stream}
We will find below that the phase-space motion induced by gravitational collapse implies, together with critical balance, that~${\gamma\geq -d}$, so the only possibility compatible with a smooth gravitational field is $\gamma = \delta = -d$.

As stated in \S \ref{subsec:cold}, our analysis focuses on the cold-dark-matter limit, where $sv_{\rm th} \ll 1$, which is equivalent to $k \ll k_{\rm c}(v_{\rm th}^{-1})$. On these scales,~$f$ is just a collection of streams, so when zooming in on one of them, passing through a phase-space point $(\mathbf{x}_0,\mathbf{v}_0)$, and applying the Raychaudhuri equation~\eqref{eqn:Raychaudhuri} in its frame of reference, one finds that the stream is unstable to gravitational collapse. This applies if the stream's density, denoted $\rho_{\mathbf{x}_0,\mathbf{v}_0}$ in \S \ref{subsec:cold}, is larger than the background density, $\rho_{\rm b}$, due to the rest of the distribution. Indeed, let $V_0$ be a spatial region around $\mathbf{x}_0$, where the mean density of the stream,
\begin{equation}
\rho_{\mathbf{x}_0,\mathbf{v}_0}(V_0) = \frac{1}{\textrm{Vol}(V_0)}\int_{V_0}\mathrm{d}^d x~\rho_{\mathbf{x}_0,\mathbf{v}_0}(\mathbf{x}),
\end{equation}
is larger than $\rho_{\rm b}$, and let $r=1/k \sim \left[\textrm{Vol}(V_0)\right]^{1/d}$ be the spatial size of $V_0$. The gravitational effect of $\rho_{\rm b}$ may be absorbed into a re-definition of the background (an effective cosmology) felt by the stream, via the `separate universe' principle \cite{Daietal2015}. If $\rho_{\mathbf{x}_0,\mathbf{v}_0}(V_0) \geq \rho_{\rm b}$, the stream is over-dense (with respect to the re-defined effective background) and will therefore collapse.
As the entire system is cold, homogeneous and isotropic, such over-densities generically exist on all scales $r \ll k_{\rm nl}^{-1}$.\footnote{Initial conditions $f_{\rm in}$ where this does not occur can be constructed, but these require additional symmetries, and are unstable---to the Jeans instability---so we ignore them here.}

All the terms in equation \eqref{eqn:Raychaudhuri} are of dimension [time]$^{-2}$, and the last two terms on the right-hand-side are $O(k^2v_{\rm th}^2)$,\footnote{Indeed, if initially the system had $\omega^i = 0$, then, in the limit $v_{\rm th} \to 0$, this will stay this way locally in phase-space \cite{BertschingerJain1994}.} while the $\mathcal{H}\theta$ term is completely negligible for~$k \gg \mathcal{H}/c$. Equation~\eqref{eqn:Raychaudhuri} then reduces to
\begin{equation}\label{eqn:Raychaudhuri approx}
\frac{\mathrm{d}\theta}{\mathrm{d}\eta} = - \frac{\theta^2}{3} + \del \cdot \mathbf{g} + O(k^2v_{\rm th}^2),
\end{equation}
which means that for $s_{\rm c}(k)v_{\rm th} \ll 1$, the gravity term $\del \cdot \mathbf{g}$ dominates and there is nothing to stop a collapse. This collapse occurs on a time scale $\tau_{\rm g}$, the \emph{gravitational time}, given by
\begin{equation}\label{eqn:tau g}
\tau_{\rm g}^{-2} \sim \abs{\del \cdot \mathbf{g}} \sim 4\pi G \delta\rho_r, 
\end{equation}
where again $r = 1/k$, and,\footnote{We take $k\gg \mathcal{H}/c$ and assume that $\tau_{\rm g} \ll \mathcal{H}^{-1}$ so that the scale factor $a$ in equation \eqref{eqn:Poisson} is effectively a constant.} analogously to \eqref{eqn:delta g r definition},
\begin{equation}\label{eqn:delta rho r definition}
\delta \rho_r^2 \sim \int_{k'>k} \!\!\!\!\!\!P(k') \mathrm{d}^d k'.
\end{equation}
This collapse is nothing but the Jeans instability occurring in every stream individually, in its own reference frame; this is not dissimilar to the instabilities described in, e.g., \cite{Henon1973,Barnesetal1986}. 
Indeed, a linear stability analysis of a spatially homogeneous distribution gives the Jeans-instability growth rate \cite{Jeans1902}
\begin{equation}\label{eqn: Jeans}
\tau_{\rm J}^{-2} = \tau_{\rm g}^{-2} - k^2v_{\rm th}^2,
\end{equation}
which is the same as can be inferred from the Raychaudhuri equation \eqref{eqn:Raychaudhuri approx}.
One special case where the collapse time may be deduced explicitly is described in appendix~\ref{appendix: EdS}.

Thus, the system experiences collapse on all spatial scales smaller than the outer scale and larger than~$1/k_{\rm c}(v_{\rm th}^{-1})$, at a rate $\tau_{\rm J}^{-1} \sim \tau_{\rm g}^{-1}$, by equation~\eqref{eqn: Jeans}, provided that $\tau_{\rm g}kv_{\rm th} \ll 1$. Gravitational collapse is a critically balanced process, so if we recall the other time scales of our system, identified in \S \ref{subsec:critical balance}, we immediately see that the gravitational time is
\begin{equation}
\tau_{\rm g} \sim \tau_{\rm c} \sim \frac{1}{\sqrt{\kappa}},
\end{equation}
where the last equality follows from equation \eqref{eqn:critical balance time smooth}, valid for a smooth gravitational field. Then, requiring consistency with Poisson's equation \eqref{eqn:tau g}, we must immediately conclude that $\delta \rho_r$ is constant in $r$. Equivalently, $\gamma = -d$ in equation \eqref{eqn: power spectrum parameterisation},
which is the main result of the paper:
\begin{equation}\label{eqn:P(k) is k^-d}
P(k) \sim F_1 k^{-d}.
\end{equation}
Then, by equation \eqref{eqn:gamma and delta for non-smooth field} and the matching condition \eqref{eqn:matching critical balance line}, $\delta = \gamma$ and $F_1 \sim F_2 \tau_{\rm g}^d$.
We have thus obtained that, for cold dark matter,
\begin{equation}\label{eqn:F asymptotics CDM}
\hat{F}(k,s) \sim F_1\begin{cases}
k^{-d}, & \mbox{if } s \ll s_{\rm c}(k) \ll v_{\rm th}^{-1},\\
\tau_{\rm g}^{-d}s^{-d}, & \mbox{if }  k \ll k_{\rm c}(s) \ll k_{\rm c}(v_{\rm th}^{-1}),
\end{cases}
\end{equation}
provided that $k \gg k_{\rm nl}$ and $s \gg v_{\rm rms}^{-1}$.
These are the leading-order asymptotics: they hold for large~$k$ at~$s\to 0$, and large~$s$ at~$k \to 0$. In general, $\hat{F}$ can also depend on the angle between~$\mathbf{s}$ and~$\mathbf{k}$, \emph{viz.}, on $\mathbf{k}\cdot \mathbf{s}$. This angular dependence arises only at the next order in $s\ll s_c(k)$ or~$k\ll k_c(s)$ (whichever obtains), because it must not exist at $s=0$, or at $k=0$.

The analysis of time scales may be recast in terms of length scales as follows. There are only two physically meaningful length scales in the system studied in this paper: $k_{\rm nl}^{-1}$, the scale below which critical balance has had enough time to establish itself since the Big Bang, and $v_{\rm th}\tau_{\rm c} \sim 1/k_{\rm c}(v_{\rm th}^{-1})$, below which $f$ may no longer be treated as a collection of streams, but rather describes `warm' dark matter (which is no longer Jeans unstable). If it were the case that $\gamma < -d$, then necessarily, these two length scales would be equal, for otherwise $\tau_{\rm g}$ would be shorter than $\tau_{\rm c}$. In this paper, however, we are interested in the case $v_{\rm th}\tau_{\rm c} \ll k_{\rm nl}^{-1}$, and in structures on scales $k^{-1}$ between these two values. Thus, $k\ll k_{\rm c}(v_{\rm th}^{-1})$ implies that $\gamma \geq -d$ (otherwise the collapse time would be much longer than $\tau_{\rm c}$ on scales where the system should be unstable), whereas $k \gg k_{\rm nl}$ necessitates---via critical balance---that $\gamma \leq -d$.

The $k^{-d}$ asymptotic \eqref{eqn:P(k) is k^-d} of the density power spectrum may be derived by an altogether different, yet systematic approach---by performing an asymptotic analysis of the an integral expression for $P(k)$ and examining its critical points. We shall do this in \S \ref{sec:saddle-point approach}, which is self-contained, but before that, let us see what kind of flow in $(\mathbf{k},\mathbf{s})$ space is engendered by equation \eqref{eqn: F transport equation Batchelor}, and see how the power spectrum \eqref{eqn:F asymptotics CDM} is supported by this flow (and \emph{vice versa}).

\subsection{Phase-space cascade}
\label{subsec:phase space cascade}
Consider first the homogeneous part of equation \eqref{eqn: F transport equation Batchelor}, ignoring $\hat{S}$ until \S \ref{subsec: source CDM}. The positive eigenvalues of $\Phi^i_j$ drive a rotation in $(\mathbf{k},\mathbf{s})$, where small-scale velocity structure interchanges with small-scale spatial structure, while the negative eigenvalues drive a flow of both to ever smaller scales. In fact, one can analyse equation \eqref{eqn:Batchelor f^2 transport equation} directly---before ensemble-averaging---while still ignoring the right-hand-side. As~$\Phi^i_j$ generically depends on time, this analysis holds locally in time (and space, on scales below the outer scale), but by critical balance, the long mode (large-scale)~$\Phi^i_j$ cannot vary on a time scale shorter than the critical-balance time. We therefore approximate it as constant (in which case there are analytical solutions), but the qualitative features described here---namely, a rotation in $(\mathbf{k},\mathbf{s})$-space and a flow to larger values of $k$ and $s$---remain also true for a time-dependent~$\Phi^i_j$. Generically, in $d>1$, $\Phi^i_j$ would have both positive and negative eigenvalues.

Consider a positive eigenvalue of $\Phi^i_j$. If $\mathbf{s}_+$ is its corresponding eigenvector, then for $\mathbf{s} \parallel \mathbf{s}_+$, equation \eqref{eqn:Batchelor f^2 transport equation} reduces to a transport equation under the action of a harmonic-oscillator potential, i.e., a rotation in the $(\mathbf{k},\mathbf{s})$ space. This ensures that the large-$s$ structure in the initial condition (cold streams) is transported to large $k$, and \emph{vice versa}.
The negative eigenvalues ensure that there is a flow to ever smaller scales, because then the solution is a linear combination of hyperbolic functions (cf. \cite{Nastacetal2024}).

In $1\mathrm{D}$, it would appear na\"{i}vely that there is only phase-space rotation when $\Phi^i_j$ is evaluated in an initially over-dense region, because there is only one, necessarily positive, eigenvalue. This, however, is misleading. For any~$d$, as the system evolves, matter moves around, and thus that region generically changes from being over-dense to under-dense (alternatively, it does so for different realisations of the initial conditions), and therefore the sign of $\tr \, \Phi^i_j$ also changes. Thus, there is a temporal sequence of phase-space rotations and stretchings (see figure \ref{fig:results time evolution})---essentially, differential phase-space rotation, leading to the generation of small-scale structure. 


Thus, phase-space rotations transfer small-scale structure from $s$ to $k$ and back, and that is supplemented by a flow of $|\hat{f}|^2$ to ever larger $k$ and $s$ (similar to the plasma echo joint with a phase-space cascade in \cite{Nastacetal2023,Nastacetal2024}).\footnote{This behaviour is generic: chaos---the exponential separation of nearby trajectories in phase-space---combined with Liouville's theorem, necessitates the formation of structure on smaller and smaller scales.} Hence, equation \eqref{eqn:Batchelor f^2 transport equation} describes a type of phase-space turbulence, where~$C_2$, being the integral of $|\hat{f}|^2$ over scales, is cascaded to smaller scales by larger-scale tidal fields, in a critically balanced manner.

\begin{figure}
\centering
\includegraphics[width=0.49\textwidth]{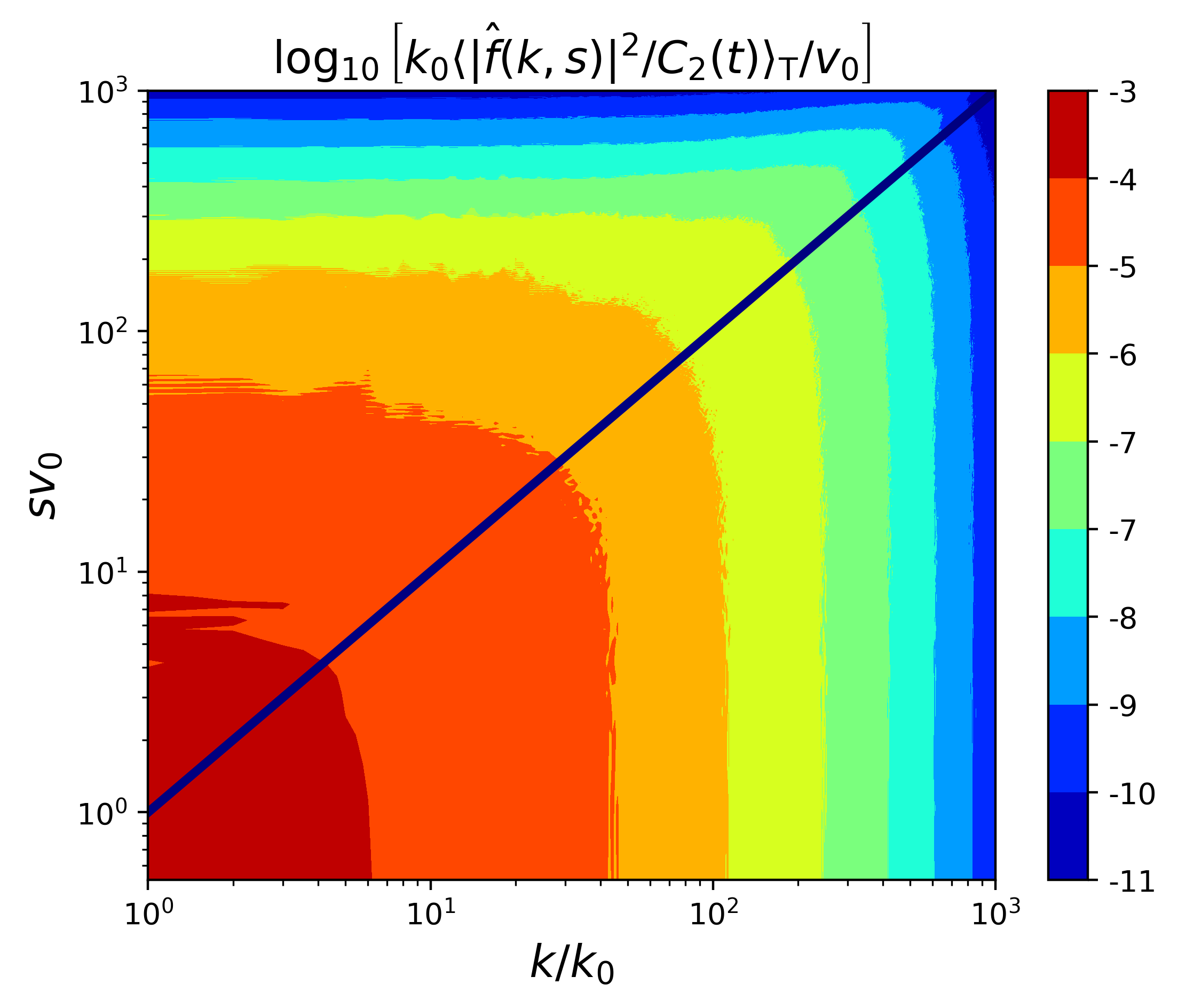}
\caption{A contour plot of the time-averaged power spectrum~$\langle |\hat{f}|^2(t,k,s)/C_2(t)\rangle_{\rm T}$, for the same simulation as in figure~\ref{fig:results time evolution} (note that $C_2$ decays because of collisions and finite-grid effects, so it is sensible to normalise the power spectrum by $C_2$ at every time). The line $s=k\tau_0$ is also plotted in blue.}
\label{fig:results colour plot of F}
\end{figure}

To test this conclusion numerically, we calculated the time-averaged phase-space power spectrum $\hat{F}$, as a function of $\abs{k}$ and $\abs{s}$ (with the negative values folded on the positive ones), for the simulation shown in figure~\ref{fig:results time evolution}. This is plotted in figure \ref{fig:results colour plot of F}, which shows the contours of the time-average (as a proxy for an ensemble-average)~$\left\langle |\hat{f}|^2/C_2(t)\right\rangle_{\rm T}$. They are arranged in a rectangular shape: the critical-balance line \eqref{eqn:critical balance smooth s_c(k)} is the diagonal of this rectangle. 
This rectangular structure is brought about by the aforementioned rotation in the $(k,s)$ space, conjoined with the effect of the negative eigenvalues of $\Phi^i_j$, which stretch structures along the critical-balance line. Both of these processes establish a symmetry with respect to this line. The line $s = k\tau_0$ is plotted in figure \ref{fig:results colour plot of F} and identified as the critical-balance line, yielding $\kappa \sim \tau_0^{-2} = 4 \pi G\overline{\rho}_{\rm m}$ for this simulation.

\subsection{Scale-dependent non-linear fluxes and sources for a cold system}
\label{subsec: source CDM}
Having described the homogeneous part of equation~\eqref{eqn:Batchelor f^2 transport equation}, let us now consider the source term on its right-hand side.
Equation \eqref{eqn: F transport equation Batchelor} is a conservation equation with a source, of the form
\begin{equation}\label{eqn:conservation equation with a source}
\frac{\partial \hat{F}}{\partial \eta} + \del_{\mathbf{k},\mathbf{s}} \cdot \boldsymbol{\Gamma} = \hat{S}.
\end{equation}
Here $\del_{\mathbf{k},\mathbf{s}}$ is a 6-dimensional phase-space gradient, and~$\boldsymbol{\Gamma}$ is a 6-dimensional flux, whose components are
\begin{align}
\label{eqn: s flux definition gamma}\Gamma^{\mathbf{s}}_i & = k_i\hat{F} \\
\Gamma^{\mathbf{k}}_i & = -s^j\left\langle \Phi_{ij} |\hat{f}|^2\right\rangle.\label{eqn: k flux definition gamma}
\end{align}
The flux $\boldsymbol{\Gamma}$ of $C_2$ towards larger $k$ and $s$ is the `cascade' that we described in \S\ref{subsec:phase space cascade}.
Let us calculate the phase-space flux $\mathcal{F}^{\mathbf{s}}$ flowing through a sphere in $\mathbf{s}$-space of radius $s$, where~$s_{\rm c}(k_{\rm nl}) \ll s \ll v_{\rm th}^{-1}$.
In steady state, integrating equation \eqref{eqn:conservation equation with a source} over~$\mathbf{k}$ yields
\begin{equation}
\frac{\partial}{\partial \mathbf{s}} \cdot \int  \Gamma^{\mathbf{s}}\mathrm{d}^dk = \int \hat{S}\mathrm{d}^d k .
\end{equation}
Integrating both sides over the ball in $\mathbf{s}$-space with radius~$s$ (at all $\mathbf{k}$) and using Gauss' theorem yields
\begin{equation}\label{eqn: integrated source}
\mathcal{F}^{\mathbf{s}} \equiv \iint s^{d-2}s^i\Gamma^{\mathbf{s}}_i \mathrm{d}^{d-1}\Omega_s\mathrm{d}^dk = \iint_{s'<s}\!\!\!\!\!\!\hat{S}\mathrm{d}^dk\mathrm{d}^ds',
\end{equation}
where the $\mathrm{d}^{d-1}\Omega_s$ integral is over the $d$-dimensional solid angle and the integral $\mathrm{d}^dk$ is over all $\mathbf{k}$. Similarly, the flux in~$\mathbf{k}$~is
\begin{equation}\label{eqn:k integrated source}
\mathcal{F}^{\mathbf{k}} \equiv \iint k^{d-2}k^i\Gamma^{\mathbf{k}}_i  \mathrm{d}^{d-1}\Omega_k\mathrm{d}^{d}s = \iint_{k'<k}\!\!\!\!\!\!\hat{S}\mathrm{d}^dk'\mathrm{d}^ds.
\end{equation}

Hence, the accumulated non-linear source $\iint \hat{S}$ plays the r\^{o}le of a mechanism that injects $C_2$, which is balanced at each $k$ by the flux passing this $C_2$ on to smaller scales.
Qualitatively, since the system is Jeans unstable at every scale $k' < k$, and since the resulting collapse time scale $\tau_{\rm g}$ is the same at all scales, the integrated source \eqref{eqn:k integrated source} should receive contributions from the collapses occurring on all of these scales, adding up coherently.

Let us consider how phase-space fluxes depend on $k$ and $s$ in a turbulent state described by the phase-space spectrum \eqref{eqn:F asymptotics CDM}. Using equations \eqref{eqn: k flux definition gamma}, \eqref{eqn:k integrated source} and the estimate $s^j \Phi^i_j/k \sim \tau_{\rm nl}^{-1}$, we have
\begin{equation}\label{eqn: source term s-cubed}
\mathcal{F}^{\mathbf{k}} \sim \frac{k^d}{\tau_{\rm nl}} \int_{s<s_{\rm c}(k)}\!\!\!\!\!\! \mathrm{d}^d s\hat{F} \sim \frac{F_1}{\tau_{\rm nl}} \left[s_{\rm c}(k)\right]^d \sim F_1 \kappa^{-(d-1)/2}k^d,
\end{equation}
the last expression having been obtained via critical balance ($\tau_{\rm nl} \sim \tau_{\rm c}$) and equations \eqref{eqn:critical balance time smooth} and \eqref{eqn:critical balance smooth s_c(k)}. Similarly, from equations \eqref{eqn: s flux definition gamma}, \eqref{eqn: integrated source}, and $s/k\sim \tau_{\rm l}$, we have
\begin{equation}
\mathcal{F}^{\mathbf{s}} \sim \frac{s^d}{\tau_{\rm l}} \int_{k<k_{\rm c}(s)}\!\!\!\!\!\! \mathrm{d}^dk \hat{F} \sim \frac{F_2}{\tau_{\rm l}} \tau_{\rm g}^{-d} \left[k_{\rm c}(s)\right]^{-d} \sim F_1 \sqrt{\kappa} s^d,
\end{equation}
using $\tau_{\rm l} \sim \tau_{\rm c}$ and equations \eqref{eqn:critical balance time smooth} and \eqref{eqn:critical balance k_c(s)}.
The balances \eqref{eqn: integrated source} and \eqref{eqn:k integrated source} imply that the accumulated source terms will have the same scalings, which indeed can be confirmed by direct estimates of the corresponding integrals, using the expression \eqref{eqn:source definition} for the non-linear source, critical balance, and the phase-space power spectrum \eqref{eqn:F asymptotics CDM} ($\mathrm{i}\mathbf{s}\cdot\hat{\mathbf{g}}$ is estimated similarly to $s^j \Phi^i_j/k$ in equation \eqref{eqn: source term s-cubed}).

Thus, due to Jeans collapses occurring at all scales, the $C_2$ fluxes accumulate to become $\propto k^d$ in $k$ space---and, by critical balance, $\propto s^d$ in $s$ space. This is in contrast to how this works in Kolmogorov or Batchelor-style turbulence \cite{Kolmogorov1941,Batchelor1959}, where the fluxes of the corresponding invariants (energy of the fluid motion or passive-scalar variance, respectively) are constant across the inertial or convective ranges, because injection of these quantities occurs only at the outer scale. In kinetic systems, including our gravitational one, such a situation obtains at $sv_{\rm th} \gg 1$ and $k \gg k_{\rm c}\left(v_{\rm th}^{-1}\right)$---for warm dark matter, or an ordinary plasma. The phase-space spectrum, derived for the latter case in Ref.~\cite{Nastacetal2024}, is then steeper in both $k$ and $s$, with the $k^{-d}$ and $s^{-d}$ scalings replaced by $k^{-2d}$ and $s^{-2d}$, respectively (see appendix \ref{subsec:simulation results})---we shall see in the next sub-section that this is indeed how our scalings \eqref{eqn:F asymptotics CDM} are cut off at the thermal scale.

\begin{figure*}
\centering
\includegraphics[width=\textwidth]{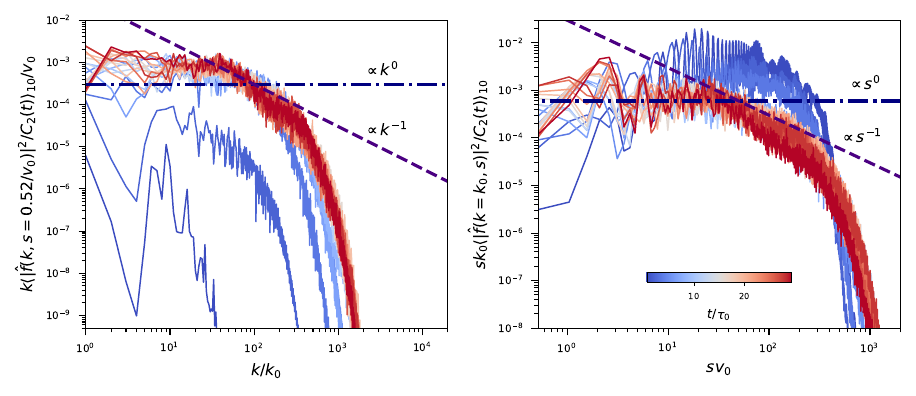}

\caption{The time evolution of the power spectra $\langle |\hat{f}|^2/C_2(t) \rangle_{10}$ for the simulation illustrated in figure \ref{fig:results time evolution} (early times: blue; late times: red). The time averaging was over $10$ simulation outputs (corresponding to time windows of duration $1.36\tau_0$). \emph{Left panel}: the $k$-spectrum at $s=0.52/v_0$. \emph{Right panel}: the $s$-spectrum at $k=k_0$. Both spectra are compensated by the expected asymptotics (multiplied by $k^{1}$ and $s^{1}$, respectively). A $k^{-1}$ power law establishes itself quickly, extending to $k\sim 200k_0$, which is of the order of $k_{\rm c}(v_{\rm th}^{-1})$, as expected; similarly, the $s^{-1}$ power law extends to $sv_0\sim 200$. The steeper spectra at smaller scales are the scalings that prevail below the thermal scale (see appendix \ref{subsec:simulation results}).}
\label{fig:spectra time evolution}
\end{figure*}

\begin{figure*}
\centering
\includegraphics[width=0.49\textwidth]{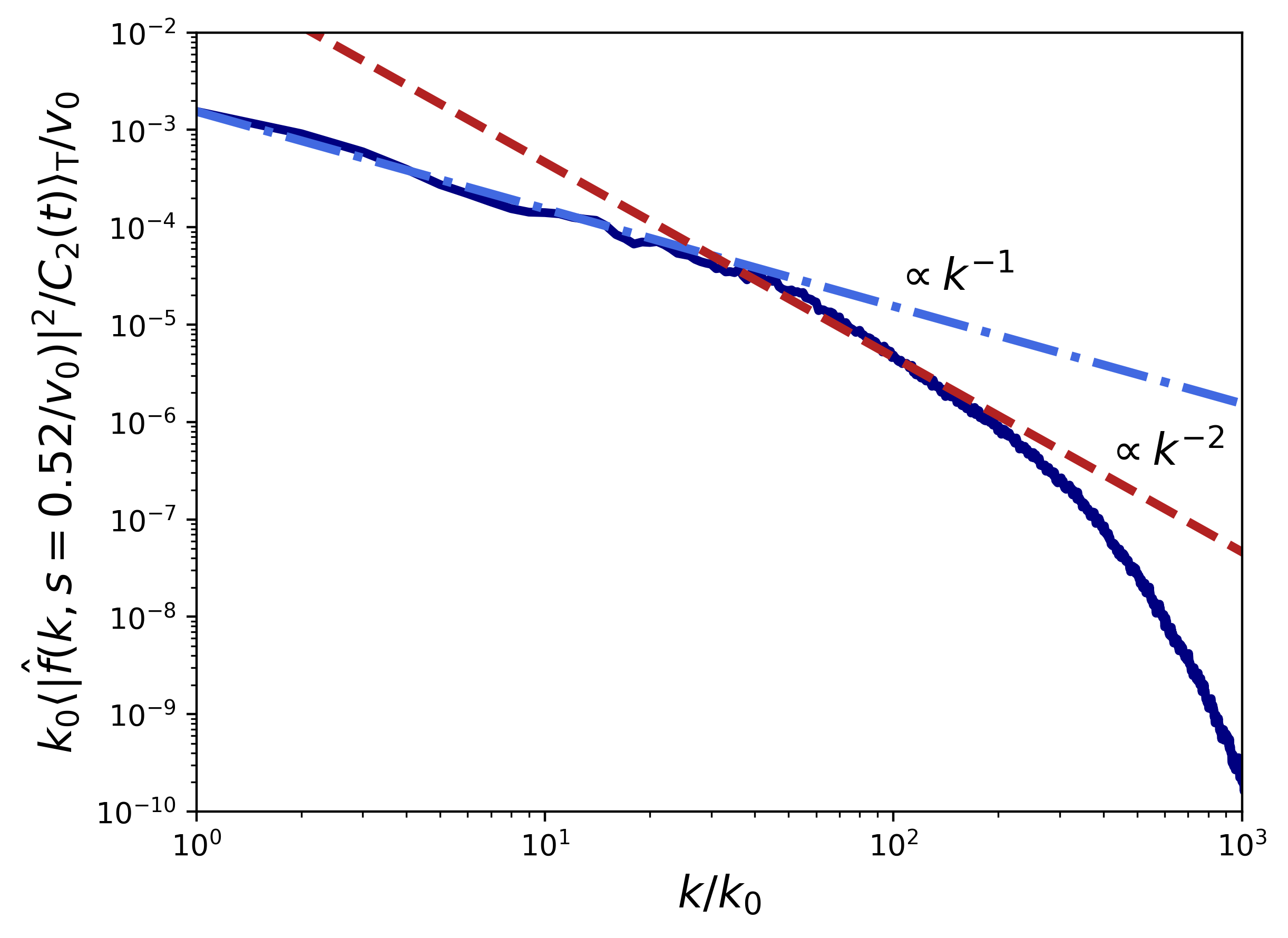}
\includegraphics[width=0.49\textwidth]{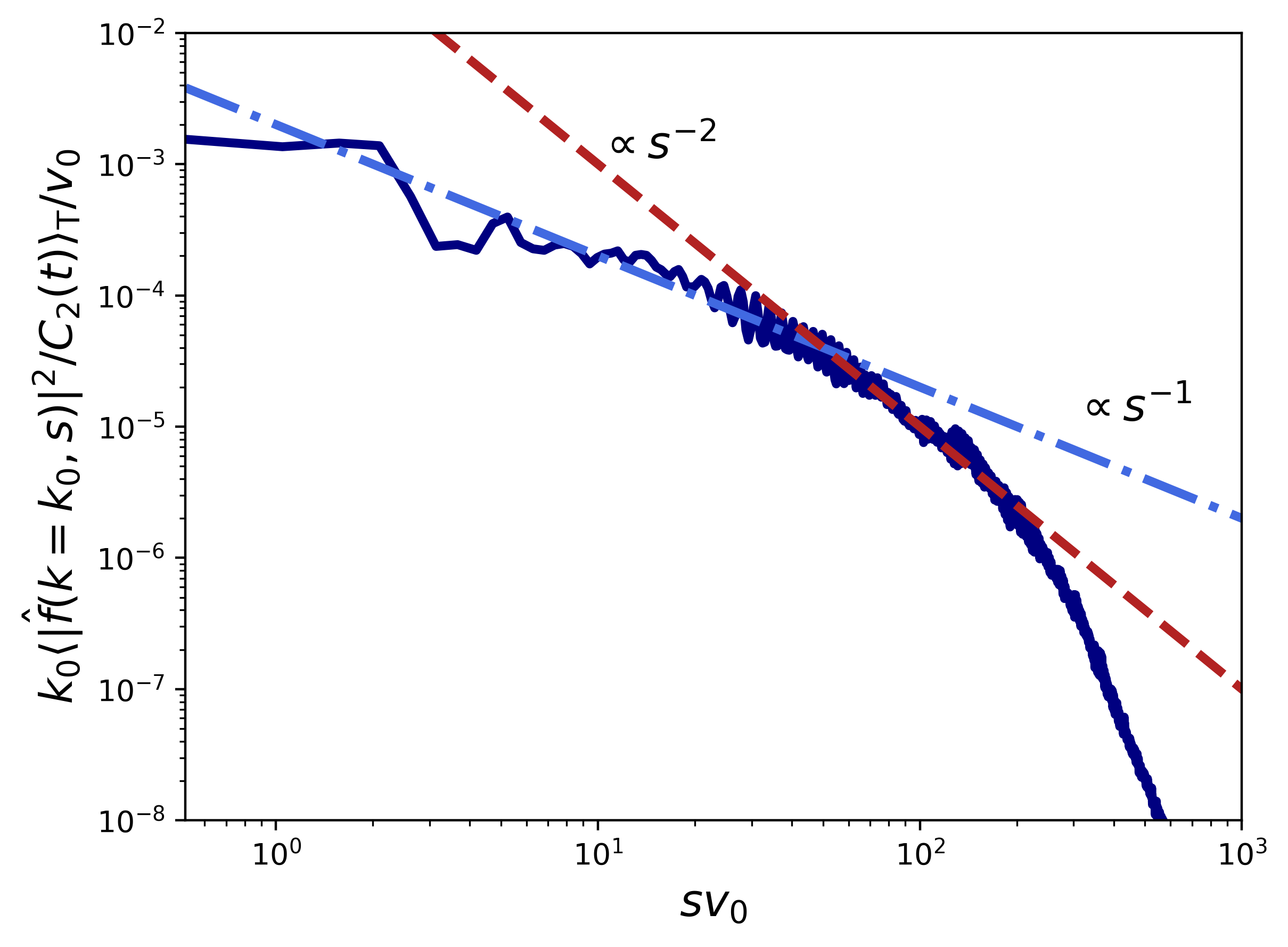}

\caption{The power spectra for the same simulation as in figure \ref{fig:results time evolution}, but now averaged over $t\in [0,30]\tau_0$, as a proxy for the ensemble average. Unlike in figure \ref{fig:spectra time evolution}, they are uncompensated, and display the theoretical $k^{-d}$ and $s^{-d}$ scalings explicitly~($d=1$).}
\label{fig:spectra averaged over time}
\end{figure*}

\subsection{Numerical results}

\begin{figure*}
\centering

\includegraphics[width=1\textwidth]{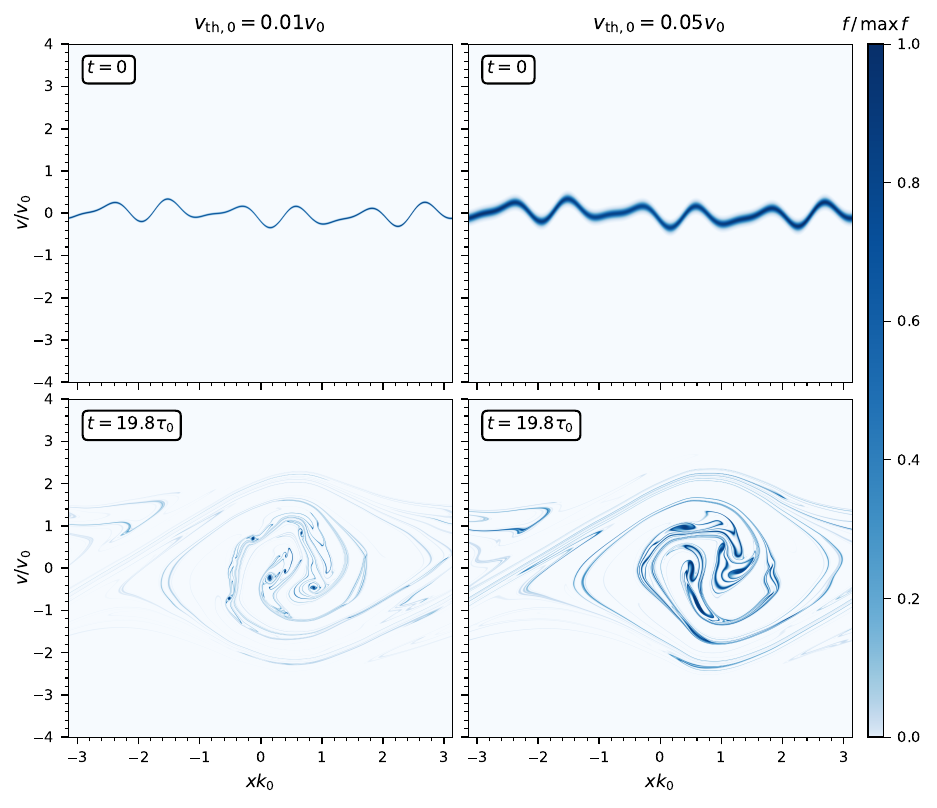}
\includegraphics[width=0.49\textwidth]{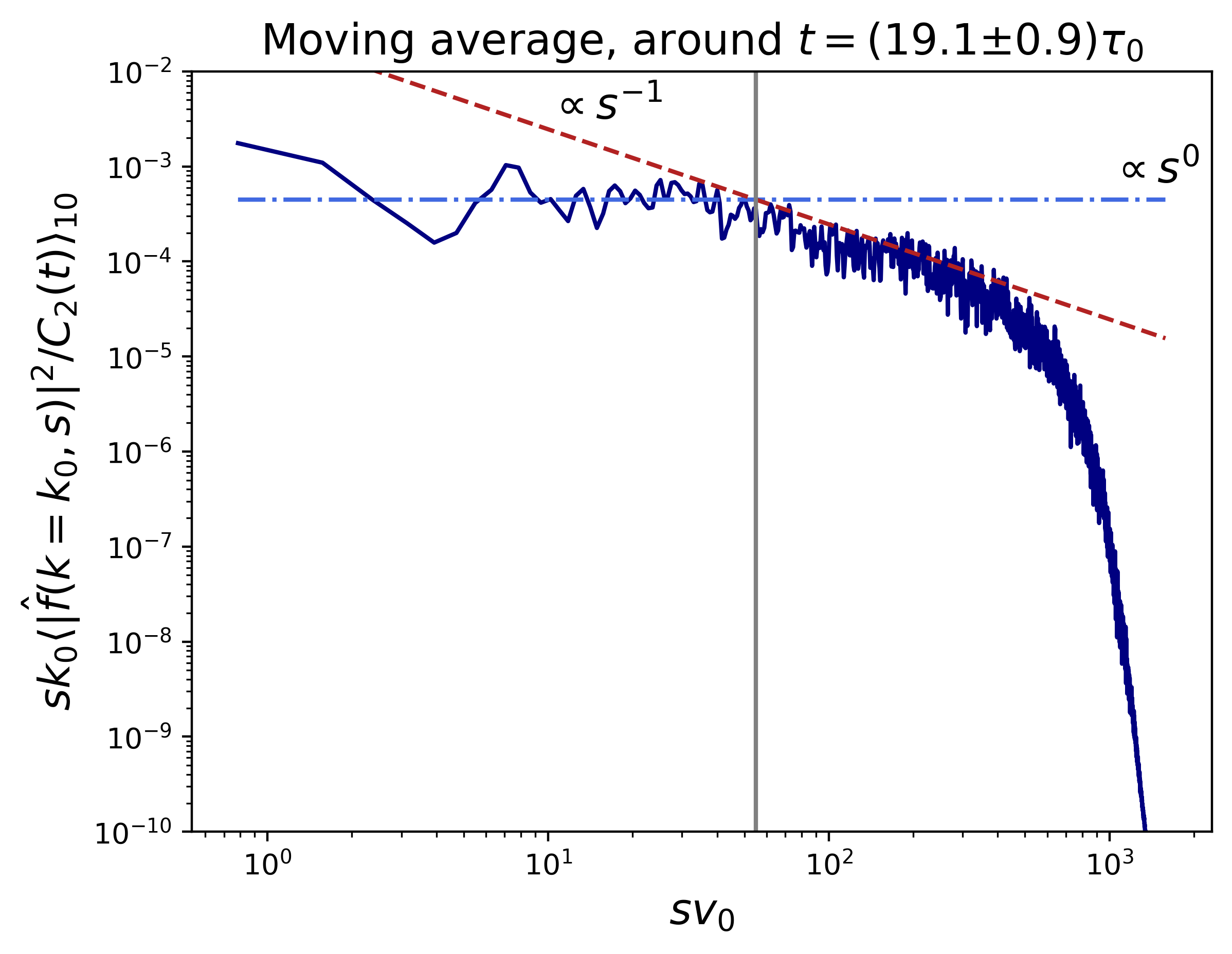}
\includegraphics[width=0.49\textwidth]{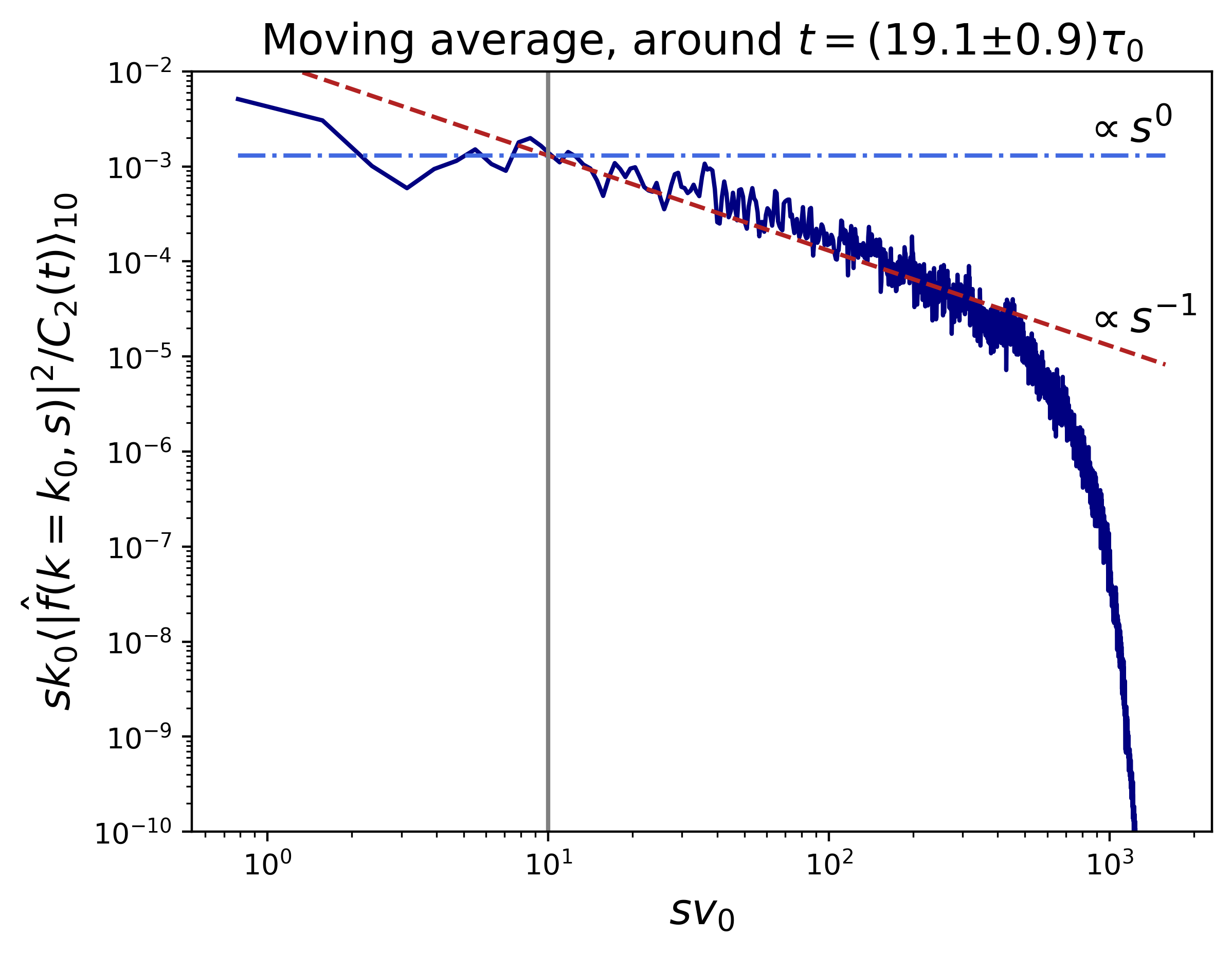}
\caption{A comparison between the evolution and the spectra of two systems with identical initial conditions, save for the value of $v_{\rm th}$. \emph{Left panels}: a colder case, $v_{\rm th} = 0.01v_0$; \emph{right panels}: a warmer case, $v_{\rm th} = 0.05v_0$. The plots in the bottom row show the $s$ spectra at the end of the simulations, compensated by~$s$. In the warmer case, the spectrum ceases to scale as~$s^{-1}$ at around~$sv_0 \sim 10$, while in the colder one, it does so at~$sv_0\sim 50$ (marked by grey lines). In both cases the $s^{-1}$ scaling is replaced at $sv_{\rm th} \gg 1$ by $s^{-2}$. See text and appendix~\ref{appendix: simulation set up} for details. Videos are available in this \href{https://ybginat.com/index.php/gravitational-turbulence/}{link}.}
\label{fig:gkeyll results two v_th's}
\end{figure*}

Let us now check whether the theoretical asymptotic scalings~\eqref{eqn:F asymptotics CDM} are reproduced in our $1$D Vlasov--Poisson simulation. The power spectra of the system illustrated in figure~\ref{fig:results time evolution} are shown in figure~\ref{fig:spectra time evolution}: a~$k^{-1}$ power law establishes itself quickly, persisting up to~$k\sim 100k_0$, which is of the order of~$k_{\rm c}(v_{\rm th}^{-1})$, as expected. The spectra that we plot are $|\hat{f}|^2/C_2(t)$, averaged over short time intervals (compared with the simulation's duration) as a proxy for the ensemble average. We re-scale $|\hat{f}|^2$ by $C_2(t)$ at every time step because $C_2$ decays due to collisions and finite-grid effects.
In figure \ref{fig:spectra averaged over time}, we average over the entire simulation, displaying cleaner power laws in the appropriate ranges of~$k$ and~$s$. While the system is not stationary, this only affects the overall amplitude---not the overall scaling---and figure \ref{fig:spectra time evolution} shows that the re-scaling by $C_2$ corrects for that; hence, we can trust figure~\ref{fig:spectra averaged over time} to give an adequate approximation for $\hat{F}$. 

As the system is sourced by the gravitational-collapse instability, which was shown in \S\ref{subsubsec: Jeans stream} to continue until~$(kv_{\rm th})^{-1}$ matches the collapse time, the scaling of the power spectrum must be truncated at $v_{\rm th}$. To test this proposition, it is necessary to see whether it is indeed the case that the asymptotics \eqref{eqn:F asymptotics CDM} persist until $v_{\rm th}$ is reached. We ran two identical simulations, differing only by the value of $v_{\rm th}$, to verify this. The result is presented in figure \ref{fig:gkeyll results two v_th's}: the left column has $v_{\rm th} = 0.01v_0$ while the right column has $v_{\rm th} = 0.05v_0$. The $s$-spectra in the bottom row show that indeed, the $s^{-1}$ power law is truncated at a lower value of $s$, by a factor that matches the ratio of~$v_{\rm th}$ for the two runs. The $s^{-2}$ (or $k^{-2}$) scaling of $\hat{F}$ at~$sv_{\rm th} \gg 1$ (or $k\gg k_{\rm c}(v_{\rm th}^{-1})$) is a known result in plasma systems \cite{Nastacetal2023,Nastacetal2024}, discussed in appendix \ref{subsec:simulation results}.

\section{Saddle-point approach}
\label{sec:saddle-point approach}
In this section, we focus on $d=3$ spatial dimensions, and derive the $k^{-d}$ scaling of the density power spectrum using a stationary-phase approximation.
For pure gravitational Newtonian evolution of $N$ identical particles, the density power spectrum is given exactly by~\cite{KonradBartelmann2022}
\begin{equation}\label{eqn: PS definition}
P(k,t) =\frac{M^2}{V} \prod_{n=1}^{N}\int \mathrm{d}^3q_n\mathrm{d}^3p_n \mathcal{P}(\set{q},\set{p}) \mathrm{e}^{\mathrm{i}\mathbf{k}\cdot\left[\mathbf{x}_1(t)-\mathbf{x}_2(t)\right]},
\end{equation}
where $(\mathbf{q}_n,\mathbf{p}_n)$ is the initial phase-space position of particle~$n$, $(\mathbf{x}_n(t),\mathbf{v}_n(t))$ is the phase-space position of particle $n$ at time $t$, and $\mathcal{P}(\set{q},\set{p})$ is the joint probability distribution of the initial phase-space positions~$(\set{q},\set{p}) \equiv \set{(\mathbf{q}_n,\mathbf{p}_n)}_{n=1}^N$ of all particles. This equation is permutation-invariant, and, therefore, the choice of two particles is arbitrary.

For \emph{cold} dark matter with Gaussian initial conditions, the initial distribution is
\begin{equation}\label{eqn:initial condition distribution}
\mathcal{P}(\set{q},\set{p}) = \frac{V^{-N}\mathcal{C}(\set{q},\set{p})}{\sqrt{(2\pi)^{3N} \det C^N_{pp}}} \mathrm{e}^{-\set{p}^T (C^N_{pp})^{-1} \set{p}/2},
\end{equation}
where $C^N_{pp} = C^N_{pp}(\set{q})$ is the $3N\times3N$ covariance matrix of~$\set{p}$, and~$\mathcal{C}$ encapsulates initial density-density and density-momentum correlations~\cite{Fabisetal2018}, both of whose functional dependence on particle positions depends on the cosmology (we take a $\Lambda$CDM background).

The usual procedure to obtain the power spectrum~$P(k)$ from equation \eqref{eqn: PS definition} would involve integrating out particles $3,\ldots,N$, leaving only a $12$-dimensional integral over the phase-sub-space of particles $1$ and $2$. We will, however, go the other way round, and integrate \emph{first} over the relative position and momentum of this pair, and only then over all other particles---we will see that this order of integration is well suited to deriving the asymptotics of~$P(k,t)$ as~$k \to \infty$. Let us change the integration variables from $\mathbf{q}_1$ and~$\mathbf{q}_2$ to~$\mathbf{q} \equiv \mathbf{q}_1 - \mathbf{q}_2$ and~$\mathbf{Q} \equiv (\mathbf{q}_1 + \mathbf{q}_2)/2$, and to their conjugate momenta,~$\mathbf{p}$ and~$\mathbf{P}$, respectively. As~$\set{p}^T (C^N_{pp})^{-1} \set{p}$ is quadratic in~$\set{p}$, we can make the $\mathbf{p}$ dependence explicit, \emph{viz.},
\begin{equation}\label{eqn:momentum correlations}
\begin{aligned}
& \set{p}^T (C^N_{pp})^{-1} \set{p} \equiv \mathbf{p}^T \Sigma^{-1}\left(\mathbf{q}\right) \mathbf{p} \\ &
-2\mathbf{a}\left(\mathbf{q},\mathbf{Q},\mathbf{P},\set{(q,p)}_3^N\right) \cdot \mathbf{p} - 2B\left(\mathbf{q},\mathbf{Q},\mathbf{P},\set{(q,p)}_3^N\right),
\end{aligned}
\end{equation}
where $\mathbf{a}$ is a linear function of momenta and $B$ a quadratic one.
Below we will require the following properties of $C_{pp}^N$, derived in appendix \ref{appendix: Initial Momentum Correlations}, based on the assumption $v_{\rm th} \to 0$ (so the conclusions again will be valid at $k\ll k_{\rm c}(v_{\rm th}^{-1})$, as in~\S\ref{sec:Phase-Space Cascade}):
(i) $\Sigma \simeq Aq^2$ at small $q$ for some order-unity matrix $A$, (ii) $\abs{\mathbf{a}} \sim O(q^{-1})$ (at most) in this limit, and (iii) at $q\to\infty$, the correlation matrix $C_{pp}^N$ tends to a constant, i.e., $\Sigma$ is order unity in the limit $q\to \infty$.

Using Duhamel's principle for Hamilton's equations of motion, one has \cite{KonradBartelmann2022}
\begin{equation}\label{eqn:Duhamel}
\mathrm{i}\mathbf{k}\cdot\left[\mathbf{x}_1(t)-\mathbf{x}_2(t)\right]= \mathrm{i}\mathbf{q}\cdot \mathbf{k} + \mathrm{i}g_{qp}\left(t,t_{\rm initial}\right)\mathbf{p}\cdot \mathbf{k} + \mathrm{i}k \psi_I,
\end{equation}
where $g_{qp}$ is a `propagator' that encapsulates the linear evolution (defined below), and the `interaction term' is
\begin{align}
& k\psi_I\left(\mathbf{P},\mathbf{Q},\set{(q,p)}_3^N\right) \\ &
\equiv \mathbf{k}\cdot \int_{0}^{t}\mathrm{d}t' g_{qp}(t',t_{\rm initial})\left(\mathbf{F}_1(t') - \mathbf{F}_2(t')\right),\nonumber
\end{align}
$\mathbf{F}_n$ being the additional acceleration of particle $n$, relative to its motion in the part already included in $g_{qp}$, caused by all other particles. For example, if $g_{qp} = t-t_{\rm initial}$, one re-obtains regular Newtonian dynamics. Using results of first-order cosmological perturbation theory (or, alternatively, re-summed kinetic field theory \cite{Lilowetal2019}), we take $g_{qp}(t) \equiv \left[D_+(t) - D_+(t_{\rm initial})\right]/\left[\mathrm{d}D_{+}/\mathrm{d}t\right]$, where $D_+$ is the $\Lambda$CDM growth factor. Then ignoring the $\psi_I$ term in \eqref{eqn:Duhamel} would just yield the Zel'dovich approximation, whose asymptotics were studied by \cite{Konradetal2020,Konradetal2022}. 

\subsection{Possible saddle points}
Using equations \eqref{eqn:momentum correlations} and \eqref{eqn:Duhamel}, we define the exponent
\begin{equation}\label{eqn:exponent}
\varphi \equiv -\frac{1}{2}\mathbf{p}^T \Sigma^{-1}\mathbf{p} + \mathbf{a}\cdot\mathbf{p} + \mathrm{i}\mathbf{q}\cdot \mathbf{k} + \mathrm{i}g_{qp}\mathbf{p}\cdot \mathbf{k} + \mathrm{i}k \psi_I + B,
\end{equation}
so the integral \eqref{eqn: PS definition} for the power spectrum turns into
\begin{equation}\label{eqn: power spectrum with phi}
P(k,t) \propto \prod_{n=3}^{N}\int \frac{\mathrm{d}^3q_n\mathrm{d}^3p_n \mathrm{d}^3Q\mathrm{d}^3P \mathrm{d}^3q\mathrm{d}^3p}{V^N\sqrt{(2\pi)^{3N} \det C^N_{pp}}}  \mathcal{C}(\set{q},\set{p}) \mathrm{e}^{\varphi}.
\end{equation}
We start by integrating over $\mathbf{p}$ and $\mathbf{q}$, applying the saddle-point approximation.
This is a movable-saddle problem, so it requires care in handling \cite{Olver1974,bleistein1975asymptotic}.\footnote{Note also that the saddle point may be complex, but that is innocuous, since the exponent can be continued analytically to the complex plane.} The function $\psi_I$ is a smooth function of $\mathbf{q}$ and $\mathbf{p}$, because it arises from the Hamiltonian flow in phase-space, generated by a smooth gravitational potential (recall that we neglect collisions and that time is bounded). This assertion is just the statement that $\lambda = 1$ as in \S\ref{sec:Phase-Space Cascade}. Letting $\boldsymbol{\psi}_q \equiv \partial \psi_I/\partial \mathbf{q}$ and $\boldsymbol{\psi}_p \equiv \partial \psi_I/\partial \mathbf{p}$, the exponent \eqref{eqn:exponent} is stationary when
\begin{equation}\label{eqn:saddle point}
\begin{aligned}
& -\Sigma^{-1} \mathbf{p} + \mathbf{a} + \mathrm{i}g_{qp}\mathbf{k} + \mathrm{i}k\boldsymbol{\psi}_p = 0, \\ &
-\frac{1}{2}p^i \frac{\partial (\Sigma^{-1})_{ij}}{\partial \mathbf{q}}p^j + p_i\frac{\partial a^i}{\partial \mathbf{q}} + \frac{\partial B}{\partial \mathbf{q}} + \mathrm{i}\mathbf{k} + \mathrm{i}k\boldsymbol{\psi}_q = 0.
\end{aligned}
\end{equation}
Let us parameterise the solution of these equations as
\begin{align}
\mathbf{p} = k^\alpha \mathbf{c}, ~~ &
\mathbf{q} = k^{\beta} \mathbf{d},
\end{align}
where $\mathbf{c}$ and $\mathbf{d}$ are complex vectors whose magnitude remains finite as $k/k_{\rm nl}\to\infty$. Together with $\alpha$, $\beta$, they are to be determined by equations~\eqref{eqn:saddle point}, by seeking a dominant balance, i.e., such a balance where the exponent \eqref{eqn:exponent} has the weakest~$k$ dependence around the stationary point that solves equations~\eqref{eqn:saddle point}.

\emph{A priori}, equations \eqref{eqn:saddle point} could permit a balance that is independent of the initial-condition distribution, i.e., that same balance would exist for uniform initial conditions. But in that case, Liouville's theorem---after changing variables from the initial phase-space positions $\set{(\mathbf{q}_n,\mathbf{p}_n)}^{N}$ to the current ones $\set{(\mathbf{x}_n,\mathbf{v}_n)}^N$---ensures that this yields a contribution to $P(k)$ proportional to $\delta^{\rm D}(\mathbf{k})$, whence this balance is not dominant at large $k$. Conversely, the dominant balance must involve the first terms on the left-hand sides of equations \eqref{eqn:saddle point}.
If $\beta > 0$, then, as $\Sigma \underset{q \to \infty}{\longrightarrow} \mbox{const}$, the first term in the first of equations \eqref{eqn:saddle point} is proportional to $p$. As this must play a part in a dominant balance, this implies that $\alpha = 1$. Substituting such a saddle point into equation \eqref{eqn:exponent} gives an exponentially suppressed contribution to $P(k)$, of order $\sim O\left[\exp(\propto -k^2)\right]$ at most.

Thus, stationary points with $\beta <0$ are dominant, as they could contribute a power-law tail to $P(k)$.
For $\beta < 0$, equations \eqref{eqn:saddle point} imply $1= \alpha -2\beta$ and $1= 2\alpha -3\beta$,
whence $\alpha = \beta = -1$. This is consistent as long as
\begin{equation}
\lim_{\overset{(q,p) \sim k^{-1}}{k\to \infty}} \left(\abs{\boldsymbol{\psi}_q }, \abs{\boldsymbol{\psi}_p}\right) = O(1),
\end{equation}
which is generically valid, and so the interaction term is potentially as important as the linear term; that both are equally important is essentially a statement of critical balance (cf. \S \ref{subsec:critical balance}).


\subsection{Evaluation of the asymptotics}

Having proven that the asymptotic expansion of the power spectrum is given, up to exponentially small (in $k$) contributions, by its dominant saddle point at $q,p \sim k^{-1}$, one is allowed to replace $\varphi$ with its expansion at small $q$ and $p$. 
This yields
\begin{widetext}
\begin{equation}
\varphi \simeq -\frac{1}{2}\mathbf{p}^T \Sigma^{-1}\mathbf{p} + \mathbf{a}\cdot\mathbf{p} + \mathrm{i}\mathbf{q}\cdot \mathbf{k} + \mathrm{i}g_{qp}\mathbf{p}\cdot \mathbf{k}
+ \mathrm{i}k\left[\mathbf{p}\cdot\boldsymbol{\psi}_p(0) + \mathbf{q}\cdot\boldsymbol{\psi}_q(0)\right] + B(q=0).
\end{equation}
With this asymptotic approximation for $\varphi$, we need to integrate equation \eqref{eqn: power spectrum with phi} over $\mathbf{p}$ and $\mathbf{q}$ to obtain its asymptotic scaling with $k$.
Observe that $\det C^N_{pp}$ factorises into $\det \Sigma$ multiplied by normalisation factors for the other momenta variables, and the only other pieces that still depend on $\mathbf{Q}$, $\mathbf{P}$, and the initial positions and momenta of the other particles are~$\mathcal{C}$,~$\mathbf{a}$,~$B$, and~$\boldsymbol{\psi}_{q,p}$.
Thus,
\begin{equation}\label{eqn:Pkt to be integrated}
P(k,t) \simeq  \frac{M^2}{V^2}\expec{\int\frac{\mathrm{d}^3q\mathrm{d}^3p~\mathcal{C}}{\left[(2\pi)^3\det\Sigma\right]^{1/2}}\exp\left\{-\frac{1}{2}\mathbf{p}^T \Sigma^{-1}\mathbf{p} + \mathbf{a}\cdot\mathbf{p} + \mathrm{i}\mathbf{q}\cdot \mathbf{k} + \mathrm{i}g_{qp}\mathbf{p}\cdot \mathbf{k} + \mathrm{i}k\left[\mathbf{p}\cdot\boldsymbol{\psi}_p(0)+\mathbf{q}\cdot\boldsymbol{\psi}_q(0) \right]\right\}},
\end{equation}
where the average is over the position and momentum of the centre of mass of particles $1$ and $2$, as well as the positions and momenta of all the other particles ($\mathrm{e}^B$ is absorbed into the average). Integrating over $\mathbf{p}$, we get
\begin{equation}
P(k,t) \simeq \frac{M^2}{V^2}\expec{\int\mathrm{d}^3q ~\mathcal{C}~\exp\left\{-\frac{k^2}{2}\left[g_{qp}\hat{\vk} + \boldsymbol{\psi}_p(0) - \frac{\mathrm{i}\mathbf{a}}{k}\right]^T \Sigma(\mathbf{q})\left[g_{qp}\hat{\vk} + \boldsymbol{\psi}_p(0)- \frac{\mathrm{i}\mathbf{a}}{k}\right] + \mathrm{i}\mathbf{q}\cdot \mathbf{k} + \mathrm{i}k\mathbf{q}\cdot\boldsymbol{\psi}_q(0)\right\}}.
\end{equation}
This integration does not introduce any powers of $k$ because both $\mathrm{d}^3p$ and $\sqrt{\det \Sigma}$ are proportional to $k^{-3}$ in the vicinity of the stationary point, so the two cancel.
Recalling that the saddle point is at~$\mathbf{q}=\mathbf{d}/k$, where~$\mathbf{d}$ is a finite constant, we change variables to~$\mathbf{y} = k\mathbf{q}$, whence
\begin{equation}\label{eqn:P asymptotic integrated over p but not yet q}
P(k,t) \simeq \frac{M^2}{V^2k^3}\expec{\int\mathrm{d}^3y ~\mathcal{C}~\mathrm{e}^{-k^2\left[g_{qp}\hat{\vk} + \boldsymbol{\psi}_p(0)- \frac{\mathrm{i}}{k}\mathbf{a}(\mathbf{y}/k)\right]^T \Sigma(\mathbf{y}/k)\left[g_{qp}\hat{\vk} + \boldsymbol{\psi}_p(0)- \frac{\mathrm{i}}{k}\mathbf{a}(\mathbf{y}/k)\right]/2 + \mathrm{i}\mathbf{y}\cdot \left[\hat{\vk} + \boldsymbol{\psi}_q(0)\right]}} \equiv \frac{F_1}{V}k^{-3}.
\end{equation}
\end{widetext}
Therefore, the integral $\mathrm{d}^3q$, as opposed to the momentum integral in \eqref{eqn:Pkt to be integrated}, is not compensated by any function that scales like $k^{-3}$, so it is this integration that yields the $k^{-3}$ scaling, which emerges from $q\sim k^{-1}$ at the stationary point. The exponential in equation \eqref{eqn:P asymptotic integrated over p but not yet q} is order unity in the limit $k \to \infty$, because $\Sigma \sim y^2/k^2$, and also because $\mathbf{a}/k$ is at most order unity,\footnote{The exponential might depend on $\hat{\vk}$, the direction of $\mathbf{k}$, before being ensemble-averaged.} whence the $y$ integral is a Gaussian integral, and the entire factor multiplying $k^{-3}$ in equation \eqref{eqn:P asymptotic integrated over p but not yet q}, denoted by $F_1/V$, is of order unity. 


Thus, we have established that for collisionless, cold dark matter in the non-relativistic limit, the asymptotic behaviour of the power spectrum as $k \to \infty$ is dominated by the contribution of the saddle point at $q, p \sim k^{-1}$. This yields $P(k) \propto k^{-3}$, with an order-unity coefficient; that this coefficient is non-zero follows from the more qualitative argument of \S \ref{sec:Phase-Space Cascade}.

\section{Discussion}
\label{sec:discussion}

\subsection{The importance of being cold}

The $k^{-d}$ power spectrum arose from a turbulent state, where the gravitational field was treated as a smooth field---this is an analogue of Batchelor turbulence in hydrodynamics \cite{Batchelor1959}, where a tracer is advected by a large-scale, smooth velocity field.
A crucial step in the derivation of the form of the power spectrum was the occurrence of gravitational collapse on all $k$ such that $\tau_{\rm g}kv_{\rm th} \ll 1$---that the initial configuration was cold, and Jeans unstable. This was also necessary in \S \ref{sec:saddle-point approach}, where the initial conditions were special to cold dark matter, with the particular functional form of the $C_{pp}^N$ matrix, $\Sigma \sim Aq^2$, reflecting the fact that if two particles started out at exactly the same spatial position, they would remain together forever.

The coldness of the system means that locally in phase-space, the distribution function is always a stream, and therefore, locally, everywhere and on every scale larger than the thermal scale, the Raychaudhuri equation~\eqref{eqn:Raychaudhuri} applies. This, in turn, implies that the stream is unstable to gravitational collapse, with the collapse time scale~\eqref{eqn:tau g}, matching the critical-balance time~\eqref{eqn:critical balance time}.
As the collapse occurs on all scales, one deduces from Poisson's equation that the only way in which critical balance could be satisfied is if the density power spectrum is $P(k) \propto k^{-d}$---otherwise, the two physical length scales $k_{\rm nl}$ and $v_{\rm th}\tau_{\rm c}$ would have to coincide, thereby violating the coldness assumption.

When the system is not cold, these assumptions fail. Indeed, the source term would not accumulate when integrated over $k$ and $s$ up to small, non-linear scales, because at $sv_{\rm th}\gg 1$ and $s_c(k)v_{\rm th} \gg 1$, there is no instability to drive the collapse, 
so one would expect the flux to be $\mathcal{F}^{\mathbf{k}} = \textrm{const}$, rather than $\propto k^d$ in equation \eqref{eqn: source term s-cubed}. In other words, for a warm system, the two length scales $k_{\rm nl}$ and $v_{\rm th} \tau_{\rm c}$ are indeed the same.

While in a collisionless setting, $v_{\rm th}$ would not change with time,\footnote{Because, if we denote $\norm{f}_{\infty} = \max f$, $C_2 \sim \norm{f}_{\infty}^2v_{\rm th}^dV$, and both $C_2$ and $\norm{f}_{\infty}$ are conserved (in fact, one can define $v_{\rm th} \equiv C_2^{1/d} \norm{f}_\infty^{-2/d}V^{-1/d}/\sqrt{\pi}$, as a measure of the thinness of the distribution). There is of course a change in $\norm{f}_{\infty}$ because of the Universe's expansion, but this is very slow for the large-$k$ limit, where~$k \gg \mathcal{H}/c$.} in reality finite-$N$ effects do increase~$v_{\rm th}$. Additionally, baryons would influence the matter power spectrum significantly on small scales, probably leading to a deviation from the power-law scalings found here. The approach used in this paper does not apply once interactions with baryons are included, and this is its main limitation. As mentioned, it only applies insofar as dark matter behaves as a \emph{cold}, classical \emph{phase-space} fluid---other dark-matter candidates might follow different scalings, and we defer this to future work.

Were it not for baryons, a full theory of the constituent particle of dark matter should predict both the initial value of~$v_{\rm th}$ and the functional form of the effective collision operator; this should allow one to find how~$C_2$ changes with time, and thence how~$v_{\rm th}$ evolves. Measuring the break in the dark-matter power spectrum---the transition from~$k^{-3}$ to a different, steeper power law of $k^{-2d}$ \cite{Nastacetal2023,Nastacetal2024} (in $d=1$, from $k^{-1}$ to a $k^{-2}$ scaling, as in figure \ref{fig:spectra averaged over time})---would hypothetically allow one to find the present-day value of $v_{\rm th}$, which is directly connected to the nature of dark matter (if indeed it held that~$u_\nu \ll v_{\rm th}$, where the collisional velocity scale~$u_\nu$ is defined in appendix \ref{appendix: simulation set up}).
For example, for WIMPs, which decouple from the photons when non-relativistic, one would have
\begin{align}
v_{\rm th}(z) & = c \frac{T(z)}{T_{\rm kd}}\sqrt{\frac{2k_{\rm B}T_{kd}}{mc^2}} \\ &
\approx 3.3\times 10^{-12} c (1+z)\left[\frac{10\textrm{ MeV}}{T_{\rm kd}}\right]^{1/2}\left[\frac{\textrm{GeV}}{m}\right]^{1/2}, \nonumber
\end{align}
where $T(z)$ is the photon temperature at redshift $z$ and $T_{\rm kd}$ is the kinetic-decoupling temperature \cite{KolbTurner1990,Greenetal2004,LoebZaldarriage2005}. This is
far too small to be observed practically, and moreover, it corresponds to (cf. appendix \ref{appendix: EdS})
\begin{align}
k_{\rm c}(v_{\rm th}^{-1}) & \sim \delta_{\rm typ}^{1/2} \frac{H}{v_{\rm th}} \\ &
\approx \frac{10^8h(z)}{1+z} \delta_{\rm typ}^{1/2} \textrm{ Mpc}^{-1} \left[\frac{m}{\textrm{GeV}}\right]^{1/2}\left[\frac{T_{\rm kd}}{10\textrm{ MeV}}\right]^{1/2}, \nonumber
\end{align}
where $\delta_{\rm typ}$ is a typical over-density and the Hubble constant is $H(z) = 100 h(z)~\textrm{km } \textrm{s}^{-1} \textrm{ Mpc}^{-1}$ (this value is of the same order as the free-streaming scale \cite{Greenetal2004} for $\delta_{\rm typ} \sim 1$). 

\subsection{Some like it hot: similarities with plasma physics}


A phase-space Batchelor-type cascade was recently proposed to be the universal r\'{e}gime of (plasma) Vlasov--Poisson turbulence at Debye and sub-Debye scales \cite{Nastacetal2023,Nastacetal2024}. This has been verified numerically in $1\mathrm{D}$ simulations of turbulence driven by an external forcing \cite{Nastacetal2024} and by the two-stream instability \cite{Ewart2024}. Like in the cold-dark-matter turbulence presented here, the (electric) field fluctuations are spatially smooth, so the phase-space mixing of the distribution function is dominated by the outer-scale fields. However, unlike cold-dark-matter turbulence, which is sourced at every Jeans-unstable scale, the cascade in the plasma case has a constant flux of $C_2$, because $C_2$ is only sourced at the outer scale; this changes the scalings of the power spectrum and of the field spectrum. This situation is analogous to the gravitational phase-space turbulence at $s \gg v_{\rm th}^{-1}$ that appears in the simulations presented in figure \ref{fig:gkeyll results two v_th's}. We will investigate this r\'{e}gime further in future work.


\subsection{Implications for dark-matter haloes}

The $P(k) \propto k^{-3}$ scaling (in~$3\mathrm{D}$) derived here for the non-linear power spectrum sheds some light on universal properties of dark matter haloes. Within the hierarchical-clustering paradigm~\cite{PressSchechter1974}, the small-scale limit of the power spectrum is dominated by the one-halo term,
\begin{equation}
P_{1h}(k) = \int \mathrm{d}M ~\frac{\mathrm{d}n}{\mathrm{d}M} [R^3\hat{\delta}(kR)]^2,
\end{equation}
where~$\mathrm{d}n/\mathrm{d}M$ is the halo mass function,~$R(M) \equiv R_{200}(M)/c(M)$,~$c(M)$ is the halo's concentration, $R_{200}$ is the radius where the density is $200$ times the critical density of the Universe, and $\hat{\delta}$ is the Fourier transform of the normalised halo density profile~$\delta(x/R)$ \cite{MaFry2000}. If we define~$\Lambda_c \equiv -\mathrm{d}\ln c/\mathrm{d}\ln M$ and~$\Lambda_n \equiv \mathrm{d} \ln \left(\mathrm{d}n/\mathrm{d}M\right)/\mathrm{d}\ln M$, both at~$M \to 0$, and also
\begin{equation}
\Lambda_\delta \equiv -\lim_{kR\to \infty} \frac{\mathrm{d}\ln \hat{\delta}(kR)}{\mathrm{d}\ln (kR)},
\end{equation}
then, as $k \to \infty$,
\begin{equation}
\begin{aligned}\label{eqn:gamma 1 halo}
P_{1h}(k) \sim k^\gamma, & \qquad \gamma = -\min\set{\frac{3(3+\Lambda_n)}{1+3\Lambda_c},2\Lambda_\delta}.
\end{aligned}
\end{equation}
Setting $\gamma = -3$, as would be required by our result, yields a relation between the universal concentration-to-mass relation and the halo mass function for cold dark matter: either $\Lambda_\delta = 3/2$ \cite[cf.][]{GoudaNakamura1989,YanoGouda2000,Navarroetal2004,DelosWhite2023} and $\Lambda_n > 3\Lambda_c-2$, or $\Lambda_\delta > 3/2$ and
\begin{equation}\label{eqn: nabla n nabla c}
\Lambda_n = 3\Lambda_c - 2.
\end{equation}
This constrains semi-analytic prescriptions for halo mass functions \cite{Asgari2023halo}. We remark that $\Lambda_\delta > 3/2$ for both the Navarro--Frenk--White \cite{NFW1997} and the Einasto profiles \cite{Einasto1965}.

\section{Summary}
\label{sec:summary}

In this paper, we have proposed a physical mechanism that produces the $k^{-d}$ asymptotic scaling of the dark-matter power spectrum naturally. This was done in two ways: by expressing $P(k)$ as a phase-space integral and analysing it with a stationary-phase method (\S \ref{sec:saddle-point approach}), and via a phenomenological study of a critically balanced phase-space turbulent state, akin to Batchelor turbulence, for a cold, collisionless, self-gravitating system (\S \ref{sec:Phase-Space Cascade}).
Both methods are phase-space based, and so remain valid even after streams cross. The fact that the phase-space distribution function is cold (i.e., that it is only non-zero in a $d$-dimensional sub-manifold of phase-space) was crucial to both approaches. Gravitational collapse sources a cascade of the quadratic Casimir invariant, which is sustained by the joint action of phase mixing and tidal forces (by smooth fields). The balance between linear free streaming and tidal forces, which ultimately dictates the form of the power spectrum, also appears in the saddle-point argument of \S \ref{sec:saddle-point approach}. In contrast with many turbulent systems, gravitational turbulence is different in that the flux of the Casimir invariant is not constant over the range of scales of interest, but there still exists a universal scaling r\'{e}gime of the phase-space power spectrum.

The validity of our approach is supported by $1$D Vlasov--Poisson simulations, which confirm our theoretical predictions.

Our determination of the small-scale asymptotics of the dark-matter power spectrum may allow for a non-trivial test of effective field theories of the large scales, either by imposing these asymptotics on them, or by using the phase-space power spectrum found here as a closure for these theories. We intend to investigate these possibilities in future work.

\acknowledgments{We wish to thank Michael Barnes, William Clarke, and Anatoly Spitkovsky for helpful discussions, and are grateful to James Juno for help and advice with the Gkeyll code. Y.B.G. was supported in part by a Leverhulme Trust International Professorship Grant to S. L. Sondhi (No. LIP-2020-014). Y.B.G. and A.A.S. were supported in part by a Simons Investigator Award to A.A.S. from the Simons Foundation. M.L.N. was supported by a Clarendon Scholarship at Oxford. R.J.E. was supported by a UK EPSRC studentship. S.K. and M.B. were supported by the Deutsche Forschungsgemeinschaft (DFG, German Research Foundation) under Germany's Excellence Strategy EXC 2181/1--390900948 (the Heidelberg STRUCTURES Excellence Cluster). This research was also supported in part by NSF grant PHY-2309135 to the Kavli Institute for Theoretical Physics (KITP).}

%

\bibliography{asymptotics}

\begin{thebibliography}{84}%
\makeatletter
\providecommand \@ifxundefined [1]{%
 \@ifx{#1\undefined}
}%
\providecommand \@ifnum [1]{%
 \ifnum #1\expandafter \@firstoftwo
 \else \expandafter \@secondoftwo
 \fi
}%
\providecommand \@ifx [1]{%
 \ifx #1\expandafter \@firstoftwo
 \else \expandafter \@secondoftwo
 \fi
}%
\providecommand \natexlab [1]{#1}%
\providecommand \enquote  [1]{``#1''}%
\providecommand \bibnamefont  [1]{#1}%
\providecommand \bibfnamefont [1]{#1}%
\providecommand \citenamefont [1]{#1}%
\providecommand \href@noop [0]{\@secondoftwo}%
\providecommand \href [0]{\begingroup \@sanitize@url \@href}%
\providecommand \@href[1]{\@@startlink{#1}\@@href}%
\providecommand \@@href[1]{\endgroup#1\@@endlink}%
\providecommand \@sanitize@url [0]{\catcode `\\12\catcode `\$12\catcode
  `\&12\catcode `\#12\catcode `\^12\catcode `\_12\catcode `\%12\relax}%
\providecommand \@@startlink[1]{}%
\providecommand \@@endlink[0]{}%
\providecommand \url  [0]{\begingroup\@sanitize@url \@url }%
\providecommand \@url [1]{\endgroup\@href {#1}{\urlprefix }}%
\providecommand \urlprefix  [0]{URL }%
\providecommand \Eprint [0]{\href }%
\providecommand \doibase [0]{https://doi.org/}%
\providecommand \selectlanguage [0]{\@gobble}%
\providecommand \bibinfo  [0]{\@secondoftwo}%
\providecommand \bibfield  [0]{\@secondoftwo}%
\providecommand \translation [1]{[#1]}%
\providecommand \BibitemOpen [0]{}%
\providecommand \bibitemStop [0]{}%
\providecommand \bibitemNoStop [0]{.\EOS\space}%
\providecommand \EOS [0]{\spacefactor3000\relax}%
\providecommand \BibitemShut  [1]{\csname bibitem#1\endcsname}%
\let\auto@bib@innerbib\@empty
\bibitem [{\citenamefont {{Peebles}}(1980)}]{Peebles1980}%
  \BibitemOpen
  \bibfield  {author} {\bibinfo {author} {\bibfnamefont {P.~J.~E.}\
  \bibnamefont {{Peebles}}},\ }\href@noop {} {\emph {\bibinfo {title} {{The
  Large-Scale Structure of the Universe}}}}\ (\bibinfo  {publisher} {Princeton
  University Press, Princeton, N.J.},\ \bibinfo {year} {1980})\BibitemShut
  {NoStop}%
\bibitem [{\citenamefont {Dodelson}(2003)}]{Dodelson:2003ft}%
  \BibitemOpen
  \bibfield  {author} {\bibinfo {author} {\bibfnamefont {S.}~\bibnamefont
  {Dodelson}},\ }\href@noop {} {\emph {\bibinfo {title} {{Modern Cosmology}}}}\
  (\bibinfo  {publisher} {Academic Press},\ \bibinfo {address} {Amsterdam},\
  \bibinfo {year} {2003})\BibitemShut {NoStop}%
\bibitem [{\citenamefont {{Desjacques}}\ \emph {et~al.}(2018)\citenamefont
  {{Desjacques}}, \citenamefont {{Jeong}},\ and\ \citenamefont
  {{Schmidt}}}]{Desjacquesetal2018}%
  \BibitemOpen
  \bibfield  {author} {\bibinfo {author} {\bibfnamefont {V.}~\bibnamefont
  {{Desjacques}}}, \bibinfo {author} {\bibfnamefont {D.}~\bibnamefont
  {{Jeong}}},\ and\ \bibinfo {author} {\bibfnamefont {F.}~\bibnamefont
  {{Schmidt}}},\ }\bibfield  {title} {\bibinfo {title} {{Large-scale galaxy
  bias}},\ }\href {https://doi.org/10.1016/j.physrep.2017.12.002} {\bibfield
  {journal} {\bibinfo  {journal} {\physrep}\ }\textbf {\bibinfo {volume}
  {733}},\ \bibinfo {pages} {1} (\bibinfo {year} {2018})},\ \Eprint
  {https://arxiv.org/abs/1611.09787} {arXiv:1611.09787 [astro-ph.CO]}
  \BibitemShut {NoStop}%
\bibitem [{\citenamefont {{Aghanim}}\ \emph {et~al.}(2020)\citenamefont
  {{Aghanim}}, \citenamefont {{Akrami}}, \citenamefont {{Ashdown}},
  \citenamefont {{Aumont}}, \citenamefont {{Baccigalupi}}, \citenamefont
  {{Ballardini}}, \citenamefont {{Banday}}, \citenamefont {{Barreiro}},
  \citenamefont {{Bartolo}}, \citenamefont {{Basak}} \emph
  {et~al.}}]{Planck2020}%
  \BibitemOpen
  \bibfield  {author} {\bibinfo {author} {\bibfnamefont {N.}~\bibnamefont
  {{Aghanim}}}, \bibinfo {author} {\bibfnamefont {Y.}~\bibnamefont {{Akrami}}},
  \bibinfo {author} {\bibfnamefont {M.}~\bibnamefont {{Ashdown}}}, \bibinfo
  {author} {\bibfnamefont {J.}~\bibnamefont {{Aumont}}}, \bibinfo {author}
  {\bibfnamefont {C.}~\bibnamefont {{Baccigalupi}}}, \bibinfo {author}
  {\bibfnamefont {M.}~\bibnamefont {{Ballardini}}}, \bibinfo {author}
  {\bibfnamefont {A.~J.}\ \bibnamefont {{Banday}}}, \bibinfo {author}
  {\bibfnamefont {R.~B.}\ \bibnamefont {{Barreiro}}}, \bibinfo {author}
  {\bibfnamefont {N.}~\bibnamefont {{Bartolo}}}, \bibinfo {author}
  {\bibfnamefont {S.}~\bibnamefont {{Basak}}}, \emph {et~al.} (\bibinfo
  {collaboration} {Planck Collaboration}),\ }\bibfield  {title} {\bibinfo
  {title} {{Planck 2018 results. VI. Cosmological parameters}},\ }\href
  {https://doi.org/10.1051/0004-6361/201833910} {\bibfield  {journal} {\bibinfo
   {journal} {\aap}\ }\textbf {\bibinfo {volume} {641}},\ \bibinfo {eid} {A6}
  (\bibinfo {year} {2020})},\ \Eprint {https://arxiv.org/abs/1807.06209}
  {arXiv:1807.06209 [astro-ph.CO]} \BibitemShut {NoStop}%
\bibitem [{\citenamefont {{Tegmark}}\ and\ \citenamefont
  {{Zaldarriaga}}(2002)}]{TegmarkZaldarriaga2002}%
  \BibitemOpen
  \bibfield  {author} {\bibinfo {author} {\bibfnamefont {M.}~\bibnamefont
  {{Tegmark}}}\ and\ \bibinfo {author} {\bibfnamefont {M.}~\bibnamefont
  {{Zaldarriaga}}},\ }\bibfield  {title} {\bibinfo {title} {{Separating the
  early universe from the late universe: Cosmological parameter estimation
  beyond the black box}},\ }\href {https://doi.org/10.1103/PhysRevD.66.103508}
  {\bibfield  {journal} {\bibinfo  {journal} {\prd}\ }\textbf {\bibinfo
  {volume} {66}},\ \bibinfo {eid} {103508} (\bibinfo {year} {2002})},\ \Eprint
  {https://arxiv.org/abs/astro-ph/0207047} {arXiv:astro-ph/0207047 [astro-ph]}
  \BibitemShut {NoStop}%
\bibitem [{\citenamefont {{Cabass}}\ \emph {et~al.}(2023)\citenamefont
  {{Cabass}}, \citenamefont {{Ivanov}}, \citenamefont {{Lewandowski}},
  \citenamefont {{Mirbabayi}},\ and\ \citenamefont
  {{Simonovi{\'c}}}}]{Cabassetal2023}%
  \BibitemOpen
  \bibfield  {author} {\bibinfo {author} {\bibfnamefont {G.}~\bibnamefont
  {{Cabass}}}, \bibinfo {author} {\bibfnamefont {M.~M.}\ \bibnamefont
  {{Ivanov}}}, \bibinfo {author} {\bibfnamefont {M.}~\bibnamefont
  {{Lewandowski}}}, \bibinfo {author} {\bibfnamefont {M.}~\bibnamefont
  {{Mirbabayi}}},\ and\ \bibinfo {author} {\bibfnamefont {M.}~\bibnamefont
  {{Simonovi{\'c}}}},\ }\bibfield  {title} {\bibinfo {title} {{Snowmass white
  paper: Effective field theories in cosmology}},\ }\href
  {https://doi.org/10.1016/j.dark.2023.101193} {\bibfield  {journal} {\bibinfo
  {journal} {Physics of the Dark Universe}\ }\textbf {\bibinfo {volume} {40}},\
  \bibinfo {eid} {101193} (\bibinfo {year} {2023})},\ \Eprint
  {https://arxiv.org/abs/2203.08232} {arXiv:2203.08232 [astro-ph.CO]}
  \BibitemShut {NoStop}%
\bibitem [{\citenamefont {Heggie}\ and\ \citenamefont
  {Hut}(2003)}]{HeggieHut2003}%
  \BibitemOpen
  \bibfield  {author} {\bibinfo {author} {\bibfnamefont {D.}~\bibnamefont
  {Heggie}}\ and\ \bibinfo {author} {\bibfnamefont {P.}~\bibnamefont {Hut}},\
  }\href {https://doi.org/10.1017/CBO9781139164535} {\emph {\bibinfo {title}
  {The Gravitational Million–Body Problem: A Multidisciplinary Approach to
  Star Cluster Dynamics}}}\ (\bibinfo  {publisher} {Cambridge University Press,
  Cambridge},\ \bibinfo {year} {2003})\BibitemShut {NoStop}%
\bibitem [{\citenamefont {{Hamilton}}\ \emph {et~al.}(1991)\citenamefont
  {{Hamilton}}, \citenamefont {{Kumar}}, \citenamefont {{Lu}},\ and\
  \citenamefont {{Matthews}}}]{Hamiltonetal1991}%
  \BibitemOpen
  \bibfield  {author} {\bibinfo {author} {\bibfnamefont {A.~J.~S.}\
  \bibnamefont {{Hamilton}}}, \bibinfo {author} {\bibfnamefont
  {P.}~\bibnamefont {{Kumar}}}, \bibinfo {author} {\bibfnamefont
  {E.}~\bibnamefont {{Lu}}},\ and\ \bibinfo {author} {\bibfnamefont
  {A.}~\bibnamefont {{Matthews}}},\ }\bibfield  {title} {\bibinfo {title}
  {{Reconstructing the primordial spectrum of fluctuations of the Universe from
  the observed nonlinear clustering of galaxies}},\ }\href
  {https://doi.org/10.1086/186057} {\bibfield  {journal} {\bibinfo  {journal}
  {\apjl}\ }\textbf {\bibinfo {volume} {374}},\ \bibinfo {pages} {L1} (\bibinfo
  {year} {1991})}\BibitemShut {NoStop}%
\bibitem [{\citenamefont {{Ginat}}(2021)}]{Ginat2021}%
  \BibitemOpen
  \bibfield  {author} {\bibinfo {author} {\bibfnamefont {Y.~B.}\ \bibnamefont
  {{Ginat}}},\ }\bibfield  {title} {\bibinfo {title} {{Multiple-scales approach
  to the averaging problem in cosmology}},\ }\href
  {https://doi.org/10.1088/1475-7516/2021/02/049} {\bibfield  {journal}
  {\bibinfo  {journal} {\jcap}\ }\textbf {\bibinfo {volume} {2021}},\ \bibinfo
  {eid} {049} (\bibinfo {year} {2021})},\ \Eprint
  {https://arxiv.org/abs/2005.03026} {arXiv:2005.03026 [gr-qc]} \BibitemShut
  {NoStop}%
\bibitem [{\citenamefont {{Baumann}}\ \emph {et~al.}(2012)\citenamefont
  {{Baumann}}, \citenamefont {{Nicolis}}, \citenamefont {{Senatore}},\ and\
  \citenamefont {{Zaldarriaga}}}]{Baumannetal2012}%
  \BibitemOpen
  \bibfield  {author} {\bibinfo {author} {\bibfnamefont {D.}~\bibnamefont
  {{Baumann}}}, \bibinfo {author} {\bibfnamefont {A.}~\bibnamefont
  {{Nicolis}}}, \bibinfo {author} {\bibfnamefont {L.}~\bibnamefont
  {{Senatore}}},\ and\ \bibinfo {author} {\bibfnamefont {M.}~\bibnamefont
  {{Zaldarriaga}}},\ }\bibfield  {title} {\bibinfo {title} {{Cosmological
  non-linearities as an effective fluid}},\ }\href
  {https://doi.org/10.1088/1475-7516/2012/07/051} {\bibfield  {journal}
  {\bibinfo  {journal} {\jcap}\ }\textbf {\bibinfo {volume} {2012}},\ \bibinfo
  {eid} {051} (\bibinfo {year} {2012})},\ \Eprint
  {https://arxiv.org/abs/1004.2488} {arXiv:1004.2488 [astro-ph.CO]}
  \BibitemShut {NoStop}%
\bibitem [{\citenamefont {Baldauf}(2020)}]{Baldauf2020}%
  \BibitemOpen
  \bibfield  {author} {\bibinfo {author} {\bibfnamefont {T.}~\bibnamefont
  {Baldauf}},\ }\bibfield  {title} {\bibinfo {title} {{Effective field theory
  of large-scale structure}},\ }in\ \href
  {https://doi.org/10.1093/oso/9780198855743.003.0007} {\emph {\bibinfo
  {booktitle} {{Effective Field Theory in Particle Physics and Cosmology:
  Lecture Notes of the Les Houches Summer School: Volume 108, July 2017}}}}\
  (\bibinfo  {publisher} {Oxford University Press, Oxford},\ \bibinfo {year}
  {2020})\BibitemShut {NoStop}%
\bibitem [{\citenamefont {Ivanov}(2023)}]{Ivanov2023}%
  \BibitemOpen
  \bibfield  {author} {\bibinfo {author} {\bibfnamefont {M.~M.}\ \bibnamefont
  {Ivanov}},\ }\bibinfo {title} {Effective field theory for large-scale
  structure},\ in\ \href {https://doi.org/10.1007/978-981-19-3079-9_5-1} {\emph
  {\bibinfo {booktitle} {Handbook of Quantum Gravity}}},\ \bibinfo {editor}
  {edited by\ \bibinfo {editor} {\bibfnamefont {C.}~\bibnamefont {Bambi}},
  \bibinfo {editor} {\bibfnamefont {L.}~\bibnamefont {Modesto}},\ and\ \bibinfo
  {editor} {\bibfnamefont {I.}~\bibnamefont {Shapiro}}}\ (\bibinfo  {publisher}
  {Springer},\ \bibinfo {address} {Singapore},\ \bibinfo {year}
  {2023})\BibitemShut {NoStop}%
\bibitem [{\citenamefont {{Springel}}\ \emph {et~al.}(2018)\citenamefont
  {{Springel}}, \citenamefont {{Pakmor}}, \citenamefont {{Pillepich}},
  \citenamefont {{Weinberger}}, \citenamefont {{Nelson}}, \citenamefont
  {{Hernquist}}, \citenamefont {{Vogelsberger}}, \citenamefont {{Genel}},
  \citenamefont {{Torrey}}, \citenamefont {{Marinacci}},\ and\ \citenamefont
  {{Naiman}}}]{Springeletal2018}%
  \BibitemOpen
  \bibfield  {author} {\bibinfo {author} {\bibfnamefont {V.}~\bibnamefont
  {{Springel}}}, \bibinfo {author} {\bibfnamefont {R.}~\bibnamefont
  {{Pakmor}}}, \bibinfo {author} {\bibfnamefont {A.}~\bibnamefont
  {{Pillepich}}}, \bibinfo {author} {\bibfnamefont {R.}~\bibnamefont
  {{Weinberger}}}, \bibinfo {author} {\bibfnamefont {D.}~\bibnamefont
  {{Nelson}}}, \bibinfo {author} {\bibfnamefont {L.}~\bibnamefont
  {{Hernquist}}}, \bibinfo {author} {\bibfnamefont {M.}~\bibnamefont
  {{Vogelsberger}}}, \bibinfo {author} {\bibfnamefont {S.}~\bibnamefont
  {{Genel}}}, \bibinfo {author} {\bibfnamefont {P.}~\bibnamefont {{Torrey}}},
  \bibinfo {author} {\bibfnamefont {F.}~\bibnamefont {{Marinacci}}},\ and\
  \bibinfo {author} {\bibfnamefont {J.}~\bibnamefont {{Naiman}}},\ }\bibfield
  {title} {\bibinfo {title} {{First results from the IllustrisTNG simulations:
  matter and galaxy clustering}},\ }\href
  {https://doi.org/10.1093/mnras/stx3304} {\bibfield  {journal} {\bibinfo
  {journal} {\mnras}\ }\textbf {\bibinfo {volume} {475}},\ \bibinfo {pages}
  {676} (\bibinfo {year} {2018})},\ \Eprint {https://arxiv.org/abs/1707.03397}
  {arXiv:1707.03397 [astro-ph.GA]} \BibitemShut {NoStop}%
\bibitem [{\citenamefont {{Taruya}}\ and\ \citenamefont
  {{Colombi}}(2017)}]{TaruyaColombi2017}%
  \BibitemOpen
  \bibfield  {author} {\bibinfo {author} {\bibfnamefont {A.}~\bibnamefont
  {{Taruya}}}\ and\ \bibinfo {author} {\bibfnamefont {S.}~\bibnamefont
  {{Colombi}}},\ }\bibfield  {title} {\bibinfo {title} {{Post-collapse
  perturbation theory in 1D cosmology---beyond shell-crossing}},\ }\href
  {https://doi.org/10.1093/mnras/stx1501} {\bibfield  {journal} {\bibinfo
  {journal} {\mnras}\ }\textbf {\bibinfo {volume} {470}},\ \bibinfo {pages}
  {4858} (\bibinfo {year} {2017})},\ \Eprint {https://arxiv.org/abs/1701.09088}
  {arXiv:1701.09088 [astro-ph.CO]} \BibitemShut {NoStop}%
\bibitem [{\citenamefont {{Jenkins}}\ \emph {et~al.}(1998)\citenamefont
  {{Jenkins}}, \citenamefont {{Frenk}}, \citenamefont {{Pearce}}, \citenamefont
  {{Thomas}}, \citenamefont {{Colberg}}, \citenamefont {{White}}, \citenamefont
  {{Couchman}}, \citenamefont {{Peacock}}, \citenamefont {{Efstathiou}},\ and\
  \citenamefont {{Nelson}}}]{Jenkinsetal1998}%
  \BibitemOpen
  \bibfield  {author} {\bibinfo {author} {\bibfnamefont {A.}~\bibnamefont
  {{Jenkins}}}, \bibinfo {author} {\bibfnamefont {C.~S.}\ \bibnamefont
  {{Frenk}}}, \bibinfo {author} {\bibfnamefont {F.~R.}\ \bibnamefont
  {{Pearce}}}, \bibinfo {author} {\bibfnamefont {P.~A.}\ \bibnamefont
  {{Thomas}}}, \bibinfo {author} {\bibfnamefont {J.~M.}\ \bibnamefont
  {{Colberg}}}, \bibinfo {author} {\bibfnamefont {S.~D.~M.}\ \bibnamefont
  {{White}}}, \bibinfo {author} {\bibfnamefont {H.~M.~P.}\ \bibnamefont
  {{Couchman}}}, \bibinfo {author} {\bibfnamefont {J.~A.}\ \bibnamefont
  {{Peacock}}}, \bibinfo {author} {\bibfnamefont {G.}~\bibnamefont
  {{Efstathiou}}},\ and\ \bibinfo {author} {\bibfnamefont {A.~H.}\ \bibnamefont
  {{Nelson}}},\ }\bibfield  {title} {\bibinfo {title} {{Evolution of Structure
  in Cold Dark Matter Universes}},\ }\href {https://doi.org/10.1086/305615}
  {\bibfield  {journal} {\bibinfo  {journal} {\apj}\ }\textbf {\bibinfo
  {volume} {499}},\ \bibinfo {pages} {20} (\bibinfo {year} {1998})},\ \Eprint
  {https://arxiv.org/abs/astro-ph/9709010} {arXiv:astro-ph/9709010 [astro-ph]}
  \BibitemShut {NoStop}%
\bibitem [{\citenamefont {{Springel}}\ \emph {et~al.}(2005)\citenamefont
  {{Springel}}, \citenamefont {{White}}, \citenamefont {{Jenkins}},
  \citenamefont {{Frenk}}, \citenamefont {{Yoshida}}, \citenamefont {{Gao}},
  \citenamefont {{Navarro}}, \citenamefont {{Thacker}}, \citenamefont
  {{Croton}}, \citenamefont {{Helly}}, \citenamefont {{Peacock}}, \citenamefont
  {{Cole}}, \citenamefont {{Thomas}}, \citenamefont {{Couchman}}, \citenamefont
  {{Evrard}}, \citenamefont {{Colberg}},\ and\ \citenamefont
  {{Pearce}}}]{Springeletal2005}%
  \BibitemOpen
  \bibfield  {author} {\bibinfo {author} {\bibfnamefont {V.}~\bibnamefont
  {{Springel}}}, \bibinfo {author} {\bibfnamefont {S.~D.~M.}\ \bibnamefont
  {{White}}}, \bibinfo {author} {\bibfnamefont {A.}~\bibnamefont {{Jenkins}}},
  \bibinfo {author} {\bibfnamefont {C.~S.}\ \bibnamefont {{Frenk}}}, \bibinfo
  {author} {\bibfnamefont {N.}~\bibnamefont {{Yoshida}}}, \bibinfo {author}
  {\bibfnamefont {L.}~\bibnamefont {{Gao}}}, \bibinfo {author} {\bibfnamefont
  {J.}~\bibnamefont {{Navarro}}}, \bibinfo {author} {\bibfnamefont
  {R.}~\bibnamefont {{Thacker}}}, \bibinfo {author} {\bibfnamefont
  {D.}~\bibnamefont {{Croton}}}, \bibinfo {author} {\bibfnamefont
  {J.}~\bibnamefont {{Helly}}}, \bibinfo {author} {\bibfnamefont {J.~A.}\
  \bibnamefont {{Peacock}}}, \bibinfo {author} {\bibfnamefont {S.}~\bibnamefont
  {{Cole}}}, \bibinfo {author} {\bibfnamefont {P.}~\bibnamefont {{Thomas}}},
  \bibinfo {author} {\bibfnamefont {H.}~\bibnamefont {{Couchman}}}, \bibinfo
  {author} {\bibfnamefont {A.}~\bibnamefont {{Evrard}}}, \bibinfo {author}
  {\bibfnamefont {J.}~\bibnamefont {{Colberg}}},\ and\ \bibinfo {author}
  {\bibfnamefont {F.}~\bibnamefont {{Pearce}}},\ }\bibfield  {title} {\bibinfo
  {title} {{Simulations of the formation, evolution and clustering of galaxies
  and quasars}},\ }\href {https://doi.org/10.1038/nature03597} {\bibfield
  {journal} {\bibinfo  {journal} {\nat}\ }\textbf {\bibinfo {volume} {435}},\
  \bibinfo {pages} {629} (\bibinfo {year} {2005})},\ \Eprint
  {https://arxiv.org/abs/astro-ph/0504097} {arXiv:astro-ph/0504097 [astro-ph]}
  \BibitemShut {NoStop}%
\bibitem [{\citenamefont {{Mocz}}\ \emph {et~al.}(2020)\citenamefont {{Mocz}},
  \citenamefont {{Fialkov}}, \citenamefont {{Vogelsberger}}, \citenamefont
  {{Becerra}}, \citenamefont {{Shen}}, \citenamefont {{Robles}}, \citenamefont
  {{Amin}}, \citenamefont {{Zavala}}, \citenamefont {{Boylan-Kolchin}},
  \citenamefont {{Bose}}, \citenamefont {{Marinacci}}, \citenamefont
  {{Chavanis}}, \citenamefont {{Lancaster}},\ and\ \citenamefont
  {{Hernquist}}}]{Moczetal2020}%
  \BibitemOpen
  \bibfield  {author} {\bibinfo {author} {\bibfnamefont {P.}~\bibnamefont
  {{Mocz}}}, \bibinfo {author} {\bibfnamefont {A.}~\bibnamefont {{Fialkov}}},
  \bibinfo {author} {\bibfnamefont {M.}~\bibnamefont {{Vogelsberger}}},
  \bibinfo {author} {\bibfnamefont {F.}~\bibnamefont {{Becerra}}}, \bibinfo
  {author} {\bibfnamefont {X.}~\bibnamefont {{Shen}}}, \bibinfo {author}
  {\bibfnamefont {V.~H.}\ \bibnamefont {{Robles}}}, \bibinfo {author}
  {\bibfnamefont {M.~A.}\ \bibnamefont {{Amin}}}, \bibinfo {author}
  {\bibfnamefont {J.}~\bibnamefont {{Zavala}}}, \bibinfo {author}
  {\bibfnamefont {M.}~\bibnamefont {{Boylan-Kolchin}}}, \bibinfo {author}
  {\bibfnamefont {S.}~\bibnamefont {{Bose}}}, \bibinfo {author} {\bibfnamefont
  {F.}~\bibnamefont {{Marinacci}}}, \bibinfo {author} {\bibfnamefont {P.-H.}\
  \bibnamefont {{Chavanis}}}, \bibinfo {author} {\bibfnamefont
  {L.}~\bibnamefont {{Lancaster}}},\ and\ \bibinfo {author} {\bibfnamefont
  {L.}~\bibnamefont {{Hernquist}}},\ }\bibfield  {title} {\bibinfo {title}
  {{Galaxy formation with BECDM - II. Cosmic filaments and first galaxies}},\
  }\href {https://doi.org/10.1093/mnras/staa738} {\bibfield  {journal}
  {\bibinfo  {journal} {\mnras}\ }\textbf {\bibinfo {volume} {494}},\ \bibinfo
  {pages} {2027} (\bibinfo {year} {2020})},\ \Eprint
  {https://arxiv.org/abs/1911.05746} {arXiv:1911.05746 [astro-ph.CO]}
  \BibitemShut {NoStop}%
\bibitem [{\citenamefont {{Chen}}\ and\ \citenamefont
  {{Pietroni}}(2020)}]{ChenPietroni2020}%
  \BibitemOpen
  \bibfield  {author} {\bibinfo {author} {\bibfnamefont {S.-F.}\ \bibnamefont
  {{Chen}}}\ and\ \bibinfo {author} {\bibfnamefont {M.}~\bibnamefont
  {{Pietroni}}},\ }\bibfield  {title} {\bibinfo {title} {{Asymptotic expansions
  for Large Scale Structure}},\ }\href
  {https://doi.org/10.1088/1475-7516/2020/06/033} {\bibfield  {journal}
  {\bibinfo  {journal} {\jcap}\ }\textbf {\bibinfo {volume} {2020}},\ \bibinfo
  {eid} {033} (\bibinfo {year} {2020})},\ \Eprint
  {https://arxiv.org/abs/2002.11357} {arXiv:2002.11357 [astro-ph.CO]}
  \BibitemShut {NoStop}%
\bibitem [{\citenamefont {{Gilbert}}(1968)}]{Gilbert1968}%
  \BibitemOpen
  \bibfield  {author} {\bibinfo {author} {\bibfnamefont {I.~H.}\ \bibnamefont
  {{Gilbert}}},\ }\bibfield  {title} {\bibinfo {title} {{Collisional relaxation
  in stellar systems}},\ }\href {https://doi.org/10.1086/149616} {\bibfield
  {journal} {\bibinfo  {journal} {\apj}\ }\textbf {\bibinfo {volume} {152}},\
  \bibinfo {pages} {1043} (\bibinfo {year} {1968})}\BibitemShut {NoStop}%
\bibitem [{\citenamefont {{Chavanis}}(2012)}]{Chavanis2012}%
  \BibitemOpen
  \bibfield  {author} {\bibinfo {author} {\bibfnamefont {P.-H.}\ \bibnamefont
  {{Chavanis}}},\ }\bibfield  {title} {\bibinfo {title} {{Kinetic theory of
  long-range interacting systems with angle-action variables and collective
  effects}},\ }\href {https://doi.org/10.1016/j.physa.2012.02.019} {\bibfield
  {journal} {\bibinfo  {journal} {Physica A}\ }\textbf {\bibinfo {volume}
  {391}},\ \bibinfo {pages} {3680} (\bibinfo {year} {2012})},\ \Eprint
  {https://arxiv.org/abs/1107.1475} {arXiv:1107.1475 [cond-mat.stat-mech]}
  \BibitemShut {NoStop}%
\bibitem [{\citenamefont {Lazarovici}\ and\ \citenamefont
  {Pickl}(2017)}]{LazaroviciPickl2017}%
  \BibitemOpen
  \bibfield  {author} {\bibinfo {author} {\bibfnamefont {D.}~\bibnamefont
  {Lazarovici}}\ and\ \bibinfo {author} {\bibfnamefont {P.}~\bibnamefont
  {Pickl}},\ }\bibfield  {title} {\bibinfo {title} {A mean field limit for the
  {V}lasov-{P}oisson system},\ }\href
  {https://doi.org/10.1007/s00205-017-1125-0} {\bibfield  {journal} {\bibinfo
  {journal} {Arch. Ration. Mech. Anal.}\ }\textbf {\bibinfo {volume} {225}},\
  \bibinfo {pages} {1201} (\bibinfo {year} {2017})}\BibitemShut {NoStop}%
\bibitem [{\citenamefont {{Halle}}\ \emph {et~al.}(2019)\citenamefont
  {{Halle}}, \citenamefont {{Colombi}},\ and\ \citenamefont
  {{Peirani}}}]{Halleetal2019}%
  \BibitemOpen
  \bibfield  {author} {\bibinfo {author} {\bibfnamefont {A.}~\bibnamefont
  {{Halle}}}, \bibinfo {author} {\bibfnamefont {S.}~\bibnamefont {{Colombi}}},\
  and\ \bibinfo {author} {\bibfnamefont {S.}~\bibnamefont {{Peirani}}},\
  }\bibfield  {title} {\bibinfo {title} {{Phase-space structure analysis of
  self-gravitating collisionless spherical systems}},\ }\href
  {https://doi.org/10.1051/0004-6361/201833460} {\bibfield  {journal} {\bibinfo
   {journal} {\aap}\ }\textbf {\bibinfo {volume} {621}},\ \bibinfo {eid} {A8}
  (\bibinfo {year} {2019})},\ \Eprint {https://arxiv.org/abs/1701.01384}
  {arXiv:1701.01384 [astro-ph.GA]} \BibitemShut {NoStop}%
\bibitem [{\citenamefont {{Rampf}}(2021)}]{Rampf2021}%
  \BibitemOpen
  \bibfield  {author} {\bibinfo {author} {\bibfnamefont {C.}~\bibnamefont
  {{Rampf}}},\ }\bibfield  {title} {\bibinfo {title} {{Cosmological
  Vlasov-Poisson equations for dark matter}},\ }\href
  {https://doi.org/10.1007/s41614-021-00055-z} {\bibfield  {journal} {\bibinfo
  {journal} {Rev. Mod. Plasma Phys.}\ }\textbf {\bibinfo {volume} {5}},\
  \bibinfo {eid} {10} (\bibinfo {year} {2021})}\BibitemShut {NoStop}%
\bibitem [{\citenamefont {{Green}}\ \emph {et~al.}(2004)\citenamefont
  {{Green}}, \citenamefont {{Hofmann}},\ and\ \citenamefont
  {{Schwarz}}}]{Greenetal2004}%
  \BibitemOpen
  \bibfield  {author} {\bibinfo {author} {\bibfnamefont {A.~M.}\ \bibnamefont
  {{Green}}}, \bibinfo {author} {\bibfnamefont {S.}~\bibnamefont {{Hofmann}}},\
  and\ \bibinfo {author} {\bibfnamefont {D.~J.}\ \bibnamefont {{Schwarz}}},\
  }\bibfield  {title} {\bibinfo {title} {{The power spectrum of SUSY-CDM on
  subgalactic scales}},\ }\href
  {https://doi.org/10.1111/j.1365-2966.2004.08232.x} {\bibfield  {journal}
  {\bibinfo  {journal} {\mnras}\ }\textbf {\bibinfo {volume} {353}},\ \bibinfo
  {pages} {L23} (\bibinfo {year} {2004})},\ \Eprint
  {https://arxiv.org/abs/astro-ph/0309621} {arXiv:astro-ph/0309621 [astro-ph]}
  \BibitemShut {NoStop}%
\bibitem [{\citenamefont {{Loeb}}\ and\ \citenamefont
  {{Zaldarriaga}}(2005)}]{LoebZaldarriage2005}%
  \BibitemOpen
  \bibfield  {author} {\bibinfo {author} {\bibfnamefont {A.}~\bibnamefont
  {{Loeb}}}\ and\ \bibinfo {author} {\bibfnamefont {M.}~\bibnamefont
  {{Zaldarriaga}}},\ }\bibfield  {title} {\bibinfo {title} {{Small-scale power
  spectrum of cold dark matter}},\ }\href
  {https://doi.org/10.1103/PhysRevD.71.103520} {\bibfield  {journal} {\bibinfo
  {journal} {\prd}\ }\textbf {\bibinfo {volume} {71}},\ \bibinfo {eid} {103520}
  (\bibinfo {year} {2005})},\ \Eprint {https://arxiv.org/abs/astro-ph/0504112}
  {arXiv:astro-ph/0504112 [astro-ph]} \BibitemShut {NoStop}%
\bibitem [{\citenamefont {{Kunz}}\ \emph {et~al.}(2016)\citenamefont {{Kunz}},
  \citenamefont {{Nesseris}},\ and\ \citenamefont {{Sawicki}}}]{Kunzetal2016}%
  \BibitemOpen
  \bibfield  {author} {\bibinfo {author} {\bibfnamefont {M.}~\bibnamefont
  {{Kunz}}}, \bibinfo {author} {\bibfnamefont {S.}~\bibnamefont {{Nesseris}}},\
  and\ \bibinfo {author} {\bibfnamefont {I.}~\bibnamefont {{Sawicki}}},\
  }\bibfield  {title} {\bibinfo {title} {{Constraints on dark-matter properties
  from large-scale structure}},\ }\href
  {https://doi.org/10.1103/PhysRevD.94.023510} {\bibfield  {journal} {\bibinfo
  {journal} {\prd}\ }\textbf {\bibinfo {volume} {94}},\ \bibinfo {eid} {023510}
  (\bibinfo {year} {2016})},\ \Eprint {https://arxiv.org/abs/1604.05701}
  {arXiv:1604.05701 [astro-ph.CO]} \BibitemShut {NoStop}%
\bibitem [{\citenamefont {{Thomas}}\ \emph {et~al.}(2016)\citenamefont
  {{Thomas}}, \citenamefont {{Kopp}},\ and\ \citenamefont
  {{Skordis}}}]{Thomasetal2016}%
  \BibitemOpen
  \bibfield  {author} {\bibinfo {author} {\bibfnamefont {D.~B.}\ \bibnamefont
  {{Thomas}}}, \bibinfo {author} {\bibfnamefont {M.}~\bibnamefont {{Kopp}}},\
  and\ \bibinfo {author} {\bibfnamefont {C.}~\bibnamefont {{Skordis}}},\
  }\bibfield  {title} {\bibinfo {title} {{Constraining the properties of dark
  matter with observations of the Cosmic Microwave Background}},\ }\href
  {https://doi.org/10.3847/0004-637X/830/2/155} {\bibfield  {journal} {\bibinfo
   {journal} {\apj}\ }\textbf {\bibinfo {volume} {830}},\ \bibinfo {eid} {155}
  (\bibinfo {year} {2016})},\ \Eprint {https://arxiv.org/abs/1601.05097}
  {arXiv:1601.05097 [astro-ph.CO]} \BibitemShut {NoStop}%
\bibitem [{\citenamefont {{Gilman}}\ \emph {et~al.}(2020)\citenamefont
  {{Gilman}}, \citenamefont {{Birrer}}, \citenamefont {{Nierenberg}},
  \citenamefont {{Treu}}, \citenamefont {{Du}},\ and\ \citenamefont
  {{Benson}}}]{Gilmanetal2020}%
  \BibitemOpen
  \bibfield  {author} {\bibinfo {author} {\bibfnamefont {D.}~\bibnamefont
  {{Gilman}}}, \bibinfo {author} {\bibfnamefont {S.}~\bibnamefont {{Birrer}}},
  \bibinfo {author} {\bibfnamefont {A.}~\bibnamefont {{Nierenberg}}}, \bibinfo
  {author} {\bibfnamefont {T.}~\bibnamefont {{Treu}}}, \bibinfo {author}
  {\bibfnamefont {X.}~\bibnamefont {{Du}}},\ and\ \bibinfo {author}
  {\bibfnamefont {A.}~\bibnamefont {{Benson}}},\ }\bibfield  {title} {\bibinfo
  {title} {{Warm dark matter chills out: constraints on the halo mass function
  and the free-streaming length of dark matter with eight quadruple-image
  strong gravitational lenses}},\ }\href
  {https://doi.org/10.1093/mnras/stz3480} {\bibfield  {journal} {\bibinfo
  {journal} {\mnras}\ }\textbf {\bibinfo {volume} {491}},\ \bibinfo {pages}
  {6077} (\bibinfo {year} {2020})},\ \Eprint {https://arxiv.org/abs/1908.06983}
  {arXiv:1908.06983 [astro-ph.CO]} \BibitemShut {NoStop}%
\bibitem [{\citenamefont {{Ili{\'c}}}\ \emph {et~al.}(2021)\citenamefont
  {{Ili{\'c}}}, \citenamefont {{Kopp}}, \citenamefont {{Skordis}},\ and\
  \citenamefont {{Thomas}}}]{Ilicetal2021}%
  \BibitemOpen
  \bibfield  {author} {\bibinfo {author} {\bibfnamefont {S.}~\bibnamefont
  {{Ili{\'c}}}}, \bibinfo {author} {\bibfnamefont {M.}~\bibnamefont {{Kopp}}},
  \bibinfo {author} {\bibfnamefont {C.}~\bibnamefont {{Skordis}}},\ and\
  \bibinfo {author} {\bibfnamefont {D.~B.}\ \bibnamefont {{Thomas}}},\
  }\bibfield  {title} {\bibinfo {title} {{Dark matter properties through cosmic
  history}},\ }\href {https://doi.org/10.1103/PhysRevD.104.043520} {\bibfield
  {journal} {\bibinfo  {journal} {\prd}\ }\textbf {\bibinfo {volume} {104}},\
  \bibinfo {eid} {043520} (\bibinfo {year} {2021})},\ \Eprint
  {https://arxiv.org/abs/2004.09572} {arXiv:2004.09572 [astro-ph.CO]}
  \BibitemShut {NoStop}%
\bibitem [{\citenamefont {{Menci}}(2002)}]{Menci2002}%
  \BibitemOpen
  \bibfield  {author} {\bibinfo {author} {\bibfnamefont {N.}~\bibnamefont
  {{Menci}}},\ }\bibfield  {title} {\bibinfo {title} {{An Eulerian perturbation
  approach to large-scale structures: extending the adhesion approximation}},\
  }\href {https://doi.org/10.1046/j.1365-8711.2002.05133.x} {\bibfield
  {journal} {\bibinfo  {journal} {\mnras}\ }\textbf {\bibinfo {volume} {330}},\
  \bibinfo {pages} {907} (\bibinfo {year} {2002})},\ \Eprint
  {https://arxiv.org/abs/astro-ph/0111228} {arXiv:astro-ph/0111228 [astro-ph]}
  \BibitemShut {NoStop}%
\bibitem [{\citenamefont {{Kozlikin}}\ \emph {et~al.}(2021)\citenamefont
  {{Kozlikin}}, \citenamefont {{Lilow}}, \citenamefont {{Fabis}},\ and\
  \citenamefont {{Bartelmann}}}]{Kozlikinetal2021}%
  \BibitemOpen
  \bibfield  {author} {\bibinfo {author} {\bibfnamefont {E.}~\bibnamefont
  {{Kozlikin}}}, \bibinfo {author} {\bibfnamefont {R.}~\bibnamefont {{Lilow}}},
  \bibinfo {author} {\bibfnamefont {F.}~\bibnamefont {{Fabis}}},\ and\ \bibinfo
  {author} {\bibfnamefont {M.}~\bibnamefont {{Bartelmann}}},\ }\bibfield
  {title} {\bibinfo {title} {{A first comparison of Kinetic Field Theory with
  Eulerian Standard Perturbation Theory}},\ }\href
  {https://doi.org/10.1088/1475-7516/2021/06/035} {\bibfield  {journal}
  {\bibinfo  {journal} {\jcap}\ }\textbf {\bibinfo {volume} {2021}},\ \bibinfo
  {eid} {035} (\bibinfo {year} {2021})},\ \Eprint
  {https://arxiv.org/abs/2012.05812} {arXiv:2012.05812 [astro-ph.CO]}
  \BibitemShut {NoStop}%
\bibitem [{\citenamefont {{Knorr}}(1977)}]{Knorr1977}%
  \BibitemOpen
  \bibfield  {author} {\bibinfo {author} {\bibfnamefont {G.}~\bibnamefont
  {{Knorr}}},\ }\bibfield  {title} {\bibinfo {title} {{Time asymptotic
  statistics of the Vlasov equation}},\ }\href
  {https://doi.org/10.1017/S0022377800020808} {\bibfield  {journal} {\bibinfo
  {journal} {J. Plasma Phys.}\ }\textbf {\bibinfo {volume} {17}},\ \bibinfo
  {pages} {553} (\bibinfo {year} {1977})}\BibitemShut {NoStop}%
\bibitem [{\citenamefont {Diamond}\ \emph {et~al.}(2010)\citenamefont
  {Diamond}, \citenamefont {Itoh},\ and\ \citenamefont
  {Itoh}}]{Diamond_Itoh_Itoh_2010}%
  \BibitemOpen
  \bibfield  {author} {\bibinfo {author} {\bibfnamefont {P.~H.}\ \bibnamefont
  {Diamond}}, \bibinfo {author} {\bibfnamefont {S.-I.}\ \bibnamefont {Itoh}},\
  and\ \bibinfo {author} {\bibfnamefont {K.}~\bibnamefont {Itoh}},\ }\href@noop
  {} {\emph {\bibinfo {title} {Modern Plasma Physics}}}\ (\bibinfo  {publisher}
  {Cambridge University Press, Cambridge},\ \bibinfo {year} {2010})\BibitemShut
  {NoStop}%
\bibitem [{\citenamefont {{Lesur}}\ and\ \citenamefont
  {{Diamond}}(2013)}]{LesurDiamond2013}%
  \BibitemOpen
  \bibfield  {author} {\bibinfo {author} {\bibfnamefont {M.}~\bibnamefont
  {{Lesur}}}\ and\ \bibinfo {author} {\bibfnamefont {P.~H.}\ \bibnamefont
  {{Diamond}}},\ }\bibfield  {title} {\bibinfo {title} {{Nonlinear
  instabilities driven by coherent phase-space structures}},\ }\href
  {https://doi.org/10.1103/PhysRevE.87.031101} {\bibfield  {journal} {\bibinfo
  {journal} {\pre}\ }\textbf {\bibinfo {volume} {87}},\ \bibinfo {eid} {031101}
  (\bibinfo {year} {2013})}\BibitemShut {NoStop}%
\bibitem [{\citenamefont {{Nastac}}\ \emph {et~al.}(2024)\citenamefont
  {{Nastac}}, \citenamefont {{Ewart}}, \citenamefont {{Sengupta}},
  \citenamefont {{Schekochihin}}, \citenamefont {{Barnes}},\ and\ \citenamefont
  {{Dorland}}}]{Nastacetal2023}%
  \BibitemOpen
  \bibfield  {author} {\bibinfo {author} {\bibfnamefont {M.~L.}\ \bibnamefont
  {{Nastac}}}, \bibinfo {author} {\bibfnamefont {R.~J.}\ \bibnamefont
  {{Ewart}}}, \bibinfo {author} {\bibfnamefont {W.}~\bibnamefont {{Sengupta}}},
  \bibinfo {author} {\bibfnamefont {A.~A.}\ \bibnamefont {{Schekochihin}}},
  \bibinfo {author} {\bibfnamefont {M.}~\bibnamefont {{Barnes}}},\ and\
  \bibinfo {author} {\bibfnamefont {W.~D.}\ \bibnamefont {{Dorland}}},\
  }\bibfield  {title} {\bibinfo {title} {{Phase-space entropy cascade and
  irreversibility of stochastic heating in nearly collisionless plasma
  turbulence}},\ }\href {https://doi.org/10.1103/PhysRevE.109.065210}
  {\bibfield  {journal} {\bibinfo  {journal} {\pre}\ }\textbf {\bibinfo
  {volume} {109}},\ \bibinfo {eid} {065210} (\bibinfo {year} {2024})},\ \Eprint
  {https://arxiv.org/abs/2310.18211} {arXiv:2310.18211 [physics.plasm-ph]}
  \BibitemShut {NoStop}%
\bibitem [{\citenamefont {{Nastac}}\ \emph {et~al.}(2025)\citenamefont
  {{Nastac}}, \citenamefont {{Ewart}}, \citenamefont {{Juno}}, \citenamefont
  {{Barnes}},\ and\ \citenamefont {{Schekochihin}}}]{Nastacetal2024}%
  \BibitemOpen
  \bibfield  {author} {\bibinfo {author} {\bibfnamefont {M.~L.}\ \bibnamefont
  {{Nastac}}}, \bibinfo {author} {\bibfnamefont {R.~J.}\ \bibnamefont
  {{Ewart}}}, \bibinfo {author} {\bibfnamefont {J.}~\bibnamefont {{Juno}}},
  \bibinfo {author} {\bibfnamefont {M.}~\bibnamefont {{Barnes}}},\ and\
  \bibinfo {author} {\bibfnamefont {A.~A.}\ \bibnamefont {{Schekochihin}}},\
  }\bibfield  {title} {\bibinfo {title} {{Universal fluctuation spectrum of
  Vlasov-Poisson turbulence}},\ }\href
  {https://doi.org/10.48550/arXiv.2503.17278} {\bibfield  {journal} {\bibinfo
  {journal} {arXiv e-prints}\ ,\ \bibinfo {eid} {arXiv:2503.17278}} (\bibinfo
  {year} {2025})},\ \Eprint {https://arxiv.org/abs/2503.17278}
  {arXiv:2503.17278 [physics.plasm-ph]} \BibitemShut {NoStop}%
\bibitem [{\citenamefont {{A. A. Schekochihin}}(2024)}]{AlexKTnotes}%
  \BibitemOpen
  \bibfield  {author} {\bibinfo {author} {\bibnamefont {{A. A.
  Schekochihin}}},\ }\href@noop {} {\bibinfo {title} {Lectures on kinetic
  theory and magnetohydrodynamics of plasmas}},\ \bibinfo {howpublished}
  {\href{https://www-thphys.physics.ox.ac.uk/people/AlexanderSchekochihin/KT/2015/KTLectureNotes.pdf}{available
  on-line}} (\bibinfo {year} {2024}),\ \bibinfo {note} {{University of Oxford
  MMathPhys course}}\BibitemShut {NoStop}%
\bibitem [{\citenamefont {{Kolmogorov}}(1941)}]{Kolmogorov1941}%
  \BibitemOpen
  \bibfield  {author} {\bibinfo {author} {\bibfnamefont {A.}~\bibnamefont
  {{Kolmogorov}}},\ }\bibfield  {title} {\bibinfo {title} {{The local structure
  of turbulence in incompressible viscous fluid for very large Reynolds'
  numbers}},\ }\href
  {https://royalsocietypublishing.org/doi/pdf/10.1098/rspa.1991.0075}
  {\bibfield  {journal} {\bibinfo  {journal} {Akad. Nauk SSSR Doklady}\
  }\textbf {\bibinfo {volume} {30}},\ \bibinfo {pages} {301} (\bibinfo {year}
  {1941})}\BibitemShut {NoStop}%
\bibitem [{\citenamefont {{Batchelor}}(1959)}]{Batchelor1959}%
  \BibitemOpen
  \bibfield  {author} {\bibinfo {author} {\bibfnamefont {G.~K.}\ \bibnamefont
  {{Batchelor}}},\ }\bibfield  {title} {\bibinfo {title} {{Small-scale
  variation of convected quantities like temperature in turbulent fluid. Part
  1. General discussion and the case of small conductivity}},\ }\href
  {https://www.cambridge.org/core/journals/journal-of-fluid-mechanics/article/smallscale-variation-of-convected-quantities-like-temperature-in-turbulent-fluid-part-1-general-discussion-and-the-case-of-small-conductivity/A8AEA175A906F98CDDB0F9ED146BB9FE}
  {\bibfield  {journal} {\bibinfo  {journal} {J. Fluid Mech.}\ }\textbf
  {\bibinfo {volume} {5}},\ \bibinfo {pages} {113} (\bibinfo {year}
  {1959})}\BibitemShut {NoStop}%
\bibitem [{\citenamefont {{Konrad}}\ and\ \citenamefont
  {{Bartelmann}}(2022{\natexlab{a}})}]{KonradBartelmann2022}%
  \BibitemOpen
  \bibfield  {author} {\bibinfo {author} {\bibfnamefont {S.}~\bibnamefont
  {{Konrad}}}\ and\ \bibinfo {author} {\bibfnamefont {M.}~\bibnamefont
  {{Bartelmann}}},\ }\bibfield  {title} {\bibinfo {title} {{Kinetic field
  theory for cosmic structure formation}},\ }\href
  {https://doi.org/10.1007/s40766-022-00037-y} {\bibfield  {journal} {\bibinfo
  {journal} {Rivista Nuovo Cimento}\ }\textbf {\bibinfo {volume} {45}},\
  \bibinfo {pages} {737} (\bibinfo {year} {2022}{\natexlab{a}})},\ \Eprint
  {https://arxiv.org/abs/2202.11077} {arXiv:2202.11077 [astro-ph.CO]}
  \BibitemShut {NoStop}%
\bibitem [{\citenamefont {{Zel'dovich}}(1970)}]{Zeldovich1970}%
  \BibitemOpen
  \bibfield  {author} {\bibinfo {author} {\bibfnamefont {Y.~B.}\ \bibnamefont
  {{Zel'dovich}}},\ }\bibfield  {title} {\bibinfo {title} {{Gravitational
  instability: an approximate theory for large density perturbations}},\
  }\href@noop {} {\bibfield  {journal} {\bibinfo  {journal} {\aap}\ }\textbf
  {\bibinfo {volume} {500}},\ \bibinfo {pages} {13} (\bibinfo {year}
  {1970})}\BibitemShut {NoStop}%
\bibitem [{\citenamefont {{Shandarin}}\ and\ \citenamefont
  {{Zel'dovich}}(1989)}]{ShandarinZeldovich1989}%
  \BibitemOpen
  \bibfield  {author} {\bibinfo {author} {\bibfnamefont {S.~F.}\ \bibnamefont
  {{Shandarin}}}\ and\ \bibinfo {author} {\bibfnamefont {Y.~B.}\ \bibnamefont
  {{Zel'dovich}}},\ }\bibfield  {title} {\bibinfo {title} {{The large-scale
  structure of the universe: Turbulence, intermittency, structures in a
  self-gravitating medium}},\ }\href
  {https://doi.org/10.1103/RevModPhys.61.185} {\bibfield  {journal} {\bibinfo
  {journal} {Rev. Mod. Phys.}\ }\textbf {\bibinfo {volume} {61}},\ \bibinfo
  {pages} {185} (\bibinfo {year} {1989})}\BibitemShut {NoStop}%
\bibitem [{\citenamefont {Gaite}(2012)}]{Gaite_2012}%
  \BibitemOpen
  \bibfield  {author} {\bibinfo {author} {\bibfnamefont {J.}~\bibnamefont
  {Gaite}},\ }\bibfield  {title} {\bibinfo {title} {A non-perturbative
  kolmogorov turbulence approach to the cosmic web structure},\ }\href
  {https://doi.org/10.1209/0295-5075/98/49002} {\bibfield  {journal} {\bibinfo
  {journal} {Europhysics Letters}\ }\textbf {\bibinfo {volume} {98}},\ \bibinfo
  {pages} {49002} (\bibinfo {year} {2012})}\BibitemShut {NoStop}%
\bibitem [{\citenamefont {{Jeans}}(1902)}]{Jeans1902}%
  \BibitemOpen
  \bibfield  {author} {\bibinfo {author} {\bibfnamefont {J.~H.}\ \bibnamefont
  {{Jeans}}},\ }\bibfield  {title} {\bibinfo {title} {{The stability of a
  spherical nebula}},\ }\href {https://doi.org/10.1098/rsta.1902.0012}
  {\bibfield  {journal} {\bibinfo  {journal} {Phil. Trans. Roy. Soc. London A}\
  }\textbf {\bibinfo {volume} {199}},\ \bibinfo {pages} {1} (\bibinfo {year}
  {1902})}\BibitemShut {NoStop}%
\bibitem [{\citenamefont {Arnold}\ \emph {et~al.}(2006)\citenamefont {Arnold},
  \citenamefont {Kozlov},\ and\ \citenamefont {Neishtadt}}]{Arnoldetal2006}%
  \BibitemOpen
  \bibfield  {author} {\bibinfo {author} {\bibfnamefont {V.~I.}\ \bibnamefont
  {Arnold}}, \bibinfo {author} {\bibfnamefont {V.~V.}\ \bibnamefont {Kozlov}},\
  and\ \bibinfo {author} {\bibfnamefont {A.~I.}\ \bibnamefont {Neishtadt}},\
  }\href@noop {} {\emph {\bibinfo {title} {Mathematical Aspects of Classical
  and Celestial Mechanics}}},\ \bibinfo {edition} {3rd}\ ed.,\ \bibinfo
  {series} {Encyclopaedia of Mathematical Sciences}, Vol.~\bibinfo {volume}
  {3}\ (\bibinfo  {publisher} {Springer-Verlag, Berlin},\ \bibinfo {year}
  {2006})\ \bibinfo {note} {[Dynamical systems. III], Translated from the
  Russian original by E. Khukhro}\BibitemShut {NoStop}%
\bibitem [{\citenamefont {Juno}\ \emph {et~al.}(2018)\citenamefont {Juno},
  \citenamefont {Hakim}, \citenamefont {TenBarge}, \citenamefont {Shi},\ and\
  \citenamefont {Dorland}}]{JUNO2018110}%
  \BibitemOpen
  \bibfield  {author} {\bibinfo {author} {\bibfnamefont {J.}~\bibnamefont
  {Juno}}, \bibinfo {author} {\bibfnamefont {A.}~\bibnamefont {Hakim}},
  \bibinfo {author} {\bibfnamefont {J.}~\bibnamefont {TenBarge}}, \bibinfo
  {author} {\bibfnamefont {E.}~\bibnamefont {Shi}},\ and\ \bibinfo {author}
  {\bibfnamefont {W.}~\bibnamefont {Dorland}},\ }\bibfield  {title} {\bibinfo
  {title} {{Discontinuous Galerkin algorithms for fully kinetic plasmas}},\
  }\href {https://doi.org/https://doi.org/10.1016/j.jcp.2017.10.009} {\bibfield
   {journal} {\bibinfo  {journal} {J. Comput. Phys.}\ }\textbf {\bibinfo
  {volume} {353}},\ \bibinfo {pages} {110} (\bibinfo {year}
  {2018})}\BibitemShut {NoStop}%
\bibitem [{\citenamefont {Frisch}(1995)}]{Frisch_1995}%
  \BibitemOpen
  \bibfield  {author} {\bibinfo {author} {\bibfnamefont {U.}~\bibnamefont
  {Frisch}},\ }\href@noop {} {\emph {\bibinfo {title} {Turbulence: The Legacy
  of A. N. Kolmogorov}}}\ (\bibinfo  {publisher} {Cambridge University Press,
  Cambridge},\ \bibinfo {year} {1995})\BibitemShut {NoStop}%
\bibitem [{\citenamefont {{Bertschinger}}\ and\ \citenamefont
  {{Jain}}(1994)}]{BertschingerJain1994}%
  \BibitemOpen
  \bibfield  {author} {\bibinfo {author} {\bibfnamefont {E.}~\bibnamefont
  {{Bertschinger}}}\ and\ \bibinfo {author} {\bibfnamefont {B.}~\bibnamefont
  {{Jain}}},\ }\bibfield  {title} {\bibinfo {title} {{Gravitational instability
  of cold matter}},\ }\href {https://doi.org/10.1086/174501} {\bibfield
  {journal} {\bibinfo  {journal} {\apj}\ }\textbf {\bibinfo {volume} {431}},\
  \bibinfo {pages} {486} (\bibinfo {year} {1994})},\ \Eprint
  {https://arxiv.org/abs/astro-ph/9307033} {arXiv:astro-ph/9307033 [astro-ph]}
  \BibitemShut {NoStop}%
\bibitem [{\citenamefont {{H\'{e}non}}(1973)}]{Henon1973}%
  \BibitemOpen
  \bibfield  {author} {\bibinfo {author} {\bibfnamefont {M.}~\bibnamefont
  {{H\'{e}non}}},\ }\bibfield  {title} {\bibinfo {title} {{Numerical
  experiments on the stability of spherical stellar systems}},\ }\href@noop {}
  {\bibfield  {journal} {\bibinfo  {journal} {\aap}\ }\textbf {\bibinfo
  {volume} {24}},\ \bibinfo {pages} {229} (\bibinfo {year} {1973})}\BibitemShut
  {NoStop}%
\bibitem [{\citenamefont {{Fillmore}}\ and\ \citenamefont
  {{Goldreich}}(1984)}]{FillmoreGoldrecih1984}%
  \BibitemOpen
  \bibfield  {author} {\bibinfo {author} {\bibfnamefont {J.~A.}\ \bibnamefont
  {{Fillmore}}}\ and\ \bibinfo {author} {\bibfnamefont {P.}~\bibnamefont
  {{Goldreich}}},\ }\bibfield  {title} {\bibinfo {title} {{Self-similar
  gravitational collapse in an expanding universe}},\ }\href
  {https://doi.org/10.1086/162070} {\bibfield  {journal} {\bibinfo  {journal}
  {\apj}\ }\textbf {\bibinfo {volume} {281}},\ \bibinfo {pages} {1} (\bibinfo
  {year} {1984})}\BibitemShut {NoStop}%
\bibitem [{\citenamefont {{Bertschinger}}(1985)}]{Bertschinger1985}%
  \BibitemOpen
  \bibfield  {author} {\bibinfo {author} {\bibfnamefont {E.}~\bibnamefont
  {{Bertschinger}}},\ }\bibfield  {title} {\bibinfo {title} {{Self-similar
  secondary infall and accretion in an Einstein-de Sitter universe}},\ }\href
  {https://doi.org/10.1086/191028} {\bibfield  {journal} {\bibinfo  {journal}
  {\apjs}\ }\textbf {\bibinfo {volume} {58}},\ \bibinfo {pages} {39} (\bibinfo
  {year} {1985})}\BibitemShut {NoStop}%
\bibitem [{\citenamefont {{Barnes}}\ \emph {et~al.}(1986)\citenamefont
  {{Barnes}}, \citenamefont {{Goodman}},\ and\ \citenamefont
  {{Hut}}}]{Barnesetal1986}%
  \BibitemOpen
  \bibfield  {author} {\bibinfo {author} {\bibfnamefont {J.}~\bibnamefont
  {{Barnes}}}, \bibinfo {author} {\bibfnamefont {J.}~\bibnamefont
  {{Goodman}}},\ and\ \bibinfo {author} {\bibfnamefont {P.}~\bibnamefont
  {{Hut}}},\ }\bibfield  {title} {\bibinfo {title} {{Dynamical instabilities in
  spherical stellar systems}},\ }\href {https://doi.org/10.1086/163786}
  {\bibfield  {journal} {\bibinfo  {journal} {\apj}\ }\textbf {\bibinfo
  {volume} {300}},\ \bibinfo {pages} {112} (\bibinfo {year}
  {1986})}\BibitemShut {NoStop}%
\bibitem [{\citenamefont {{Binney}}(2004)}]{Binney2004}%
  \BibitemOpen
  \bibfield  {author} {\bibinfo {author} {\bibfnamefont {J.}~\bibnamefont
  {{Binney}}},\ }\bibfield  {title} {\bibinfo {title} {{Discreteness effects in
  cosmological N-body simulations}},\ }\href
  {https://doi.org/10.1111/j.1365-2966.2004.07699.x} {\bibfield  {journal}
  {\bibinfo  {journal} {\mnras}\ }\textbf {\bibinfo {volume} {350}},\ \bibinfo
  {pages} {939} (\bibinfo {year} {2004})},\ \Eprint
  {https://arxiv.org/abs/astro-ph/0311155} {arXiv:astro-ph/0311155 [astro-ph]}
  \BibitemShut {NoStop}%
\bibitem [{\citenamefont {{Colombi}}\ and\ \citenamefont
  {{Touma}}(2014)}]{ColombiTouma2014}%
  \BibitemOpen
  \bibfield  {author} {\bibinfo {author} {\bibfnamefont {S.}~\bibnamefont
  {{Colombi}}}\ and\ \bibinfo {author} {\bibfnamefont {J.}~\bibnamefont
  {{Touma}}},\ }\bibfield  {title} {\bibinfo {title} {{Vlasov-Poisson in 1D:
  waterbags}},\ }\href {https://doi.org/10.1093/mnras/stu739} {\bibfield
  {journal} {\bibinfo  {journal} {\mnras}\ }\textbf {\bibinfo {volume} {441}},\
  \bibinfo {pages} {2414} (\bibinfo {year} {2014})}\BibitemShut {NoStop}%
\bibitem [{\citenamefont {{White}}(2014)}]{White2014}%
  \BibitemOpen
  \bibfield  {author} {\bibinfo {author} {\bibfnamefont {M.}~\bibnamefont
  {{White}}},\ }\bibfield  {title} {\bibinfo {title} {{The Zel'dovich
  approximation}},\ }\href {https://doi.org/10.1093/mnras/stu209} {\bibfield
  {journal} {\bibinfo  {journal} {\mnras}\ }\textbf {\bibinfo {volume} {439}},\
  \bibinfo {pages} {3630} (\bibinfo {year} {2014})},\ \Eprint
  {https://arxiv.org/abs/1401.5466} {arXiv:1401.5466 [astro-ph.CO]}
  \BibitemShut {NoStop}%
\bibitem [{\citenamefont {{Adkins}}\ and\ \citenamefont
  {{Schekochihin}}(2018)}]{AdkinsSchekochihin2018}%
  \BibitemOpen
  \bibfield  {author} {\bibinfo {author} {\bibfnamefont {T.}~\bibnamefont
  {{Adkins}}}\ and\ \bibinfo {author} {\bibfnamefont {A.~A.}\ \bibnamefont
  {{Schekochihin}}},\ }\bibfield  {title} {\bibinfo {title} {{A solvable model
  of Vlasov-kinetic plasma turbulence in Fourier-Hermite phase space}},\ }\href
  {https://doi.org/10.1017/S0022377818000089} {\bibfield  {journal} {\bibinfo
  {journal} {J. Plasma Phys.}\ }\textbf {\bibinfo {volume} {84}},\ \bibinfo
  {eid} {905840107} (\bibinfo {year} {2018})},\ \Eprint
  {https://arxiv.org/abs/1709.03203} {arXiv:1709.03203 [physics.plasm-ph]}
  \BibitemShut {NoStop}%
\bibitem [{\citenamefont {{Sreenivasan}}(2019)}]{Sreenivasan2019}%
  \BibitemOpen
  \bibfield  {author} {\bibinfo {author} {\bibfnamefont {K.~R.}\ \bibnamefont
  {{Sreenivasan}}},\ }\bibfield  {title} {\bibinfo {title} {{Turbulent mixing:
  A perspective}},\ }\href {https://doi.org/10.1073/pnas.1800463115} {\bibfield
   {journal} {\bibinfo  {journal} {PNAS}\ }\textbf {\bibinfo {volume} {116}},\
  \bibinfo {pages} {18175} (\bibinfo {year} {2019})}\BibitemShut {NoStop}%
\bibitem [{\citenamefont {{Goldreich}}\ and\ \citenamefont
  {{Sridhar}}(1995)}]{GoldreichSridhar1995}%
  \BibitemOpen
  \bibfield  {author} {\bibinfo {author} {\bibfnamefont {P.}~\bibnamefont
  {{Goldreich}}}\ and\ \bibinfo {author} {\bibfnamefont {S.}~\bibnamefont
  {{Sridhar}}},\ }\bibfield  {title} {\bibinfo {title} {{Toward a theory of
  interstellar turbulence. II. Strong Alfv\'{e}nic turbulence}},\ }\href
  {https://doi.org/10.1086/175121} {\bibfield  {journal} {\bibinfo  {journal}
  {\apj}\ }\textbf {\bibinfo {volume} {438}},\ \bibinfo {pages} {763} (\bibinfo
  {year} {1995})}\BibitemShut {NoStop}%
\bibitem [{\citenamefont {{Nazarenko}}\ and\ \citenamefont
  {{Schekochihin}}(2011)}]{NazarenkoSchekochihin2011}%
  \BibitemOpen
  \bibfield  {author} {\bibinfo {author} {\bibfnamefont {S.~V.}\ \bibnamefont
  {{Nazarenko}}}\ and\ \bibinfo {author} {\bibfnamefont {A.~A.}\ \bibnamefont
  {{Schekochihin}}},\ }\bibfield  {title} {\bibinfo {title} {{Critical balance
  in magnetohydrodynamic, rotating and stratified turbulence: towards a
  universal scaling conjecture}},\ }\href
  {https://doi.org/10.1017/S002211201100067X} {\bibfield  {journal} {\bibinfo
  {journal} {J. Fluid Mech.}\ }\textbf {\bibinfo {volume} {677}},\ \bibinfo
  {pages} {134} (\bibinfo {year} {2011})},\ \Eprint
  {https://arxiv.org/abs/0904.3488} {arXiv:0904.3488 [physics.flu-dyn]}
  \BibitemShut {NoStop}%
\bibitem [{\citenamefont {{Schekochihin}}(2022)}]{Schekochihin2022}%
  \BibitemOpen
  \bibfield  {author} {\bibinfo {author} {\bibfnamefont {A.~A.}\ \bibnamefont
  {{Schekochihin}}},\ }\bibfield  {title} {\bibinfo {title} {{MHD turbulence: a
  biased review}},\ }\href {https://doi.org/10.1017/S0022377822000721}
  {\bibfield  {journal} {\bibinfo  {journal} {J. Plasma Phys.}\ }\textbf
  {\bibinfo {volume} {88}},\ \bibinfo {eid} {155880501} (\bibinfo {year}
  {2022})},\ \Eprint {https://arxiv.org/abs/2010.00699} {arXiv:2010.00699
  [physics.plasm-ph]} \BibitemShut {NoStop}%
\bibitem [{\citenamefont {Hawking}\ and\ \citenamefont
  {Ellis}(1973)}]{Hawking_Ellis_1973}%
  \BibitemOpen
  \bibfield  {author} {\bibinfo {author} {\bibfnamefont {S.~W.}\ \bibnamefont
  {Hawking}}\ and\ \bibinfo {author} {\bibfnamefont {G.~F.~R.}\ \bibnamefont
  {Ellis}},\ }\href@noop {} {\emph {\bibinfo {title} {The Large Scale Structure
  of Space-Time}}},\ Cambridge Monographs on Mathematical Physics\ (\bibinfo
  {publisher} {Cambridge University Press, Cambridge},\ \bibinfo {year}
  {1973})\BibitemShut {NoStop}%
\bibitem [{\citenamefont {{Dai}}\ \emph {et~al.}(2015)\citenamefont {{Dai}},
  \citenamefont {{Pajer}},\ and\ \citenamefont {{Schmidt}}}]{Daietal2015}%
  \BibitemOpen
  \bibfield  {author} {\bibinfo {author} {\bibfnamefont {L.}~\bibnamefont
  {{Dai}}}, \bibinfo {author} {\bibfnamefont {E.}~\bibnamefont {{Pajer}}},\
  and\ \bibinfo {author} {\bibfnamefont {F.}~\bibnamefont {{Schmidt}}},\
  }\bibfield  {title} {\bibinfo {title} {{On separate universes}},\ }\href
  {https://doi.org/10.1088/1475-7516/2015/10/059} {\bibfield  {journal}
  {\bibinfo  {journal} {\jcap}\ }\textbf {\bibinfo {volume} {2015}},\ \bibinfo
  {pages} {059} (\bibinfo {year} {2015})},\ \Eprint
  {https://arxiv.org/abs/1504.00351} {arXiv:1504.00351 [astro-ph.CO]}
  \BibitemShut {NoStop}%
\bibitem [{\citenamefont {{Fabis}}\ \emph {et~al.}(2018)\citenamefont
  {{Fabis}}, \citenamefont {{Kozlikin}}, \citenamefont {{Lilow}},\ and\
  \citenamefont {{Bartelmann}}}]{Fabisetal2018}%
  \BibitemOpen
  \bibfield  {author} {\bibinfo {author} {\bibfnamefont {F.}~\bibnamefont
  {{Fabis}}}, \bibinfo {author} {\bibfnamefont {E.}~\bibnamefont {{Kozlikin}}},
  \bibinfo {author} {\bibfnamefont {R.}~\bibnamefont {{Lilow}}},\ and\ \bibinfo
  {author} {\bibfnamefont {M.}~\bibnamefont {{Bartelmann}}},\ }\bibfield
  {title} {\bibinfo {title} {{Kinetic field theory: exact free evolution of
  Gaussian phase-space correlations}},\ }\href
  {https://doi.org/10.1088/1742-5468/aab850} {\bibfield  {journal} {\bibinfo
  {journal} {J. Stat. Mech.}\ }\textbf {\bibinfo {volume} {4}},\ \bibinfo
  {pages} {043214} (\bibinfo {year} {2018})},\ \Eprint
  {https://arxiv.org/abs/1710.01611} {arXiv:1710.01611 [cond-mat.stat-mech]}
  \BibitemShut {NoStop}%
\bibitem [{\citenamefont {{Lilow}}\ \emph {et~al.}(2019)\citenamefont
  {{Lilow}}, \citenamefont {{Fabis}}, \citenamefont {{Kozlikin}}, \citenamefont
  {{Viermann}},\ and\ \citenamefont {{Bartelmann}}}]{Lilowetal2019}%
  \BibitemOpen
  \bibfield  {author} {\bibinfo {author} {\bibfnamefont {R.}~\bibnamefont
  {{Lilow}}}, \bibinfo {author} {\bibfnamefont {F.}~\bibnamefont {{Fabis}}},
  \bibinfo {author} {\bibfnamefont {E.}~\bibnamefont {{Kozlikin}}}, \bibinfo
  {author} {\bibfnamefont {C.}~\bibnamefont {{Viermann}}},\ and\ \bibinfo
  {author} {\bibfnamefont {M.}~\bibnamefont {{Bartelmann}}},\ }\bibfield
  {title} {\bibinfo {title} {{Resummed Kinetic Field Theory: general formalism
  and linear structure growth from Newtonian particle dynamics}},\ }\href
  {https://doi.org/10.1088/1475-7516/2019/04/001} {\bibfield  {journal}
  {\bibinfo  {journal} {\jcap}\ }\textbf {\bibinfo {volume} {2019}},\ \bibinfo
  {eid} {001} (\bibinfo {year} {2019})},\ \Eprint
  {https://arxiv.org/abs/1809.06942} {arXiv:1809.06942 [astro-ph.CO]}
  \BibitemShut {NoStop}%
\bibitem [{\citenamefont {{Konrad}}\ and\ \citenamefont
  {{Bartelmann}}(2022{\natexlab{b}})}]{Konradetal2020}%
  \BibitemOpen
  \bibfield  {author} {\bibinfo {author} {\bibfnamefont {S.}~\bibnamefont
  {{Konrad}}}\ and\ \bibinfo {author} {\bibfnamefont {M.}~\bibnamefont
  {{Bartelmann}}},\ }\bibfield  {title} {\bibinfo {title} {{On the asymptotic
  behaviour of cosmic density-fluctuation power spectra}},\ }\href
  {https://doi.org/10.1093/mnras/stac1795} {\bibfield  {journal} {\bibinfo
  {journal} {\mnras}\ }\textbf {\bibinfo {volume} {515}},\ \bibinfo {pages}
  {2578} (\bibinfo {year} {2022}{\natexlab{b}})},\ \Eprint
  {https://arxiv.org/abs/2110.07427} {arXiv:2110.07427 [astro-ph.CO]}
  \BibitemShut {NoStop}%
\bibitem [{\citenamefont {{Konrad}}\ \emph {et~al.}(2022)\citenamefont
  {{Konrad}}, \citenamefont {{Ginat}},\ and\ \citenamefont
  {{Bartelmann}}}]{Konradetal2022}%
  \BibitemOpen
  \bibfield  {author} {\bibinfo {author} {\bibfnamefont {S.}~\bibnamefont
  {{Konrad}}}, \bibinfo {author} {\bibfnamefont {Y.~B.}\ \bibnamefont
  {{Ginat}}},\ and\ \bibinfo {author} {\bibfnamefont {M.}~\bibnamefont
  {{Bartelmann}}},\ }\bibfield  {title} {\bibinfo {title} {{On the asymptotic
  behaviour of cosmic density-fluctuation power spectra of cold dark matter}},\
  }\href {https://doi.org/10.1093/mnras/stac2064} {\bibfield  {journal}
  {\bibinfo  {journal} {\mnras}\ }\textbf {\bibinfo {volume} {515}},\ \bibinfo
  {pages} {5823} (\bibinfo {year} {2022})},\ \Eprint
  {https://arxiv.org/abs/2202.08059} {arXiv:2202.08059 [astro-ph.CO]}
  \BibitemShut {NoStop}%
\bibitem [{\citenamefont {Olver}(1974)}]{Olver1974}%
  \BibitemOpen
  \bibfield  {author} {\bibinfo {author} {\bibfnamefont {F.~W.~J.}\
  \bibnamefont {Olver}},\ }\href@noop {} {\emph {\bibinfo {title} {{Asymptotics
  and Special Functions}}}},\ Computer Science and Applied Mathematics\
  (\bibinfo  {publisher} {Academic Press, Cambridge, Massachussetts},\ \bibinfo
  {year} {1974})\BibitemShut {NoStop}%
\bibitem [{\citenamefont {Bleistein}\ and\ \citenamefont
  {Handelsman}(1986)}]{bleistein1975asymptotic}%
  \BibitemOpen
  \bibfield  {author} {\bibinfo {author} {\bibfnamefont {N.}~\bibnamefont
  {Bleistein}}\ and\ \bibinfo {author} {\bibfnamefont {R.~A.}\ \bibnamefont
  {Handelsman}},\ }\href@noop {} {\emph {\bibinfo {title} {Asymptotic
  Expansions of Integrals}}},\ \bibinfo {edition} {2nd}\ ed.\ (\bibinfo
  {publisher} {Dover, New York},\ \bibinfo {year} {1986})\BibitemShut {NoStop}%
\bibitem [{\citenamefont {{Kolb}}\ and\ \citenamefont
  {{Turner}}(1990)}]{KolbTurner1990}%
  \BibitemOpen
  \bibfield  {author} {\bibinfo {author} {\bibfnamefont {E.~W.}\ \bibnamefont
  {{Kolb}}}\ and\ \bibinfo {author} {\bibfnamefont {M.~S.}\ \bibnamefont
  {{Turner}}},\ }\href@noop {} {\emph {\bibinfo {title} {{The Early
  Universe}}}}\ (\bibinfo  {publisher} {CRC Press, London},\ \bibinfo {year}
  {1990})\BibitemShut {NoStop}%
\bibitem [{\citenamefont {Ewart}\ \emph {et~al.}(2025)\citenamefont {Ewart},
  \citenamefont {Nastac}, \citenamefont {Bilbao}, \citenamefont {Silva},
  \citenamefont {Silva},\ and\ \citenamefont {Schekochihin}}]{Ewart2024}%
  \BibitemOpen
  \bibfield  {author} {\bibinfo {author} {\bibfnamefont {R.~J.}\ \bibnamefont
  {Ewart}}, \bibinfo {author} {\bibfnamefont {M.~L.}\ \bibnamefont {Nastac}},
  \bibinfo {author} {\bibfnamefont {P.~J.}\ \bibnamefont {Bilbao}}, \bibinfo
  {author} {\bibfnamefont {T.}~\bibnamefont {Silva}}, \bibinfo {author}
  {\bibfnamefont {L.~O.}\ \bibnamefont {Silva}},\ and\ \bibinfo {author}
  {\bibfnamefont {A.~A.}\ \bibnamefont {Schekochihin}},\ }\bibfield  {title}
  {\bibinfo {title} {Relaxation to universal non-maxwellian equilibria in a
  collisionless plasma},\ }\href {https://doi.org/10.1073/pnas.2417813122}
  {\bibfield  {journal} {\bibinfo  {journal} {Proceedings of the National
  Academy of Sciences}\ }\textbf {\bibinfo {volume} {122}},\ \bibinfo {pages}
  {e2417813122} (\bibinfo {year} {2025})},\ \Eprint
  {https://arxiv.org/abs/https://www.pnas.org/doi/pdf/10.1073/pnas.2417813122}
  {https://www.pnas.org/doi/pdf/10.1073/pnas.2417813122} \BibitemShut {NoStop}%
\bibitem [{\citenamefont {{Press}}\ and\ \citenamefont
  {{Schechter}}(1974)}]{PressSchechter1974}%
  \BibitemOpen
  \bibfield  {author} {\bibinfo {author} {\bibfnamefont {W.~H.}\ \bibnamefont
  {{Press}}}\ and\ \bibinfo {author} {\bibfnamefont {P.}~\bibnamefont
  {{Schechter}}},\ }\bibfield  {title} {\bibinfo {title} {{Formation of
  galaxies and clusters of galaxies by self-similar gravitational
  condensation}},\ }\href {https://doi.org/10.1086/152650} {\bibfield
  {journal} {\bibinfo  {journal} {\apj}\ }\textbf {\bibinfo {volume} {187}},\
  \bibinfo {pages} {425} (\bibinfo {year} {1974})}\BibitemShut {NoStop}%
\bibitem [{\citenamefont {{Ma}}\ and\ \citenamefont {{Fry}}(2000)}]{MaFry2000}%
  \BibitemOpen
  \bibfield  {author} {\bibinfo {author} {\bibfnamefont {C.-P.}\ \bibnamefont
  {{Ma}}}\ and\ \bibinfo {author} {\bibfnamefont {J.~N.}\ \bibnamefont
  {{Fry}}},\ }\bibfield  {title} {\bibinfo {title} {{Deriving the nonlinear
  cosmological power spectrum and bispectrum from analytic dark matter halo
  profiles and mass functions}},\ }\href {https://doi.org/10.1086/317146}
  {\bibfield  {journal} {\bibinfo  {journal} {\apj}\ }\textbf {\bibinfo
  {volume} {543}},\ \bibinfo {pages} {503} (\bibinfo {year} {2000})},\ \Eprint
  {https://arxiv.org/abs/astro-ph/0003343} {arXiv:astro-ph/0003343 [astro-ph]}
  \BibitemShut {NoStop}%
\bibitem [{\citenamefont {{Gouda}}\ and\ \citenamefont
  {{Nakamura}}(1989)}]{GoudaNakamura1989}%
  \BibitemOpen
  \bibfield  {author} {\bibinfo {author} {\bibfnamefont {N.}~\bibnamefont
  {{Gouda}}}\ and\ \bibinfo {author} {\bibfnamefont {T.}~\bibnamefont
  {{Nakamura}}},\ }\bibfield  {title} {\bibinfo {title} {{Non-Linear Growth of
  One-Dimensional Cosmological Density Fluctuation and Catastrophe Theory}},\
  }\href {https://doi.org/10.1143/PTP.81.633} {\bibfield  {journal} {\bibinfo
  {journal} {Progress of Theoretical Physics}\ }\textbf {\bibinfo {volume}
  {81}},\ \bibinfo {pages} {633} (\bibinfo {year} {1989})}\BibitemShut
  {NoStop}%
\bibitem [{\citenamefont {{Yano}}\ and\ \citenamefont
  {{Gouda}}(2000)}]{YanoGouda2000}%
  \BibitemOpen
  \bibfield  {author} {\bibinfo {author} {\bibfnamefont {T.}~\bibnamefont
  {{Yano}}}\ and\ \bibinfo {author} {\bibfnamefont {N.}~\bibnamefont
  {{Gouda}}},\ }\bibfield  {title} {\bibinfo {title} {{A Universal Profile of
  the Dark Matter Halo and the Two-Point Correlation Function}},\ }\href
  {https://doi.org/10.1086/309246} {\bibfield  {journal} {\bibinfo  {journal}
  {\apj}\ }\textbf {\bibinfo {volume} {539}},\ \bibinfo {pages} {493} (\bibinfo
  {year} {2000})},\ \Eprint {https://arxiv.org/abs/astro-ph/9906375}
  {arXiv:astro-ph/9906375 [astro-ph]} \BibitemShut {NoStop}%
\bibitem [{\citenamefont {{Navarro}}\ \emph {et~al.}(2004)\citenamefont
  {{Navarro}}, \citenamefont {{Hayashi}}, \citenamefont {{Power}},
  \citenamefont {{Jenkins}}, \citenamefont {{Frenk}}, \citenamefont {{White}},
  \citenamefont {{Springel}}, \citenamefont {{Stadel}},\ and\ \citenamefont
  {{Quinn}}}]{Navarroetal2004}%
  \BibitemOpen
  \bibfield  {author} {\bibinfo {author} {\bibfnamefont {J.~F.}\ \bibnamefont
  {{Navarro}}}, \bibinfo {author} {\bibfnamefont {E.}~\bibnamefont
  {{Hayashi}}}, \bibinfo {author} {\bibfnamefont {C.}~\bibnamefont {{Power}}},
  \bibinfo {author} {\bibfnamefont {A.~R.}\ \bibnamefont {{Jenkins}}}, \bibinfo
  {author} {\bibfnamefont {C.~S.}\ \bibnamefont {{Frenk}}}, \bibinfo {author}
  {\bibfnamefont {S.~D.~M.}\ \bibnamefont {{White}}}, \bibinfo {author}
  {\bibfnamefont {V.}~\bibnamefont {{Springel}}}, \bibinfo {author}
  {\bibfnamefont {J.}~\bibnamefont {{Stadel}}},\ and\ \bibinfo {author}
  {\bibfnamefont {T.~R.}\ \bibnamefont {{Quinn}}},\ }\bibfield  {title}
  {\bibinfo {title} {{The inner structure of {\ensuremath{\Lambda}}CDM haloes -
  III. Universality and asymptotic slopes}},\ }\href
  {https://doi.org/10.1111/j.1365-2966.2004.07586.x} {\bibfield  {journal}
  {\bibinfo  {journal} {\mnras}\ }\textbf {\bibinfo {volume} {349}},\ \bibinfo
  {pages} {1039} (\bibinfo {year} {2004})},\ \Eprint
  {https://arxiv.org/abs/astro-ph/0311231} {arXiv:astro-ph/0311231 [astro-ph]}
  \BibitemShut {NoStop}%
\bibitem [{\citenamefont {{Delos}}\ and\ \citenamefont
  {{White}}(2023)}]{DelosWhite2023}%
  \BibitemOpen
  \bibfield  {author} {\bibinfo {author} {\bibfnamefont {M.~S.}\ \bibnamefont
  {{Delos}}}\ and\ \bibinfo {author} {\bibfnamefont {S.~D.~M.}\ \bibnamefont
  {{White}}},\ }\bibfield  {title} {\bibinfo {title} {{Inner cusps of the first
  dark matter haloes: formation and survival in a cosmological context}},\
  }\href {https://doi.org/10.1093/mnras/stac3373} {\bibfield  {journal}
  {\bibinfo  {journal} {\mnras}\ }\textbf {\bibinfo {volume} {518}},\ \bibinfo
  {pages} {3509} (\bibinfo {year} {2023})},\ \Eprint
  {https://arxiv.org/abs/2207.05082} {arXiv:2207.05082 [astro-ph.CO]}
  \BibitemShut {NoStop}%
\bibitem [{\citenamefont {Asgari}\ \emph {et~al.}(2023)\citenamefont {Asgari},
  \citenamefont {Mead},\ and\ \citenamefont {Heymans}}]{Asgari2023halo}%
  \BibitemOpen
  \bibfield  {author} {\bibinfo {author} {\bibfnamefont {M.}~\bibnamefont
  {Asgari}}, \bibinfo {author} {\bibfnamefont {A.~J.}\ \bibnamefont {Mead}},\
  and\ \bibinfo {author} {\bibfnamefont {C.}~\bibnamefont {Heymans}},\
  }\bibfield  {title} {\bibinfo {title} {The halo model for cosmology: a
  pedagogical review},\ }\bibfield  {journal} {\bibinfo  {journal} {Open J.
  Astrophys.}\ }\textbf {\bibinfo {volume} {6}},\ \href
  {https://doi.org/10.21105/astro.2303.08752} {10.21105/astro.2303.08752}
  (\bibinfo {year} {2023})\BibitemShut {NoStop}%
\bibitem [{\citenamefont {{Navarro}}\ \emph {et~al.}(1997)\citenamefont
  {{Navarro}}, \citenamefont {{Frenk}},\ and\ \citenamefont
  {{White}}}]{NFW1997}%
  \BibitemOpen
  \bibfield  {author} {\bibinfo {author} {\bibfnamefont {J.~F.}\ \bibnamefont
  {{Navarro}}}, \bibinfo {author} {\bibfnamefont {C.~S.}\ \bibnamefont
  {{Frenk}}},\ and\ \bibinfo {author} {\bibfnamefont {S.~D.~M.}\ \bibnamefont
  {{White}}},\ }\bibfield  {title} {\bibinfo {title} {{A Universal Density
  Profile from Hierarchical Clustering}},\ }\href
  {https://doi.org/10.1086/304888} {\bibfield  {journal} {\bibinfo  {journal}
  {\apj}\ }\textbf {\bibinfo {volume} {490}},\ \bibinfo {pages} {493} (\bibinfo
  {year} {1997})},\ \Eprint {https://arxiv.org/abs/astro-ph/9611107}
  {arXiv:astro-ph/9611107 [astro-ph]} \BibitemShut {NoStop}%
\bibitem [{\citenamefont {{Einasto}}(1965)}]{Einasto1965}%
  \BibitemOpen
  \bibfield  {author} {\bibinfo {author} {\bibfnamefont {J.}~\bibnamefont
  {{Einasto}}},\ }\bibfield  {title} {\bibinfo {title} {{On the Construction of
  a Composite Model for the Galaxy and on the Determination of the System of
  Galactic Parameters}},\ }\href@noop {} {\bibfield  {journal} {\bibinfo
  {journal} {Trudy Astrofizicheskogo Instituta Alma-Ata}\ }\textbf {\bibinfo
  {volume} {5}},\ \bibinfo {pages} {87} (\bibinfo {year} {1965})}\BibitemShut
  {NoStop}%
\bibitem [{\citenamefont {{Marsh}}\ \emph {et~al.}(2024)\citenamefont
  {{Marsh}}, \citenamefont {{Ellis}},\ and\ \citenamefont
  {{Mehta}}}]{Marshetal2024}%
  \BibitemOpen
  \bibfield  {author} {\bibinfo {author} {\bibfnamefont {D.~J.~E.}\
  \bibnamefont {{Marsh}}}, \bibinfo {author} {\bibfnamefont {D.}~\bibnamefont
  {{Ellis}}},\ and\ \bibinfo {author} {\bibfnamefont {V.~M.}\ \bibnamefont
  {{Mehta}}},\ }\href@noop {} {\emph {\bibinfo {title} {{Dark Matter: Evidence,
  Theory, and Constraints}}}}\ (\bibinfo  {publisher} {Princeton University
  Press, Princeton, N.J.},\ \bibinfo {year} {2024})\BibitemShut {NoStop}%
\bibitem [{\citenamefont {Dougherty}(1964)}]{Dougherty1964}%
  \BibitemOpen
  \bibfield  {author} {\bibinfo {author} {\bibfnamefont {J.~P.}\ \bibnamefont
  {Dougherty}},\ }\bibfield  {title} {\bibinfo {title} {{Model Fokker-Planck
  equation for a plasma and its solution}},\ }\href
  {https://doi.org/10.1063/1.2746779} {\bibfield  {journal} {\bibinfo
  {journal} {Phys. Fluids}\ }\textbf {\bibinfo {volume} {7}},\ \bibinfo {pages}
  {1788} (\bibinfo {year} {1964})}\BibitemShut {NoStop}%
\bibitem [{\citenamefont {{Hakim}}\ \emph {et~al.}(2020)\citenamefont
  {{Hakim}}, \citenamefont {{Francisquez}}, \citenamefont {{Juno}},\ and\
  \citenamefont {{Hammett}}}]{Hakimetal2020}%
  \BibitemOpen
  \bibfield  {author} {\bibinfo {author} {\bibfnamefont {A.}~\bibnamefont
  {{Hakim}}}, \bibinfo {author} {\bibfnamefont {M.}~\bibnamefont
  {{Francisquez}}}, \bibinfo {author} {\bibfnamefont {J.}~\bibnamefont
  {{Juno}}},\ and\ \bibinfo {author} {\bibfnamefont {G.~W.}\ \bibnamefont
  {{Hammett}}},\ }\bibfield  {title} {\bibinfo {title} {{Conservative
  discontinuous Galerkin schemes for nonlinear Dougherty-Fokker-Planck
  collision operators}},\ }\href {https://doi.org/10.1017/S0022377820000586}
  {\bibfield  {journal} {\bibinfo  {journal} {J. Plasma Phys.}\ }\textbf
  {\bibinfo {volume} {86}},\ \bibinfo {eid} {905860403} (\bibinfo {year}
  {2020})},\ \Eprint {https://arxiv.org/abs/1903.08062} {arXiv:1903.08062
  [physics.comp-ph]} \BibitemShut {NoStop}%
\bibitem [{\citenamefont {{Tatsuno}}\ \emph {et~al.}(2009)\citenamefont
  {{Tatsuno}}, \citenamefont {{Dorland}}, \citenamefont {{Schekochihin}},
  \citenamefont {{Plunk}}, \citenamefont {{Barnes}}, \citenamefont {{Cowley}},\
  and\ \citenamefont {{Howes}}}]{Tatsunoetal2009}%
  \BibitemOpen
  \bibfield  {author} {\bibinfo {author} {\bibfnamefont {T.}~\bibnamefont
  {{Tatsuno}}}, \bibinfo {author} {\bibfnamefont {W.}~\bibnamefont
  {{Dorland}}}, \bibinfo {author} {\bibfnamefont {A.~A.}\ \bibnamefont
  {{Schekochihin}}}, \bibinfo {author} {\bibfnamefont {G.~G.}\ \bibnamefont
  {{Plunk}}}, \bibinfo {author} {\bibfnamefont {M.}~\bibnamefont {{Barnes}}},
  \bibinfo {author} {\bibfnamefont {S.~C.}\ \bibnamefont {{Cowley}}},\ and\
  \bibinfo {author} {\bibfnamefont {G.~G.}\ \bibnamefont {{Howes}}},\
  }\bibfield  {title} {\bibinfo {title} {{Nonlinear phase mixing and
  phase-space cascade of entropy in gyrokinetic plasma turbulence}},\ }\href
  {https://doi.org/10.1103/PhysRevLett.103.015003} {\bibfield  {journal}
  {\bibinfo  {journal} {\prl}\ }\textbf {\bibinfo {volume} {103}},\ \bibinfo
  {eid} {015003} (\bibinfo {year} {2009})},\ \Eprint
  {https://arxiv.org/abs/0811.2538} {arXiv:0811.2538 [physics.plasm-ph]}
  \BibitemShut {NoStop}%
\bibitem [{\citenamefont {{Davis}}\ and\ \citenamefont
  {{Peebles}}(1977)}]{DavisPeebles1977}%
  \BibitemOpen
  \bibfield  {author} {\bibinfo {author} {\bibfnamefont {M.}~\bibnamefont
  {{Davis}}}\ and\ \bibinfo {author} {\bibfnamefont {P.~J.~E.}\ \bibnamefont
  {{Peebles}}},\ }\bibfield  {title} {\bibinfo {title} {{On the integration of
  the BBGKY equations for the development of strongly nonlinear clustering in
  an expanding universe}},\ }\href {https://doi.org/10.1086/190456} {\bibfield
  {journal} {\bibinfo  {journal} {\apjs}\ }\textbf {\bibinfo {volume} {34}},\
  \bibinfo {pages} {425} (\bibinfo {year} {1977})}\BibitemShut {NoStop}%
\end{thebibliography}%

\appendix

%
%
%

\onecolumngrid

\section{Simulation methods}
\label{appendix: simulation set up}
Here we describe the details of the Vlasov--Poisson simulations whose results were shown in \S \ref{sec:Phase-Space Cascade} and figures \ref{fig:results time evolution}--\ref{fig:gkeyll results two v_th's}. The simulations were conducted using the Gkeyll code\footnote{\url{https://gkeyll.readthedocs.io/}}, which is an Eulerian solver in phase-space, originally designed for the Vlasov-Maxwell system.
Gkeyll uses a discontinuous Galerkin algorithm to compute distribution functions \cite{JUNO2018110}. We set the vacuum permittivity $\eps_0 = -1$, which turns electrostatic interactions into gravitational ones---and, for unit particle charges, it is equivalent to choosing units so that $4\pi G = 1$. Our simulations were $1$D in position space (so the phase-space is $2$D), with periodic spatial boundary conditions; the units of length were such that the box size was $L = 2\pi$. We always took the particle mass to be unity, which implied that the system's total mass was normalised to $M = 2\pi$. These three choices specify the units for the simulations, and are equivalent to choosing $\tau_0 = k_0 = v_0 = 1$ (these are defined in \S \ref{subsec:cold}). Thus, the outer scale $k_{\rm nl} \sim k_0$ is of order unity.

\subsection{Initial conditions}
We simulated a single-species distribution function, with the initial condition \eqref{eqn:initial condition f}, where we chose $v_{\rm th} = 0.005v_0$ (figure \ref{fig:results time evolution}), $0.01v_0$, or $0.05v_0$ (the left and right columns of figure \ref{fig:gkeyll results two v_th's}, respectively), and
\begin{align}
\rho_{\rm in}(x)\frac{L}{M} & = 1-\sum_{n=1}^5 \frac{a_n}{k_n} \sin\left(k_n x + \phi_n\right), \\
u_{\rm in}(x) & = v_0\sum_{n=1}^{5} b_n \cos \left(k_n x + \phi_n\right),
\end{align}
where $\phi_n \in [0,2\pi]$ are random phases, the amplitudes~$a_n/k_0$ were uniformly sampled from the interval $[0,0.2]$, and $k_n = k_0 U_n$, with~$U_n \in \set{1,2,\ldots,10}$---uniformly distributed random integers. For runs with one initial stream (figure~\ref{fig:gkeyll results two v_th's}),~$b_n=a_n/k_0$, while for runs with three initial streams (figure~\ref{fig:results time evolution}), we set~$b_n=0.05a_n/k_0$ and tripled the initial conditions \eqref{eqn:initial condition f} by shifting~$u_{\rm in}(x)$ by~$\pm 2v_0$ and duplicating for two additional streams.
We ensured that the resolution was sufficiently fine for a Maxwellian of width $v_{\rm th}$ still to be resolved: we used an $N_x \times N_v$ phase-space grid with $N_x = N_v = 4032$ for the single-stream initial condition and $N_x = N_v =  7680$ for the multi-stream one. 
The simulation box contains $v_0/v_{\rm th}$ Jeans lengths, corresponding to $200$, $100$ and $20$ Jeans lengths for the three choices of $v_{\rm th}$ above.

Because of numerical errors, $f$ can become slightly negative in very small and isolated areas. This is because the discontinuous Galerkin algorithm that Gkeyll uses is not positivity-preserving, and errors arise from over-shooting due to large derivatives. These problems are somewhat alleviated by the small collision operator (see appendix \ref{coll_numerics}), which smooths large velocity derivatives. In any case, these regions do not cause the simulation to become unstable because they are isolated and $\abs{f}$ is still very small there; the total mass occupied by negative $f$ is $\iint f\Theta(-f)\mathrm{d}x\mathrm{d}v < 10^{-5}M$ for the simulation in figure \ref{fig:results time evolution} and $<0.02M$ for figure \ref{fig:gkeyll results two v_th's}. Therefore, this does not invalidate any of our conclusions in this paper, and we have set $f\mapsto f\Theta(f)$ for the purpose of plotting figures \ref{fig:results time evolution} and \ref{fig:gkeyll results two v_th's}.

\subsection{Collisions}
\label{coll_numerics}
One would like to run collisionless simulations, but the finite resolution of any numerical representation of phase-space induces an unavoidable effective collisionality due to the grid. Many dark-matter models imply that dark matter in the Universe consists of a finite number of particles \cite{Marshetal2024}---and finite-$N$ effects also induce a effective collisionality. To control the cut-off scale in velocity space and the dissipation of $C_2$, we added a collision operator $\left(\partial f/\partial t\right)_c \propto \nu$ with a small collision frequency $\nu$. The collision operator used was the Dougherty operator---a type of Fokker-Planck operator in velocity space \cite{Dougherty1964,Hakimetal2020}:
\begin{equation}\label{eqn:collision operator}
\left(\frac{\partial f}{\partial t}\right)_c = \nu
\frac{\partial}{\partial v}\left[\left(v-u\right)f
+v_{t}^2\frac{\partial f}{\partial v}\right],
\end{equation}
where $u(x)$ is first velocity moment of $f$ and $v_t^2 \equiv \int \mathrm{d}v~(v-u)^2f/\rho$.
Details can be found in \cite{Hakimetal2020}, 
which describes the implementation of equation \eqref{eqn:collision operator} in Gkeyll. We used a Dorland number \cite{Tatsunoetal2009} of $\mathrm{Do} \equiv (\nu\tau_0)^{-1} = 10^{6}$ for all runs.

The inevitability of some collisions---whether due to the grid, finite-$N$ effects, or a collision operator---implies the existence of another scale in the problem.
The collision time scale is \cite{Hakimetal2020,Nastacetal2023} 
\begin{equation}
\tau_{\nu} \sim \frac{1}{\nu s^2v_{\rm rms}^2},
\end{equation}
where $v_{\rm rms}^2$ is the second velocity cumulant.
The velocity scale where collisions become competitive with the cascade rate ($\tau_\nu \sim \tau_{\rm g}$) is, therefore,
\begin{equation}\label{eqn:u nu}
s_\nu = \frac{1}{v_{\rm rms}\sqrt{\nu \tau_{\rm g}}}.
\end{equation}
By critical balance (see \S \ref{subsec:critical balance}), this also gives a length scale
\begin{equation}\label{eqn:l nu}
k_\nu = \frac{1}{ L\sqrt{\nu \tau_{\rm g}}} \sim \frac{k_{\rm nl}}{\sqrt{\nu \tau_{\rm g}}},
\end{equation}
which is the collisional cut-off in position space.
For a sufficiently small $\nu$, one can have\footnote{In reality, it might be that $s_\nu v_{\rm th}\lesssim 1$, for both parameters depend sensitively on the nature of the constituent particle(s) of dark matter. But as long as $s$ is smaller than both $s_\nu$ and $v_{\rm th}^{-1}$, none of the conclusions of the this paper are invalidated by this possibility.}
\begin{align}
& k_{\rm nl} \ll k \ll k_{\rm c}(v_{\rm th}^{-1}) \ll k_\nu, \\ &
v_{\rm rms}^{-1} \ll s \ll v_{\rm th}^{-1} \ll s_\nu,
\end{align}
where $k_{\rm c}(s)$ is the critical-balance scale defined by equation \eqref{eqn:critical balance k_c(s)}.
While this hierarchy justifies neglecting collisions in this paper, as phase-space structure cascades to ever smaller scales by \eqref{eqn: F transport equation Batchelor}, the collisional scales, where it is erased, are eventually reached \cite{Nastacetal2023,Nastacetal2024}. 
Likewise, in reality, a particle-noise floor $\hat{F} \approx M^2/N$ (which is not present in our simulations) is eventually reached, too. Determining whether this occurs on scales smaller or larger than $l_{\nu}$ is deferred to future work \cite[cf.][]{Nastacetal2024}.

\subsection{Phase-space power spectra}
\label{subsec:simulation results}

In the $1$D set-up described above, our prediction \eqref{eqn:F asymptotics CDM} for the phase-space power spectrum becomes
\begin{equation}
\hat{F}(k,s) \sim \begin{cases}
F_1 k^{-1}, & \mbox{if } \max\set{k_{\rm nl}, k_{\rm c}(s)} \ll k \ll \min\set{k_{\rm c}(v_{\rm th}^{-1}),k_\nu}, \\
F_2 s^{-1}, & \mbox{if } \max\set{v_{\rm rms}^{-1},s_{\rm c}(k)} \ll s \ll \min\set{v_{\rm th}^{-1},s_\nu}.
\end{cases}
\end{equation}
We are not concerned here with the asymptotics of the power spectrum~$\hat{F}$, defined in equation~\eqref{eqn: F definition}, at values of~$s$ and~$k$ above~$v_{\rm th}^{-1}$ and~$k_{\rm c}(v_{\rm th}^{-1})$, respectively, even if they lie below the collisional cut-offs $s_\nu$ and $k_\nu$ (see equations \eqref{eqn:u nu} and \eqref{eqn:l nu}). In this region, gravitational collapse halts by equation~\eqref{eqn:Raychaudhuri approx}, but if one extrapolates the findings of Ref.~\cite{Nastacetal2024}---who studied turbulence in electrostatic plasmas with~$v_{\rm th}$ of order unity---from electrostatics to gravity, then one should have, assuming~$s_\nu v_{\rm th} \gg 1$,
\begin{equation}
\hat{F}(k,s) \sim \begin{cases}
F_1 k^{-1}, & \mbox{if } \max\set{k_{\rm nl}, k_{\rm c}(s)} \ll k \ll k_{\rm c}(v_{\rm th}^{-1}), \\
F_2 s^{-1}, & \mbox{if } \max\set{v_{\rm rms}^{-1},s_{\rm c}(k)} \ll s \ll v_{\rm th}^{-1}, \\
F_3 k^{-2}, & \mbox{if } \max\set{k_{\rm nl}, k_{\rm c}(s),k_{\rm c}(v_{\rm th}^{-1})} \ll k \ll k_\nu, \\
F_4 s^{-2}, & \mbox{if } \max\set{v_{\rm rms}^{-1},s_{\rm c}(k),v_{\rm th}^{-1}} \ll s \ll s_\nu.
\end{cases}
\end{equation}
The spectrum is truncated exponentially at $k > k_\nu$ or $s>s_\nu$ (or submerged into particle noise as discussed in \cite{Nastacetal2024}).

Figure~\ref{fig:results time evolution} shows an example of the time evolution of a system that started with three streams with~$v_{\rm th} = 0.005v_0$ and had~$\mathrm{Do} = 10^{6}$. This corresponds to~$u_\nu \sim 10^{-3}v_0$, so scales are separated in the way that we have assumed. The grid size is~$N_x = N_v = 7680$ and the velocity box size is~$12v_0$, so the Nyquist velocity scale is~$s_{\rm Ny} \approx 2011v_0^{-1}$. Figure~\ref{fig:results time evolution} shows that the streams collapse quickly, by rotating and twisting in phase-space. In figure \ref{fig:spectra time evolution}, we show the evolution of the power spectra for this simulation.

\section{Einstein-de-Sitter universe}
\label{appendix: EdS}

Although the analysis of \S \ref{sec:Phase-Space Cascade}, resulting in the prediction \eqref{eqn:F asymptotics CDM} for the phase-space power spectrum, is valid for $\Lambda$CDM, and in general for any cosmology where there is a separation of scales between $\mathcal{H}/c$ and the non-linear scale $k_{\rm nl}$, in the special case of an Einstein-de-Sitter (EdS) cosmology, there exists a scale invariance (see, e.g., \cite{DavisPeebles1977,Peebles1980}) that provides additional insight, which we sketch in this appendix. This allows us to propose a global analogue of the local analysis of the main text. We focus on the case of $d=3$ for simplicity.

In an EdS space-time, the Vlasov--Poisson system is invariant under the transformation
\begin{equation}
(t,\mathbf{x},\mathbf{v},f) \mapsto (\zeta t, \zeta^{\varpi} \mathbf{x},\zeta^{\varpi-1/3} \mathbf{v},\zeta^{3\varpi+1}f),
\end{equation}
where $\varpi$ is fixed by assuming the scaling of the density power spectrum.
This means that for every (inverse) length scale $k$, there exists a (cosmic) time scale $t \propto k^{1/\varpi}$.
Hence, one obtains a (conformal) time scale, which is just the collapse conformal time in the standard spherical-collapse model,
\begin{equation}\label{eqn:collapse time}
\tau_{\rm g}(k) \equiv \frac{2}{H_0}\left(\frac{\pi}{2H_0t_0}\right)^{1/3}\left(\frac{3}{5\delta_{\rm rms}(k)}\right)^{1/2},
\end{equation}
where $H_0$ and $t_0$ are the present-day Hubble constant and cosmic time, and $\delta_{\rm rms}(k)$ is the root-mean-square amplitude of the dark-matter density fluctuation at scale $k$, normalised by the EdS growth factor $D_+\left(t_{\rm initial}\right) = a\left(t_{\rm initial}\right)$. The time scale \eqref{eqn:collapse time} turns out to be nothing but the standard gravitational time scale $\tau_{\rm g} \sim 1/\sqrt{G\delta\rho}$ (cf. equation \eqref{eqn:tau g}), where $\delta\rho \sim \delta_{\rm rms} \rho_{\rm crit}$, with the critical density defined by $\rho_{\rm crit} = 3H_0^2/(8\pi G)$.
Relating $\delta_{\rm rms}(k)$ to the over-density power spectrum~$P_{\delta}(k) = P_0 (k/k_{\rm p})^n$ for some constant $k_{\rm p}$, via $\delta_{\rm rms}(k) = \sqrt{k^3P_{\delta}(k)}$, yields
\begin{equation}\label{eqn:collapse time with p(k)}
\tau_{\rm g} \sim \frac{2\sqrt{3}}{\sqrt{5}H_0}\left(\frac{\pi}{2 H_0t_0}\right)^{1/3}\left(\frac{k_{\rm p}^{n}}{P_0 k^{3+n}}\right)^{1/4}.
\end{equation}

By critical balance (\S \ref{subsec:critical balance}), the collapse time scale $\tau_{\rm g}$ and the linear phase-mixing time $\tau_{\rm l}$ must be the same, whence
\begin{equation}
s_{\rm c}(k) \propto k^{(1-n)/4}.
\end{equation}
Consequently, a $C_2$ flux $\propto k^d$ (cf. \S\ref{subsec: source CDM}) would imply
\begin{equation}\label{eqn: analogue of cascade argument}
k^{4+\gamma} \propto s_{\rm c}(k).
\end{equation}
Using equation \eqref{eqn:collapse time with p(k)}, one concludes from equation \eqref{eqn: analogue of cascade argument} that the exponent of the density power spectrum in \eqref{eqn: power spectrum parameterisation} is $\gamma = -(15+n)/4$. In a steady state, the power spectrum must be invariant under the collapse process, and so in a self-sustaining scenario, $\gamma = n$, which is solved by $n=-3$. 
This is unsurprising, because an EdS space-time is just a matter-dominated FLRW space-time; as this paper is focused on small scales, locally, everything is matter dominated in any cosmology.

\section{Initial momentum correlations}
\label{appendix: Initial Momentum Correlations}
The purpose of this appendix is to prove, for \S \ref{sec:saddle-point approach}, that (i) $\Sigma \sim Aq^2$ and (ii) $\Sigma\mathbf{a} = O(q)$ as $q \to 0$, and that (iii) in the opposite limit, $q \to \infty$, $\Sigma$ is order unity. These quantities are defined in equation \eqref{eqn:momentum correlations}. What follows holds for $v_{\rm th} = 0$.

Consider
equation \eqref{eqn:momentum correlations}. Since $2\mathbf{p}_1 = \mathbf{p}+2\mathbf{P}$ and $2\mathbf{p}_2 = 2\mathbf{P}-\mathbf{p}$, the part involving $\mathbf{a}$ comes solely from the cross-correlations between $\mathbf{p}_{1}$, $\mathbf{p}_{2}$ and $\left(\mathbf{p}_3,\ldots,\mathbf{p}_N\right)$, \emph{viz.},
\begin{equation}
\begin{aligned}
& \mathbf{p}_1^T (C_{pp}^{N})^{-1}_{1n}\mathbf{p}_n + \mathbf{p}_2^T(C_{pp}^{N})^{-1}_{2n}\mathbf{p}_n = \mathbf{P}\left[(C_{pp}^{N})^{-1}_{1n} + (C_{pp}^{N})^{-1}_{2n}\right]\mathbf{p}_n + \frac{1}{2}\mathbf{p}\left[(C_{pp}^{N})^{-1}_{1n} - (C_{pp}^{N})^{-1}_{2n}\right]\mathbf{p}_n \\ &
+ \mathbf{P}\left[(C_{pp}^{N})^{-1}_{11} - (C_{pp}^{N})^{-1}_{22}\right]\mathbf{p} + 2\mathbf{P}\left[(C_{pp}^{N})^{-1}_{11} + (C_{pp}^{N})^{-1}_{21} + (C_{pp}^{N})^{-1}_{12} + (C_{pp}^{N})^{-1}_{22}\right]\mathbf{P},
\end{aligned}
\end{equation}
where $(C_{pp}^{N})^{-1}_{mn}$ specifies the sub-matrix of $(C_{pp}^N)^{-1}$ that pertains to particles $m$ and $n$, and the Einstein summation convention is used. Now,~$(C_{pp}^{N})^{-1}_{mn}$ is a function of the positions of the particles only. When~$q\equiv \abs{\mathbf{q}_1-\mathbf{q}_2} \to 0$,~$\mathbf{q}_1$ and~$\mathbf{q}_2$ both tend to~$\mathbf{Q} = \left(\mathbf{q}_1+\mathbf{q}_2\right)/2$, and, by the indistinguishability of the constituent particles,
\begin{align}
& (C_{pp}^{N})^{-1}_{1n}(\mathbf{q}_1 = \mathbf{Q},\mathbf{q}_2 = \mathbf{Q},\mathbf{q}_3,\ldots,\mathbf{q}_N) = (C_{pp}^{N})^{-1}_{2n}(\mathbf{q}_1 = \mathbf{Q},\mathbf{q}_2 = \mathbf{Q},\mathbf{q}_3,\ldots,\mathbf{q}_N), \\ &
(C_{pp}^{N})^{-1}_{11}(\mathbf{q}_1 = \mathbf{Q},\mathbf{q}_2 = \mathbf{Q},\mathbf{q}_3,\ldots,\mathbf{q}_N) = (C_{pp}^{N})^{-1}_{22}(\mathbf{q}_1 = \mathbf{Q},\mathbf{q}_2 = \mathbf{Q},\mathbf{q}_3,\ldots,\mathbf{q}_N).
\end{align}
Consequently, for $q\to 0$ and finite $\mathbf{Q},\mathbf{q}_3,\ldots,\mathbf{q}_N,\mathbf{P},\mathbf{p}_3,\ldots,\mathbf{p}_N$, we have
\begin{align}
& \left[(C_{pp}^{N})^{-1}_{1n} - (C_{pp}^{N})^{-1}_{2n}\right] =  O(q), \\ &
\left[(C_{pp}^{N})^{-1}_{11} - (C_{pp}^{N})^{-1}_{22}\right] = O(q).
\end{align}
Therefore, in the limit $q \to 0$, the matrix $(C_{pp}^N)^{-1}$, \emph{qua} a linear operator in momentum space, splits into a block matrix in the basis $(\mathbf{p},\mathbf{P},\mathbf{p}_3,\ldots,\mathbf{p}_N)$, with one block for $\mathbf{p}$ and another for $\left(\mathbf{P},\mathbf{p}_3,\ldots,\mathbf{p}_N\right)$. Hence its inverse $C_{pp}^N$ is also a block matrix in this basis.
Namely,

{ \begin{equation}\label{eqn:Cpp explicit appendix}
C_{pp}^N = \left(\begin{array}{c|cc}
(C_{pp}^N)_{11} + (C_{pp}^N)_{22}-(C_{pp}^N)_{12}-(C_{pp}^N)_{21} & 0 & (C_{pp}^N)_{1n} - (C_{pp}^N)_{2n} \\ \hline
0 & (C_{pp}^N)_{11} + (C_{pp}^N)_{22}+(C_{pp}^N)_{12}+(C_{pp}^N)_{21} & (C_{pp}^N)_{1n} + (C_{pp}^N)_{2n} \\
(C_{pp}^N)_{1n} - (C_{pp}^N)_{2n} & (C_{pp}^N)_{1n} + (C_{pp}^N)_{2n} & (C_{pp}^N)_{nm}
\end{array}\right).
\end{equation}}

The matrix $\Sigma^{-1}$ is the top-left block of $(C_{pp}^N)^{-1}$:
\begin{equation}
\Sigma = \left[(C_{pp}^N)_{11} + (C_{pp}^N)_{22}-(C_{pp}^N)_{12}-(C_{pp}^N)_{21}\right] - \left[(C_{pp}^N)_{1n} - (C_{pp}^N)_{2n}\right](C_{pp}^N)^{-1}_{nm}\left[(C_{pp}^N)_{1m} - (C_{pp}^N)_{2m}\right].
\end{equation}
This, in conjunction with results on the initial correlation sub-matrix of $C_{pp}^N$ pertaining to particles $1$ and $2$ \cite{Konradetal2020,KonradBartelmann2022,Konradetal2022} to the effect that, when $v_{\rm th}=0$,
$(C_{pp}^N)_{11} + (C_{pp}^N)_{22}-(C_{pp}^N)_{12}-(C_{pp}^N)_{21} = O(q^2)$ (with the coefficient determined by the cosmology, which we take to be standard $\Lambda$CDM), implies that indeed $\Sigma \sim Aq^2$ as $q\to 0$. Besides, the components $\mathrm{a}_k$ of $\mathbf{a}$ satisfy
\begin{equation}
\mathrm{a}_k \propto -\left(\Sigma^{-1}\right)_{kn}\left[(C_{pp}^N)_{1n} - (C_{pp}^N)_{2n}\right](C_{pp}^N)^{-1}_{nm}\mathbf{p}_m,
\end{equation}
so, as promised,
\begin{equation}
\Sigma \mathbf{a} \sim \left[(C_{pp}^N)_{1n} - (C_{pp}^N)_{2n}\right] = O(q).
\end{equation}

Additionally, the results of~\cite[\S 3.1.2]{KonradBartelmann2022}, in conjunction with equation~\eqref{eqn:Cpp explicit appendix}, imply that the entire~$C_{pp}^N$ matrix tends to a constant as~$q \to \infty$, which means that~$\Sigma$,~$\mathbf{a}$ and~$B$ become~$O(1)$ in that limit, too.

\end{document}